\documentclass[12pt,preprint]{aastex}

\usepackage{rotating}
\usepackage{subfigure}
\usepackage{lscape}

\newcommand{\ammonia}{\mbox{{\rm NH}$_3$}}
\newcommand{\form}{H$_2$CO}

\newcommand{\myemail}{{\tt jaguirre@sas.upenn.edu}}

\newcommand{\ie}{{i.e.\/}}
\newcommand{\eg}{{e.g.\/}}

\def\msol{\ifmmode {\>M_\odot}\else {$M_\odot$}\fi}

\newcommand{\hii}{H{\sc ii}}

\newcommand{\mum}{\ensuremath{\mu \mathrm{m}}}
\newcommand{\flux}{flux density}

\newcommand{\lon}{\ensuremath{l}}
\newcommand\jyb{Jy~beam$^{-1}$}
\newcommand\mjysr{\ensuremath{{\rm MJy~sr}^{-1}}}
\newcommand\Ros{R10}

\newcommand{\epsi}{\varepsilon}

\def\bgps{BGPS}
\newcommand{\bgpsarea}{170}
\newcommand{\bcamfwhm}{31\arcsec}
\newcommand{\bcamfwhmeff}{33\arcsec}
\newcommand{\ncores}{8454}
\newcommand{\ncoresapprox}{8400}
\newcommand{\bgpsdepthlow}{11}
\newcommand{\bgpsdepthhigh}{53}
\def\spitzer{{\em Spitzer}}

\def\Figure#1#2#3#4{
\begin{figure}[htb]
\epsscale{#4}
\plotone{#1}
\caption{#2}
\label{#3}
\end{figure}
}

\def\FigureTwo#1#2#3#4#5{
\begin{figure}[htb]
\epsscale{#5}
\plottwo{#1}{#2}
\caption{#3}
\label{#4}
\end{figure}
}

\def\Table#1#2#3#4#5{
\begin{deluxetable}{#1}
\tablewidth{0pt}
\tablecaption{#2}
\tablehead{#3}
\startdata
\label{#4}
#5
\enddata
\end{deluxetable}
}

\def\TableNote#1#2#3#4#5#6{
\begin{deluxetable}{#1}
\tablewidth{0pt}
\tablecaption{#2}
\tablehead{#3}
\startdata
\label{#4}
#5
\enddata
#6
\end{deluxetable}
}

\newcommand{\penn}{1}
\newcommand{\casa}{2}
\newcommand{\utexas}{3}
\newcommand{\virginia}{4}
\newcommand{\ucf}{5}
\newcommand{\wisc}{6}
\newcommand{\jpl}{7}
\newcommand{\ubc}{8}
\newcommand{\hawaiihilo}{9}
\newcommand{\hawaiimain}{10}

\slugcomment{DRAFT: \today}

\shorttitle{BGPS}
\shortauthors{Aguirre et al.}

\begin{document}

\newcounter{subfig}
\setcounter{subfig}{1}

\title{The Bolocam Galactic Plane Survey: Survey Description and Data
Reduction}

\author{James E.~Aguirre\altaffilmark{\penn},
        Adam G. Ginsburg\altaffilmark{\casa},
        Miranda K.~Dunham\altaffilmark{\utexas},
	Meredith M. Drosback\altaffilmark{\virginia},
        John Bally\altaffilmark{\casa}, 
	Cara Battersby\altaffilmark{\casa},
	Eric Todd Bradley\altaffilmark{\ucf},
	Claudia Cyganowski\altaffilmark{\wisc},
	Darren Dowell\altaffilmark{\jpl},
	Neal J. Evans II\altaffilmark{\utexas},
        Jason Glenn\altaffilmark{\casa},
        Paul Harvey\altaffilmark{\utexas,\casa},
        Erik Rosolowsky\altaffilmark{\ubc},
        Guy S. Stringfellow\altaffilmark{\casa},
        Josh Walawender\altaffilmark{\hawaiihilo}, and 
        Jonathan P.~Williams\altaffilmark{\hawaiimain}
}

\affil{{$^\penn$}{\it{Department of Physics and Astronomy, University of
      Pennsylvania, Philadelphia, PA }}}
\email{\myemail}

\affil{{$^\casa$}{\it{CASA, University of Colorado, 389-UCB, Boulder, CO 80309}}}

\affil{{$^\utexas$}{\it{Department of Astronomy, University of Texas,
      1 University Station C1400, Austin, TX 78712}}}

\affil{{$^\virginia$}{\it{Department of Astronomy, University of
Virginia, P.O. Box 400325, Charlottesville, VA 22904}}}

\affil{{$^\ucf$}{\it{Department of Physics, University of Central
Florida}}}

\affil{{$^\wisc$}{\it{Department of Astronomy, University of Wisconsin,
       Madison, WI 53706}}}

\affil{{$^\jpl$}{\it{Jet Propulsion Laboratory, California Institute
   of Technology, 4800 Oak Grove Dr., Pasadena, CA 91104}}}

\affil{{$^{\ubc}$}{\it{Department of Physics and Astronomy, University
       of British Columbia, Okanagan }}}

\affil{{$^{\hawaiihilo}$}{\it{Institute for Astronomy, University of
        Hawaii,640 N. Aohoku Pl., Hilo, HI 96720}}}

\affil{{$^{\hawaiimain}$}{\it{Institute for Astronomy, University of
Hawaii, 2680 Woodlawn Drive, Honolulu, HI 96822}}}

\begin{abstract}

We present the Bolocam Galactic Plane Survey (BGPS), a 1.1 mm
continuum survey at 33\arcsec\ effective resolution of \bgpsarea\
square degrees of the Galactic Plane visible from the northern
hemisphere.  The BGPS is one of the first large area, systematic
surveys of the Galactic Plane in the millimeter continuum without
pre-selected targets. The survey is contiguous over the range $-10.5
\le \lon \le 90.5$, $|b| \le 0.5$.  Towards the Cygnus X spiral arm,
the coverage was flared to $|b| \le 1.5$ for $75.5 \le \lon \le 87.5$.
In addition, cross-cuts to $|b| \le 1.5$ were made at $\lon =$~3, 15,
30 and 31.  The total area of this section is 133 square degrees.
With the exception of the increase in latitude, no pre-selection
criteria were applied to the coverage in this region.  In addition to
the contiguous region, four targeted regions in the outer Galaxy were
observed: IC1396 (9 square degrees, $97.5 \le \lon \le 100.5$, $2.25
\le b \le 5.25$), a region towards the Perseus Arm (4 square degrees
centered on $\lon = 111$, $b=0$ near NGC7538), W3/4/5 (18 square
degrees, $132.5 \le l \le 138.5$) and Gem OB1 (6 square degrees,
$187.5 \le l \le 193.5$).

The survey has detected approximately \ncoresapprox\ clumps over the
entire area to a limiting non-uniform 1-$\sigma$ noise level in the
range \bgpsdepthlow\ to \bgpsdepthhigh\ m\jyb\ in the inner Galaxy.
The BGPS source catalog is presented in a companion paper
\citep{rosolowsky10}.  This paper details the survey observations and
data reduction methods for the images.  We discuss in detail the
determination of astrometric and flux density calibration
uncertainties and compare our results to the literature.  Data
processing algorithms that separate astronomical signals from
time-variable atmospheric fluctuations in the data time-stream are
presented.  These algorithms reproduce the structure of the
astronomical sky over a limited range of angular scales and produce
artifacts in the vicinity of bright sources. Based on simulations, we
find that extended emission on scales larger than about 5\farcm9 is
nearly completely attenuated ($>90\%$) and the linear scale at which
the attenuation reaches 50\% is 3\farcm8.  Comparison with other
millimeter-wave data sets implies a possible systematic offset in flux
calibration, for which no cause has been discovered.

This presentation serves as a companion and guide to the public data
release\footnote{\tt
http://irsa.ipac.caltech.edu/Missions/bolocam.html} through NASA's
Infrared Processing and Analysis Center (IPAC) Infrared Science
Archive (IRSA).  New data releases will be provided through IPAC-IRSA
with any future improvements in the reduction.  The BGPS provides a
complementary long-wavelength spectral band for the ongoing ATLASGAL
and {\it Herschel}-SPIRE surveys, and an important database and
context for imminent observations with SCUBA-2 and ALMA.

\end{abstract}

\keywords{
methods: data analysis --
surveys -- 
ISM: clouds --
submillimeter: ISM --
stars: formation --
stars: massive
}

\section{Introduction}
\label{sec:Introduction}

Millimeter-wavelength continuum surveys of the Galactic plane provide
the most efficient way to find molecular clumps that are the likely
formation sites of massive stars and star clusters. The development of
detector arrays has made blind surveys of large areas possible.  Such
surveys bypass the need for selection based on the presence of
embedded stars or star clusters, infrared sources, masers, or radio
continuum emission. In particular, they can locate molecular clumps
before stars form, providing vital information on the initial
conditions of star formation.  These surveys can also provide valuable
constraints on the physical properties of the clumps, especially
masses and mean densities, when combined with distance information.

Galaxy-wide surveys are essential for measuring the impacts of the
environment on clump properties and star formation activity.  Do clump
properties vary with Galactocentric distance, or with location with
respect to spiral arms?  Do they depend on the level of nearby star
formation activity?  Answering these questions in our Galaxy will
provide the essential ``ground truth'' required for the analysis of
distant galaxies where individual clouds and clumps are not resolved
and only galaxy-wide average quantities can be measured.

Surveys of the galaxy in low-lying rotational transitions of the CO
molecule \citep{dame01,jackson06} have identified the locations of
large molecular clouds and traced the transition from atomic gas in
the outer galaxy to molecular gas in the inner galaxy.  Surveys in CO
have detected a strong preference for GMCs to form along spiral arms
\citep{stark06}.  Recent improved measurements of kinematic distances
\citep[e.g.][]{RotationCurve} and accurate distances from VLBI
\citep{reid09} have refined our view of the spiral structure of the
Milky Way.  This new increase in data now allows a more complete
investigation of the relationship between spiral density waves and
star formation.

Locating the early stages of star formation is quite difficult, and
often relies on the serendipitous location of a cold, dark cloud which
appears as an IRDC.  While dark clouds were known from optically
obscured regions, the ``discovery'' of IRDCs from the MSX (Mid-course
Space Experiment) satellite was presented by \citet{egan98} and
\citet{carey98}.  IRDCs have now been extensively cataloged with MSX
\citep{simon06a} and GLIMPSE 8 \mum\ data \citep{peretto09}.  In the
past decade, compelling evidence has emerged which suggests that IRDCs
are the pre-cursors to massive stars, and therefore, stellar clusters
(see, e.g. \citet{rathborne06,rathborne08}).  IRDCs, however, require the
favorable viewing condition of being in front of a bright, mid-IR
background for detection.  Millimeter wave surveys, however, provide
an efficient means of identifying cold, dense dust throughout the
Galaxy independent of the Galactic background, and thus may be seen at
large distances.  Moreover, the millimeter-wave emission is optically
thin, allowing the properties of the cloud to be determined.
Millimeter wave surveys such as BGPS are essential for understanding
massive star and cluster formation on a Galactic scale, from the inner
to the outer Galaxy.

Studies of nearby clouds have demonstrated that star formation is far
from uniform over molecular clouds, but is concentrated in unusually
dense regions called clumps or cores, which fill a small fraction of
the area of most clouds \citep{lada91,enoch07,evans09}.  Following
\citet{williams00} and \citet{mckee07}, we use the term `core' to
refer to a very dense region destined to form an individual star or
small multiple star system and the term `clump' to refer to a region
likely to form a group or cluster of stars. While clumps forming
massive clusters may be even denser than cores forming low mass stars
\citep{mueller02}, in general, clumps are larger and have a lower mean
density \citep{mckee07}.

To delineate clumps from their parent clouds as traced in CO, line
emission from molecules which are high-density tracers, such as
\ammonia, \form, CS, HCN, etc. are extremely useful.  Spectral lines
also provide excellent diagnostics of line-of-sight motions,
temperatures, and densities.  However, the interpretation of these gas
tracers is complicated by variations in tracer abundances caused by
freeze-out onto grains, sublimation caused by star formation, and
subsequent complex chemical processing
\citep{vandishoeck98}. Interpretation is further complicated by
uncertainties in optical depths and excitation conditions and the
impacts of radiation fields and shocks. These complications make the
derivation from such molecular line data of column densities, masses,
and other physical properties of star-forming clumps very difficult.
A further limitation of spectral line surveys is that the tracers of
denser gas tend to be weak, and multi-element receiver arrays have only
a modest number of elements, making blind surveys very time-consuming.
Instruments are beginning to be developed to address this limitation,
such as NRAO's K-band Focal Plane Array\footnote{\tt
https://safe.nrao.edu/wiki/bin/view/Kbandfpa/WebHome} and JCMT's
HARP-B \citep{buckle09}.

In contrast, focal-plane arrays containing hundreds of individual
bolometers sensitive to millimeter and sub-millimeter (sub-mm)
radiation are now available. This advance in technology enables blind
surveys of the Galactic plane, providing a uniform inventory of
massive star-forming and starless clumps.  Catalogs resulting from
these surveys will provide the data base for subsequent observations
in tracers such as hard-to-excite molecular lines and higher
resolution studies with instruments like CARMA and ALMA.  A
significant advantage of continuum observations of dust at long
wavelengths is the low optical depth of the dust \citep{johnstone06},
permitting a straight-forward estimate of the mass of the emitting
region.

Submillimeter observations of Galactic sources have been conducted by
a variety of different groups focusing on small regions.  The
Submillimeter Common-User Bolometer Array (SCUBA; \citet{holland99})
on the 15-m James Clerk Maxwell Telescope mapped many significant
regions in the Galactic plane over its 8 year lifetime (see
\citet{difrancesco08} for a summary), but never completed a contiguous
survey of the Galactic plane.  The Bolocam instrument has been used on
the 10.4-m Caltech Submillimeter Observatory (CSO\footnote{The Caltech
Submillimeter Observatory is supported by the NSF.}) to map nearby
large molecular clouds at 1.1 mm and identify hundreds of cores
\citep{enoch07}.  MAMBO on the 30-m IRAM telescope has been used at
1.2 mm to map both small, nearby cores \citep{kauffmann08} and more
distant regions of more massive star formation
\citep{rathborne06,motte07}.  None of these surveys covered a
significant fraction of the Galactic plane.

This situation has begun to change dramatically.  Recent wide-area
surveys have been carried out in the submillimeter bands 250 \mum\ -
500 \mum\ by the Balloon-borne Large Aperture Submillimeter Telescope
(BLAST) \citep{olmi09,roy10,netterfield09,chapin08}.  This work
provides an unprecedented view of a wide range of star forming
environments and demonstrates the richness of submillimeter datasets,
creating great anticipation for the imminent results from instruments
aboard {\em Herschel}, particularly the SPIRE camera.  From the
ground, the ATLASGAL survey \citep{schuller09}, using the LABOCA
instrument at 870 \micron\ with 19\farcs2 resolution on the APEX
telescope in Chile, has covered 95 square degrees in the Galactic
Plane, $-30 \le l \le 11.5$ and $15 \le l \le 21$ with $|b| \le 1$.
The survey will eventually encompass $-60 \le l \le 60$ with $|b| \le
1.5$.

The BGPS is the first millimeter survey of a substantial fraction of
the Galactic Plane in the northern hemisphere.  The BGPS maps in the
first quadrant overlap with the VLA Galactic Plane Survey (VGPS) in
H{\sc i} and radio continuum \citep{stil06}, the {\it Spitzer}-GLIMPSE
and GLIMPSE-II fields \citep{benjamin03}, MIPSGAL \citep{carey09}, the
BU-FCRAO $^{13}$CO $J=1 \to 0$ Galactic Ring Survey \citep{jackson06},
and ATLASGAL \citep{schuller09}, among others.  The BGPS coverage will
also overlap with the upcoming {\it Herschel} HiGAL \citep{molinari10}
and SCUBA-2/JCMT Plane Surveys \citep{difrancesco08b}.  The \bgps\
provides a long-wavelength dust continuum data point, complementing
the spectral energy distribution provided by other surveys. By
covering the both the Galactic center and anti-center regions with a
single instrument, a consistent comparison over the widest possible
range of Galactocentric radius is possible.  The BGPS gives a view of
massive star and cluster formation throughout the Galaxy, and provides
an important finder chart for future high-resolution observations with
facilities such as CARMA and ALMA.

This paper presents the imaging data from the Bolocam Galactic Plane
Survey (BGPS).  The source catalog is described in a companion paper
(\citet{rosolowsky10}; hereafter \Ros).  The outline of this paper is
as follows. Section \ref{sec:Observations} describes the instrument
and the observations.  The data analysis is described in Sections
\ref{sec:Astrometry} (astrometry), \ref{sec:Mapping} (the map-making
algorithm), and \ref{sec:Calibration} (flux calibration).  Key
discussions are in Section \ref{sec:Photometry}, which explains the
effect of the data processing on the accurate recovery of extended
structure, and in Section \ref{sec:FluxComparison}, which compares the
BGPS flux with other surveys, including a discussion of a possible
systematic offset.  The final image data products and the public
release are described in Section \ref{sec:FinalMaps}.  We conclude in
Section \ref{sec:Discussion} with a brief discussion of the broad
features of the BGPS in comparison with other surveys.

\section{Observations}
\label{sec:Observations}

We used Bolocam\footnote{{\tt http://www.cso.caltech.edu/bolocam}} to
survey more than \bgpsarea\ square degrees of the northern Galactic
Plane.  Bolocam is the facility 144-element bolometer array camera
mounted at the Cassegrain focus of the 10.4-m mirror of the CSO on the
summit of Mauna Kea.  We used the filter configuration with a band
center of 271.1~GHz (hereafter 1.1~mm) and fractional bandwidth
$\Delta \nu/\nu = 0.17$ (46 GHz).  The detectors are silicon nitride
micromesh absorbers with NTD germanium thermistors, operated at a
temperature of $\sim$250 mK.  The Bolocam array field-of-view is
7\farcm5, with individual detectors having nearly Gaussian beams of
\bcamfwhm\ FWHM.  The spacing between the individual detector beams on
the sky is 38\arcsec, so the focal plane is not instantaneously fully
sampled.  The Bolocam instrument is described in greater detail in
\citet{glenn03}.

The maps presented here were acquired during six separate observing
sessions at the CSO over the course of two years between June 2005 and
September 2007. The observing epochs are given in Table
\ref{tab:Observing}.  Bolocam observations were typically scheduled
when $\tau_{225} > 0.06$.  Between observing epochs, Bolocam was
removed from its mount at the re-imaged Cassegrain focus and stored
warm.  Thus the flux calibration and pointing model were re-computed
for each epoch to allow for variations in the instrument and optics.
The final data products were aligned to the well-constrained pointing
model created from the Epoch V and VI data (Section
\ref{sec:Astrometry}).  We found that the flux calibration did not in
fact differ significantly between epochs; see Section
\ref{sec:FluxCalibration} for more details.

Figure \ref{fig:Coverage} shows the coverage of the \bgps.  The survey
is contiguous over the range $-10.5 \le \lon \le 90.5, |b| \le 0.5$.
Towards the Cygnus X spiral arm, the coverage was flared to $|b| \le
1.5$ for $75.5 \le \lon \le 87.5$.  In addition, cross-cuts to $|b|
\le 1.5$ were made at $\lon = 3, 15, 30$ and 31.  The total area of
this section is 133 square degrees.  With the exception of the
increase in latitude, no pre-selection criteria were applied to the
coverage in this region.  In addition to the contiguous region, four
targeted regions in the outer Galaxy were observed: IC1396 (9 square
degrees, $97.5 \le \lon \le 100.5$, $2.25 \le b \le 5.25$), a region
towards the Perseus Arm (4 square degrees centered on $\lon = 111$,
$b=0$ near NGC7538), W3/4/5 (18 square degrees, $132.5 \le l \le
138.5$) and Gem OB1 (6 square degrees, $187.5 \le l \le 193.5$).  The
total area of good coverage for the BGPS is \bgpsarea\ square degrees.

Our basic observing strategy was to raster scan a field by moving the
primary mirror of the CSO.  This was done to modulate the
astrophysical signal faster than fluctuations in atmospheric opacity.
Each field was scanned in a pattern with alternating rasters along
lines of constant $l$, followed by a series of rasters along lines of
constant $b$.  Fields were observed several times with such raster
scans to improve signal-to-noise (S/N).  Thus each point in the final
map of a field was observed multiple times with the array moving in
different directions with respect to the astrophysical emission,
helping to separate it from atmospheric emission (see Section
\ref{sec:Mapping} for further details on this process).

In Epochs I and II, the fundamental observing block was a region
$1\arcdeg \times 1\arcdeg$, covered by 23 rasters along each of \lon\
or $b$.  Starting with Epoch III, the fundamental block was changed to
$3\arcdeg \times 1\arcdeg$ for increased mapping efficiency (i.e.,
less time spent in turnarounds at the end of a raster).  Each
$3\arcdeg \times 1\arcdeg$ block was covered using either 23 rasters
along lines of constant $b$, or 67 rasters along lines of constant
$l$.  In both cases, the spacing between adjacent rasters was
162\arcsec.  The total time for observing the entire block was 39 (48)
minutes for rasters along constant $b$ ($l$).  The data were
electronically sampled at 10 Hz along the scan direction, slightly
higher than the Nyquist rate for the scan speed of 120\arcsec\
s$^{-1}$.

These fundamental observing blocks were stitched together to make 29
large, contiguous maps.  These images define the ``fields'' used for
comparison.  Each ``field'' includes all observations covering a
region between 1 and 6 square degrees.

Because the Bolocam detector beams do not fully overlap on the sky, if
the array is scanned along detector rows, gaps will be left in the
resulting map.  To ameliorate this, we began using in Epoch II a field
rotator to adjust the rotation angle of the array with respect to the
scan direction so that as the scan proceeded the full extent of the
FOV in the direction orthogonal to the scan was sampled by at least
one detector.  This is shown in Figure \ref{fig:Array}.  Achieving a
uniform sampling along the scan-orthogonal direction is further
complicated by a number of missing bolometers.  A simulation was
performed to determine the optimal angle to rotate the array with
respect to the scan direction to account for both the effects of the
beam spacing and missing bolometers.  During the turnaround following
a raster, the field rotator is adjusted to this optimum angle for the
subsequent raster.  Without the field rotator, the coverage shows
variations of 100\% from pixel to pixel in a single raster of a field.
With the rotator, this is reduced to $\sim40\%$.

\section{Astrometry}
\label{sec:Astrometry}

\subsection{Absolute Reference Sources and Pointing Model}
\label{sec:PointingModel}

We constructed an absolute reference system for the BGPS by observing
bright quasars and blazars near the Galactic Plane with small raster
maps.  The sources were chosen from the SMA Submillimeter Calibrator
List\footnote{{\tt http://sma1.sma.hawaii.edu/callist/callist.html}},
since they have positions reliably determined from interferometric
measurements and are point sources at the scale of the Bolocam beam.
The distribution of absolute pointing sources over the sky is shown in
Figure \ref{fig:PointingCalibrators}.

Pointing observations were performed approximately once every two
hours over each night.  The elevation offsets showed a deviation from
zero which was empirically well-modeled by a quadratic function of
altitude.  No systematic deviation was observed for the azimuth
offsets.  The derived model is shown in Figure
\ref{fig:PointingModel}.  An RMS scatter of $\sim 6\arcsec$ for the
model over the entire observed range was achieved, with somewhat worse
scatter in altitude than in azimuth.

\subsection{Pixel Positions in the FOV}

In addition to the model for the pointing center, it is necessary to
empirically determine the actual projected pattern of the array on the
sky, and measure the rotation angle between the focal plane and the sky
coordinate system.  This is done by making observations which track
all detectors across a bright source (e.g. a planet) and making maps
from each bolometer individually.

In our first data release, the projected positions of the detectors on
the sky are made using the nominal positions without correction for
optical distortion.  This results in a slightly reduced effective
resolution (\bcamfwhmeff\ instead of \bcamfwhm).  The effective beam
size was measured from fully-sampled beam-mapping observations of
planets, and is a good description of the blurring of the beam for the
portions of a field observed by all the detectors.  Within about one
FOV of the field edges, however, only a subset of the array observes
the field, and so the effective beam size varies slightly at these
positions, and may be asymmetric.  Because this distortion is
asymmetric, it can also result in a small pointing offset.  However,
this effect is strongly mitigated by the low number of hits at field
edges: this data is generally flagged out.  The residual offset is
found to be within the overall pointing model error, but caution
should be used when measuring source locations at the field edges in
the first data release.  Future releases will have the distortion
correction applied, removing this effect and improving the effective
resolution.

\subsection{Relative Alignment and Mosaicking}

A master pointing field was constructed from observations taken in the
epoch with the best-constrained pointing model for that field (either
Epoch V or VI) for every field observed by the \bgps.  Relative
alignment between the master and other observations of the same field
was performed by finding the peak of the cross-correlation between
images. The cross-correlation peak was fit with a Gaussian profile and
the difference between the Gaussian peak and the image center was used
as the pointing offset.  The offset may be determined accurately to
within the error in finding the peak of the Gaussian, typically
$<1\arcsec$.  To create the final image, all observations of a field
were merged into a single timestream with these pointing offsets
applied.

This method of alignment makes use of all the information available in
the maps, and avoids the ambiguities inherent in using extracted
sources to align fields, as the BGPS sources are rarely point-like.
It further avoids the slightly larger effective beam and loss of peak
flux density which would result if the maps of individual observations
were combined using the pointing model alone, where each individual
observation would be co-added with the 6\arcsec\ RMS uncertainty of
the model.  In some fields with few sources, particularly those in the
range \lon=65 to 75, there was not enough signal to acquire a pointing
offset using cross-correlation.  In these fields, the pointing model
was used directly, and so the above mentioned effects affect these
fields.

\subsection{Comparison to the SCUBA Legacy Catalog}
\label{sec:SCUBAPointingComparison}

All SCUBA 850 and 450 \mum\ data has been re-processed in a uniform
manner and made publicly available as the SCUBA Legacy Catalog
\citep[][hereafter SLC]{difrancesco08}.  These maps allow a
cross-check of the accuracy of the BGPS pointing model.  We have
applied the same cross-correlation procedure used to obtain the
relative alignment of BGPS observations to compare the BGPS to the SLC
850 \mum\ images.  \citet{difrancesco08} claim a typical pointing
accuracy for SCUBA of $\sim3\arcsec$, with offsets as large as
$\sim6\arcsec$ occurring occasionally.  No further adjustment to the
nominal JCMT pointing model was performed in producing the SLC maps,
and thus we assume these numbers are typical.

There are a number of differences in the Bolocam and SLC maps which
might lead to difficulties in comparing their astrometry.  In
particular, different scan strategies were used (SCUBA observations
were typically taken in jiggle-map mode, which tends to remove
extended structure), and of course there is the difference in
wavelength and beam size between the instruments.  Nevertheless, the
by-eye morphological comparison between SLC 850 \mum\ and BGPS sources
is generally excellent, as exemplified in Figure
\ref{fig:SCUBAPointingComparison-a}.  A cross-correlation technique
was used in order to minimize these limitations, since it does not
require the establishment of a single position for an object.

The SLC has the advantage of having observations over a large portion
of the Plane, allowing the pointing to be checked (in spots) for
deviations as a function of position.  To test this, 21 of the 29 BGPS
fields were used which overlapped with the SLC.  For these fields, we
find $\Delta l = 1\farcs8 \pm 1\farcs2$, $\Delta b = -0\farcs4 \pm
0\farcs8$, nearly consistent with zero, and an RMS dispersion about
the mean BGPS position of 6\farcs2, consistent with the errors derived
for the Bolocam pointing model.  The offset for each field is shown in
Figure \ref{fig:SCUBAPointingComparison-b}.  We have {\it not}
corrected the Bolocam pointing based on the SLC or any external data
set, but have produced our final maps with the internally derived
model described in Section \ref{sec:PointingModel}.

\section{Mapping Algorithm}
\label{sec:Mapping}

At wavelengths near 1 mm, the emission from the atmosphere dominates
any astrophysical emission.  For an atmospheric zenith optical depth
of $\tau$ = 0.1, the sky brightness temperature is around 30 to 40
Kelvin at typical zenith angles.  This background loading determines
the instrument calibration (see Section \ref{sec:FluxCalibration}),
but the primary time-variable signal is due to few percent
fluctuations in atmospheric opacity.  The conversion $J$ from \jyb\
to Kelvin for Bolocam is
\begin{equation} 
\label{eq:JyPerK}
J = 1\times10^{26} \frac{2k}{A_{\rm eff}} = 
58 \, {\rm \frac{{\rm Jy\,beam}^{-1}}{{\rm K}}}
\end{equation}
where $k$ is Boltzmann's constant and $A_{\rm eff}$ is the effective
collecting area of the telescope.  Thus atmosphere fluctuations are
10's of \jyb, as compared to the typical brightness of Galactic
sources, which lie in the range of 0.1 - 1 \jyb.  The essential
signal-processing problem to be solved by the mapping algorithm, then,
is the estimation of the astrophysical emission in the presence of the
fluctuating atmosphere signal, without {\it a priori} knowledge of
either.

For this work, a custom pipeline was developed to address issues
specific to the BGPS.  It is based on, but significantly improves, the
method of \citet{enoch06} for generating maps and characterizing their
properties, and is similar in many respects to the methods used by
other bolometer cameras, {e.g.}, \citet{cotton09} (MUSTANG/GBT) or
\citet{kovacs08} (SHARC-2/CSO).  It also incorporates some of the
technical developments of \citet{sayers09} developed for Bolocam data
at 2.1 mm.  A key element of the reduction is the iterative estimation
of the atmospheric fluctuations and the astrophysical signal.  The
atmospheric model is developed from a set of principal components of
the bolometer signal time series under the assumption that the bulk of
the correlated signal is atmospheric.  After subtraction of the
atmosphere model, the astrophysical emission is estimated, then
subtracted, and the process is repeated.

\subsection{Algorithm}

We assume the raw timestream data $d$ for each bolometer (indexed by
$i$) at discrete time points (indexed by $t$) can be written as
\begin{equation}
\label{eq:TimeDomainModel}
d_{it} = s_{it}+a_{it}+c_{it}+e_{it}+p_{it}+\epsi_{it}
\end{equation}
where $s$ is the astrophysical signal, $a$ is the median atmospheric
fluctuation seen by all detectors, $c$ are terms correlated between
bolometers that are distinct from $a$, $e$ are non-random signals due
to the instrument itself, $p$ are long timescale fluctuations not
modeled by $a$ or $c$, and $\epsi$ is irreducible Gaussian noise due
to photon fluctuations and detector noise.  (Each of these terms is
discussed in greater detail in the following sections.)

In the presence of purely Gaussian noise, the maximum likelihood map
$m$ can be obtained by minimizing the goodness-of-fit statistic
\begin{equation}
\label{eq:ChiSquared}
\chi^2 = (d - Am)^T W (d - Am)
\end{equation}
Here $d$ is the concatenated data from all bolometers (of dimension
$N_{tod}$=number time samples of all detectors).  $A$ provides the
mapping between a given sample in the time-ordered data and the pixel
in the sky map to which it corresponds.  $A$ is a matrix of dimension
$N_{tod} \times N_{pix} =$number of map pixels.  Each element of $A$
is either 1 or 0.  Note that we are considering the data $d$ and the
map $m$ as vectors (the mapping to a two-dimensional map being
implicit in the matrix $A$).  $W$ is an $N_{tod} \times N_{tod}$
matrix which is the inverse of the covariance matrix $N$ of the time
domain noise $\epsi$,
\begin{equation}
\label{eq:NoiseCovariance}
N_{tt'} = \langle \epsi_t \epsi_{t'} \rangle,
\end{equation}
where $\langle \rangle$ denotes an ensemble average over many
realizations of the random time\-stream noise process $\epsi_t$.
Further detail about the way in which $W$ was estimated is given in
Section \ref{sec:Noise}.

The result of the $\chi^2$ minimization is
\begin{equation}
\label{eq:LSMap}
m = (A^T W A) A^T W d
\end{equation}
Thus, the mapping from data to map is a linear operator, $M =(A^T W A)
A^T W$.  For compactness, Equation \ref{eq:LSMap} will be written as
$M[d] = m$.  Note that the mapping operator is not invertible.
However, given a map $m$ and an observing matrix $A$, there exists a
linear operation which makes predictions about the observed
timestreams, namely $d = A m$; this will be denoted $T[m] = d$. 

Because the model $Am$ in Equation \ref{eq:ChiSquared} only includes
the celestial contribution to the observed data $d$, direct use of
Equation \ref{eq:LSMap} will produce a map containing artifacts due to
the unmodeled components $a$, $c$, $e$, and $p$.  One approach would
be to include templates for these terms as additional rows in $A$ and
proceed with a simultaneous fit.  In general, however, the correct
forms to use are not known {\it a priori}.  The goal, then, is to
produce a time series for each bolometer which as closely as possible
approximates $s + \epsi$ so that we may produce the best estimate of
the astrophysical signal $m = M[s+\epsi] = S + N$.  We proceed
iteratively, estimating each term in Equation \ref{eq:TimeDomainModel}
in the order of its relative strength.  In the following, we denote
the best estimate of a time series $x$ in the $n^{th}$ iteration as
$\tilde{x}^{(n)}$.

In the first iteration, we estimate the largest signal, $a$, according
to the method in Section \ref{sec:AtmosphereModel}, and assume
$\tilde{s}^{(0)}_{it}=\tilde{c}^{(0)}_{it}=\tilde{p}^{(0)}_{it}=0$.
We then form
\begin{equation}
\label{eq:SignalEstimate}
\xi^{(n)}_{it} = d_{it} - \tilde{a}^{(n)}_{it} - \tilde{c}^{(n)}_{it}
- \tilde{e}^{(n)}_{it} - \tilde{p}^{(n)}_{it} \approx s_{it} +
\epsi_{it}
\end{equation}
This is the best estimate of signal plus irreducible noise at
iteration step $n$, and is made into a map
\begin{equation}
M[\xi^{(n)}_{it}] = \tilde{m}^{(n)}
\end{equation}
The current best map $\tilde{m}^{(n)}$ is then deconvolved to provide
a relatively low-noise, smooth map from which to generate a timestream
\begin{equation}
\label{eq:MapToTime}
T[{\cal D}[\tilde{m}^{(n)}]] = \tilde{s}^{(n)}_{it}
\end{equation}
where ${\cal D}$ represents the deconvolution operation.  At this
stage, we can subtract both $\tilde{a}^{(n)}_{it}$ and
$\tilde{s}^{(n)}_{it}$ from the original data, and estimate
$\tilde{c}^{(n)}_{it}$, $\tilde{e}^{(n)}_{it}$, and
$\tilde{p}^{(n)}_{it}$.  The iterative process begins again with
Equation \ref{eq:SignalEstimate}.  An example time series showing the
successive removal of these models is shown in Figure
\ref{fig:IterativeMapping}.  We discuss the estimation of each of the
various terms $a$, $c$, $e$, and $p$ in the following sections.

It is useful at each step to produce a ``residual map''
\begin{equation}
\label{eq:ResidualMap}
E = M[d_{it} - \tilde{s}^{(n)}_{it} - \tilde{a}^{(n)}_{it} 
- \tilde{c}^{(n)}_{it}
- \tilde{e}^{(n)}_{it}
- \tilde{p}^{(n)}_{it} ]
\end{equation}
This residual map serves as a visualization of the progress of the
iterative method, since in a perfect process it would be map of
$\epsi$.  

\subsection{Atmosphere Fluctuation Model}
\label{sec:AtmosphereModel}

Since atmosphere fluctuations are the largest signal in the raw data,
they are modeled first.  By construction, the Bolocam beams as they
pass through the dominant layer of atmospheric water vapor a few km
above the telescope are still highly overlapped, and therefore sample
a nearly identical region of atmosphere.  The simplest model for the
atmosphere fluctuations makes use of this fact by assuming that the
largest correlated signals between detectors are due to to the
atmosphere fluctuations; this is the term $a$ in Equation
\ref{eq:TimeDomainModel}.  We construct a model of the atmosphere
fluctuations which are seen in common by all detectors, which we
denote the ``median atmosphere model''.  In each iteration $n$, we
construct the non-atmosphere-fluctuation part of the time series as
\begin{equation}
\tilde{a}^{(n)}_{it} = 
d_{it} - \tilde{s}^{(n)}_{it} - 
\tilde{c}^{(n)}_{it} - \tilde{e}^{(n)}_{it} - \tilde{p}^{(n)}_{it}
\end{equation}
The median atmosphere model is then simply
\begin{equation}
\label{eq:MeanAtmosphereModel}
\tilde{a}^{(n)}_t = 
{\rm median}_i[{\tilde{a}^{(n)}_{it}}]
\end{equation}
Each bolometer's timestream is then fit to the median atmosphere model,
\ie, we find $r_i^{(n)}$ such that $\tilde{a}^{(n)}_{it}$ most closely
approximates $r_i^{(n)} \tilde{a}^{(n)}_t$.  This defines the
relative gains of the detectors as
\begin{equation}
\label{eq:RelativeGains}
r_i^{(n)} = \frac{\sum_{t'} \tilde{a}^{(n)}_{it'} \tilde{a}^{(n)}_{t'}}
{\sum_{t'} (\tilde{a}^{(n)}_{t'})^2}
\end{equation}
The $r_i$ converge fairly rapidly.  (See the discussion in Section
\ref{sec:FluxCalibration}.)  

The simple common mode model of Equation \ref{eq:MeanAtmosphereModel}
does not remove all of the signal correlated at zero time lag between
the bolometers.  Some of this remaining correlated signal may also be
atmosphere.  \citet{sayers10} found for the 2.1 mm Bolocam data that
the subtraction of a second order spatial polynomial across the array
accounted for nearly all of their residual noise; however, the 1.1 mm
data seem to have additional correlated components that are not
completely described by such a model.  Therefore, to further remove
the atmosphere (and any other correlated instrument noise) without a
detailed physical model, we use a principal component analysis (PCA)
as given in \citet{laurent05}.  While this method was developed for
point source detection in deep, high-redshift surveys, the technique
works equally well for this survey, with the caveat that it removes
large-scale structure on the size of the array FOV.  This effect is
mitigated by the iterative process (the \citet{laurent05} analysis did
not iterate).  Figure \ref{fig:PCA_Graphical} shows an example of the
modes removed by the PCA model.  These are a graphical visualization
of the terms $c$ in Equation \ref{eq:TimeDomainModel}.

\subsection{Instrument Error Signals}
\label{sec:InstrumentErrors}

Most of the instrument error signals have characteristic features
which aid in their removal.  In particular, unlike the atmosphere
emission and correlated signals described above, these signals are
{\it not} correlated between detector channels.  The error signals
include the following:

\begin{enumerate}

\item Pickup from the 60 Hz AC power.  This appears as narrow lines in
the power spectral density (PSD) of the data.  The second harmonic of
60 Hz is aliased via beating against the 130 Hz bolometer bias
frequency into 10 Hz with sidebands split at $\sim1$ Hz.  It is
removed by replacing components at these frequencies in the Fourier
transform with Gaussian noise matched to the local mean amplitude.

\item Spikes in voltage due to cosmic ray strikes on the bolometers
(``glitches'').  An example of a glitch in the time series is shown in
Figure \ref{fig:Glitches}, along with the extracted distribution.

\item Microphonic pickup due to vibrations of the receiver.  The
microphonic effect is due to a change in the capacitive coupling of
the readout wires to the circuit ground, which is converted by the
high impedance of the detectors into a measurable signal, much like a
condenser microphone.  The most noticeable microphonics occur at the
end of each scan when the telescope is turning around and the field
rotator is adjusting.  This leads to broad spikes in the time series,
whose long decay must be removed from the data, particularly during
the beginnings of scans.

\end{enumerate}

Fortunately, most of the AC powerline pick-up occurs at frequencies
where there is no astrophysical signal, given the beam size and scan
speed (c.f. Figure 1 of \citet{sayers09}).  Thus this error is dealt
with by first notch filtering at the line frequencies and then
low-pass filtering the data.  Because of the low correlation with
astrophysical signal, this step is performed only once and is not
iterated.

Both glitches and microphonic pickup from the scan turnarounds are
degenerate with astrophysical signal and must be estimated as part of
the iterative process.  Glitches are identified as large excursions
from the RMS level after subtraction of the best atmosphere and bright
source model, and the data there is excluded in
subsequent maps.  The turnaround microphonics are explicitly modeled
as decaying exponentials at the beginnings and ends of scans.
Residuals from this model may also be removed by the PCA cleaning.

After iteratively subtracting the astrophysical model, each scan also
has a fifth order polynomial fit in time removed to deal with the
longest time scale modes ($p$ in Equation \ref{eq:TimeDomainModel}).
These modes are not correlated between detectors.  They correspond to
spatial scales much larger than those to which Bolocam remains
sensitive after subtraction of the PCA model, so their subtraction
produces maps with fewer ``stripes'', without appreciably degrading
the signal.

\subsection{Data Flagging}

Due to the large volume of data generated by the survey, it was
necessary to develop new tools to quickly visualize the data and
ensure data quality.  A common tool in radio astronomy is the
so-called ``waterfall'' plot, which is an image with frequency on the
x-axis, time on the y, and an intensity proportional to the
interferometer visibility amplitude or phase.  We have used a variant
on this to visualize the Bolocam data by displaying bolometer number
(related to the bolometer position in the focal plane) on the x-axis
and time on the y, with the intensity given by the bolometer's
response (in Jy) at that time.  An example of this is shown in Figure
\ref{fig:Flagger}.  Anomalies such as the glitch shown can be detected
manually (with practice) and interactively flagged out.

An automated flagger was also created that flags out outlier data on a
per-map-pixel basis.  In order to make a robust measurement of the
variance of the fluxes assigned to each pixel, we used the median
average deviation (MAD) over the data in the pixel and rejected high
and low outliers at the 3-sigma level.  Pixels with too little data to
compute a deviation, i.e. those with $<3$ data points, were also
flagged out - these scan-edge pixels are the dominant contribution to
the total number of flagged data points.  The fraction of the data
flagged by all methods was about 0.08\%.

\subsection{Creation of the Astrophysical Model}

The timestream data is made into a spatial map using the pointing data
corresponding to each time point for each bolometer.  The data is
weighted by inverse variance across a single scan and then drizzled
into a map with $7\farcs2$ pixels using a nearest-neighbor algorithm.
The nearest-neighbor matching allows the map to be returned to a
timestream in the same manner, but with the S/N improved by averaging
over all hits on a given pixel.  It has the disadvantage that it
accentuates the unevenness of the cross-scan sampling, since each
timestream point's value is assigned to an area much smaller than the
beam.  In principle this can be addressed by various gridding
algorithms, and may be a part of future releases of the BGPS data.

The resulting map is then subjected to a maximum-entropy clean
algorithm \citep{hollis92} using a specified kernel to produce a
deconvolved image which is mapped back into the timestream to be
subtracted (Equation \ref{eq:MapToTime}).  This has several advantages
over using the original map directly, including rejecting artifacts
smaller than the kernel and decreasing the noise of the resulting
timestream.  It is also better than a ``threshold'' method, which
tends to produce discontinuities and include noise outliers.

In some fields, flux from some sources spread out over the course of
the iterative process using certain kernel sizes.  The cause of this
artifact is not yet well-understood, but using a different kernel size
was an effective workaround and did little to change the properties of
the final map.  The standard kernel size was a 14\farcs4 FWHM
Gaussian; a 21\farcs6 kernel was used when artifacts were encountered.
Simulations indicate that this change in kernel has a negligible
effect on the flux of the source; see Section
\ref{sec:FluxCalibration}.

We also investigated whether the choice of deconvolution kernel
produced a noticeable effect on the recovered flux.  Figure
\ref{fig:Deconvolution} demonstrates the effects of using various
kernel sizes.  The best results, as determined by the flux recovered
in simulations and examination of the residual maps, were produced by
a kernel smaller than the beam but larger than the pixel size used.

\subsection{Noise Estimation and Residuals}
\label{sec:Noise}

The weighting of the time stream data used to produce a map is given
in Equation \ref{eq:NoiseCovariance}.  If the noise were uncorrelated
between timestream points, then $N_{tt'}$ would be diagonal.  If the
noise {\it is} correlated between time samples, but is stationary,
\ie, the noise properties are time-translation invariant
\begin{equation}
N_{tt'} = N(|t-t'|),
\end{equation}
and the weighting can be done using the Fourier transform of the
correlation, which is diagonal:
\begin{equation}
\tilde{N}_{\omega \omega'} = 
\langle |\tilde{\epsi}(\omega)|^2 \rangle \delta(\omega - \omega'),
\end{equation}
where $|\tilde{\epsi}(\omega)|^2$ is the power spectrum of the noise
$\epsi_t$.  
Finally, if the noise is Gaussian, then the covariance matrix
provides all the necessary statistical information about the noise.

In practice, few of these assumptions hold for the \bgps\ data: the
noise is not stationary, due to atmosphere variations, and the
unsubtracted astrophysical signal causes the residuals from which we
estimate the noise to be non-Gaussian.  Further, the true remaining
correlations between detectors due to unsubtracted atmosphere are
difficult to estimate.  Our approach, then, is to use a non-optimal
but reasonable weighting, estimating the noise from the integral over
the power spectrum of the residual timestream for each bolometer and
each raster separately.  This accounts for the non-stationarity but
ignores the correlations, which are small after the PCA subtraction.

The residual maps $E$ (Equation \ref{eq:ResidualMap}) form the basis
for the noise estimation in the map domain, and also provide a way of
estimating the systematic error resulting from imperfect subtraction
of bright sources.  These maps are produced for each region from the
data residuals after removing the terms $s$, $a$, $c$, $p$, and $e$ in
Equation \ref{eq:TimeDomainModel}. An example of a residual map is
shown in Figure \ref{fig:Deconvolution}.  In an ideal case, the
residual map $E$ represents a map of a realization of the underlying
irreducible noise $\epsi$.  However, the imperfect estimation of the
signal results in ``ghosts'' of bright sources.  These features are a
guide to the number of iterations necessary and an estimate of the
remaining systematic error.  In constructing the \bgps\ catalog, \Ros\
found that smoothed versions of the residual maps provide a reliable
means of determining the pixel-to-pixel error which accounts for the
local variations in the noise, including those due to such artifacts
(c.f. \Ros\ Figure 2).  Because observing conditions varied widely
during the survey, the RMS noise level varies over the BGPS.  However,
the noise within a given $1\arcdeg \times 1\arcdeg$ field is fairly
uniform in the absence of very bright sources.  The noise level varies
between \bgpsdepthlow\ and \bgpsdepthhigh\ m\jyb\ in the inner Galaxy.
We show the variation of the depth as a function of Galactic longitude
in Figure \ref{fig:NoiseVsLongitude}.  This noise level interacts with
the source density to produce the completeness of source extraction.
The variation of completeness as a function of Galactic longitude,
determined by simulations of the source extraction, is shown in \Ros\
Figure 9.

\section{Calibration and Photometry}
\label{sec:Calibration}

\subsection{Flux Density Calibration}
\label{sec:FluxCalibration}

The absolute flux calibration is derived from observations of Mars and
Uranus (the ``primary calibrators'').  The millimeter-wave flux of
these planets is known to $\sim 5\%$ \citep{orton86,griffin93}.  The
fluxes from these models were extracted using the JCMT's {\tt
FLUXES}\footnote{{\tt
http://docs.jach.hawaii.edu/star/sun213.htx/sun213.html}} program with
a central frequency 271.1 GHz and bandwidth 46.0 GHz (Table
\ref{tab:ColorCorrections}).  

Calibrations at (sub)millimeter wavelengths are strongly affected by
atmospheric opacity corrections.  Further, the detector responsivity
of Bolocam's bolometers is a non-linear function of the mean
atmospheric loading. To address both of these problems and relate
observations of the primary calibrators to observations of the BGPS
fields, we make use of the following relation.  The calibration ${\cal
C}$, referenced to the detectors, is given by
\begin{equation}
\label{eq:Calibration}
{\cal C} \, \left[\mathrm{\frac{V}{Jy}}\right] = 
{\cal R}(\tau) \eta A \exp{(-\tau)} \Delta\nu
\end{equation}
where ${\cal R}$ is the bolometer responsivity ([V/W]), $\eta$ is the
system optical efficiency, $A$ is the effective telescope collecting
area, $\Delta \nu$ is the bandwidth, and $\tau$ is the line-of-sight,
in-band atmosphere opacity.  Under the assumption that the only power
variation on the detectors is due to the power from the atmosphere
(i.e.  that astronomical sources are faint relative to the
atmosphere), which may be parametrized by $\tau$, ${\cal C}$ is a
single-valued function of $\tau$.  We have used the measured potential
difference across the detector thermistor $V_{DC}$ as a proxy for
$\tau$, since the bolometer resistance is a single-valued function of
loading.  This quantity is monitored continuously for all
observations.  Note that this calibration curve folds in both the
effects of changing atmospheric transmission and the changes in the
detector response with optical loading.

The calibrator field observations were obtained and reduced
differently from the science fields.  Calibrators were observed in
$\sim8\arcmin$ radius fields with two sets of observations, scanned in
orthogonal directions.  In these small fields, the iterative process
tended to diverge with more iterations because of instabilities in the
deconvolution procedure.  Therefore, the calibration curve was
determined from non-iteratively-mapped calibrator maps.  A future work
will explore applying a more consistent application of the processing
to both calibrator and science fields.

A fit to the observed values ($V_{DC}$ vs. ${\cal C}$) for the primary
calibrators was performed for each epoch separately.  The
agreement between epochs was good, so a single combined calibration
was used for all data.  The calibration curve is shown in Figure
\ref{fig:CalibrationCurves}.  The resulting error on the calibration
curve fit is less than $8\%$ (statistical) over the observed range of
$\tau$.  

The above flux density calibration only accounts for the average
calibration of bolometer Volts to Janskys.  It is also necessary to
account for the variation of bolometer response across the focal plane
(often referred to as ``flat-fielding'').  We do this by monitoring
the response to the atmosphere emission in all bolometers
(astrophysical signal is negligible in comparison to the sky except
towards SgrB2).  Being in the near field, the atmosphere is common to
all detectors, and thus this serves as a common relative reference to
calibrate out variation in the individual detector responsivities and
optical efficiencies.  (For more detail regarding the properties of
atmospheric noise above Mauna Kea as inferred from the Bolocam data,
see \citet{sayers10}.)  The change in relative response with loading
for a typical detector is shown in Figure \ref{fig:CalibrationCurves}.
The actual application of the method to the time series of the
bolometer data is shown in Figure \ref{fig:Flatfield}, where it is
clear that correlation of the timestreams is indeed improved by the
application of a single multiplicative factor (the $r_i$ of Equation
\ref{eq:RelativeGains}).

\subsection{Point Spread Function, Aperture Corrections,
and Surface Brightness Calibration}

The Bolocam point spread function (PSF, or ``beam'') is measured using
the planets Uranus and Neptune, which are nearly point sources for
Bolocam.  By stacking all observations of planets, we obtain a high
S/N profile of the beam; this is shown in Figure \ref{fig:PSF}.  The
main lobe is fit to a Gaussian profile.  This yields the effective
beam FWHM of \bcamfwhmeff, corresponding to a sigma of 14\farcs2 and
equivalent tophat radius of 19\farcs8.  The corresponding solid angle
is $2.9\times10^{-8}$ steradians.

The calibration from \jyb\ into surface brightness (MJy sr$^{-1}$) is
obtained once the beam area is determined.  For the BGPS, the
conversion factor is
\begin{equation}
1 ~\rm{Jy~beam}^{-1} = 34.5~\rm{MJy~ster}^{-1} = 0.042~{\rm Jy~pix}^{-1}
\end{equation}
for the standard 7\farcs2 pixel size.  The uncertainty in the mean
beam diameter is $3\%$, leading to a corresponding $6\%$ uncertainty
in the beam area (and in the surface brightness calibration).  This
factor is in addition to the uncertainty in the best fit $V_{DC}$
versus ${\cal C}$ curve.

Due to the sidelobes of the beam, aperture photometry may not capture
all of the flux from a point source if the aperture is too small.
\Ros\ used a standard set of apertures, and the multiplicative
correction factors to be applied to fluxes obtained from these
apertures for a point source are given in Table
\ref{tab:ApertureCorrections}.  The correction is only significant for
apertures $\le80\arcsec$, beyond which effectively all of the flux is
included.  For extended objects, the situation is more complicated
because of the effects of the mapping algorithm.  This is discussed
further in the next section.

\subsection{Effects of the Data Reduction Procedures on Photometry}
\label{sec:Photometry}

To do accurate and meaningful photometry on the Bolocam maps requires
a good estimate of the noise (Section \ref{sec:Noise}) but also an
understanding of the spatial filtering imposed by the observing
strategy and the cleaning and mapping of the data.  The additional
complexities of defining ``sources'' for the complicated morphologies
actually observed are discussed in \citet{rosolowsky10}.  Here we
restrict our discussion to the limitations on performing photometry
imposed by the data reduction.

The fundamental feature of the data is that there is a degeneracy
between large spatial scale astrophysical emission and the atmosphere
and other signals with long spatial or temporal variations.
Consequently, there are limits to the ability of the algorithm to
separate the signal $s$ from the atmosphere $a$ and time-correlated
terms $c$ in Equation \ref{eq:SignalEstimate}.  Inevitably, some of
the signal which should be present in $s$ is mixed into these terms
and is thus not present in the final map $m$.  Because $s$ is
generated from a deconvolution algorithm, it has non-zero mean by
construction, but importantly, it does {\it not} contain an unbiased
estimate of the astrophysical signal on all spatial scales: large
spatial scales are preferentially attenuated.

The use of a constant factor to convert to surface brightness
implicitly assumes that the response to all angular scales larger than
the beam is the same.  In fact, the data reduction process acts
differently on sources of different angular extent.  In performing
photometry, the \flux\ in an aperture should in fact be written as
\begin{equation}
F = f(\Omega_{src})\int_{aper}{S / \Omega_b d \Omega}
\end{equation}
where $S$ is the \flux\ in each pixel in \jyb, $\Omega_b$ is the beam
solid angle in steradians, and $f(\Omega_{src})$ is a correction
factor depending on the true angular size of the source.  The factor
$f$ can deviate significantly from unity for extended sources, and as
$\Omega_{src}$ is generally unknown, this leads to the largest
systematic uncertainty in obtaining photometric fluxes.

To determine the factor $f(\Omega_{src})$ (the flux recovered as a
function of angular scale of the source), we performed a series of
simulations.  Gaussian sources with a range of peak flux densities and
sizes were inserted on top of a background consisting of real BGPS
time series data with the astrophysical source model removed.  The
simulated time series were then processed identically to the real
data.  The recovered fluxes were measured in ellipses that included
single well-separated sources.  Figure \ref{fig:PCA_Filter} shows the
fraction of flux recovered in the map as a function of source size for
a range of different PCA components subtracted in the cleaning.  The
final reduction used 13 PCA components because this produced a
reasonable compromise between attenuating extended structure and
cleaning the atmospheric contribution.  Note that this is a
substantially more aggressive cleaning than that used in the reduction
of the Bolocam observation of the {\em Spitzer} Cores-to-Disks fields
\citep{enoch06,young06,enoch07}, largely because our mean integration
time per pixel was $\sim$seconds, whereas theirs was $\sim$minutes.
The recovered flux does depend on the source size, but importantly
does {\it not} depend on the source strength except for the strongest
sources (e.g., SgrB2).

While the precise effect of the flux reduction requires a full
simulation of the data reduction pipeline to ascertain, the flux
reduction effect can be reproduced with a relatively simple
prescription which is intuitively motivated.  Because the atmosphere
model $a$ and PCA model $c$ remove from the timestream ``signals''
which are correlated spatially across the array at the same time, this
effectively acts as a spatial filter on the map, removing spatial
scales comparable to the array FOV.  We have found that a simple
``brick wall'' filter in Fourier space which nulls all modes below 0.1
inverse arcminutes (spatial scales larger than 10\arcmin) reproduces
very well the attenuation seen in the simulation.  The comparison
between the flux attention produced by this filter and that found from
simulation is shown in Figure \ref{fig:PCA_Filter}.  The implication
is that the BGPS is not sensitive at all to scales larger than
10\arcmin.

\subsection{Bandpass Effects: Color Corrections and Line Contamination}

The width of the Bolocam passband is such that the effective band
center, or equivalently, the flux density referred to a fixed band
center, will change somewhat depending on the source spectral energy
distribution (SED).  Atmospheric transmission variations may also
affect the effective passband.  Figure \ref{fig:Bandpass} shows the
passband.  Bolocam is largely insensitive to variations in the strong
water absorption lines at 183 and 325 GHz.  We compute color
corrections due to changes in the source spectrum in Appendix
\ref{app:ColorCorrections}.

Note that we have made no correction for the contamination of the
continuum flux densities by emission from lines.  The Bolocam passband
was specifically constructed to exclude the strong $^{12}\mathrm{CO}(2
\to 1)$ emission line at 230 GHz.  We can estimate the fractional
contribution of a line to the continuum as
\begin{eqnarray}
\label{eq:LineContamination}
\nonumber \frac{L}{C} & = & 
\frac{J \int T \phi(\nu) t(\nu) \,d \nu}{\int S_\nu t(\nu) \,d \nu} 
\approx \frac{J T \Delta v \nu_c}{S_\nu \Delta \nu c} \\
& = & 0.01~\left(\frac{T \Delta v}{10~\rm{K~km~s}^{-1}}\right) 
\left(\frac{\rm{1~Jy}}{S_\nu}\right)
\end{eqnarray}
where $T$ is the line strength in Kelvin, $\phi(\nu)$ is the line
shape (of width $\Delta v$ in km s$^{-1}$), $t(\nu)$ is the Bolocam
passband (with equivalent bandwidth $\Delta \nu$ and bandcenter
$\nu_c$), and $J$ is given by Equation \ref{eq:JyPerK}.  The
approximation holds for lines in the center of the band.  The scaling
values are chosen to be typical of the line-to-continuum actually
seen.  Thus any given line is likely to at most contribute a few
percent to the measured flux, but in the most extreme star forming
environments (\eg, Sgr B2) we may expect that lines may contribute a
substantial fraction of the total measured flux density due to the
integrated effect of the lines.  For an example, see the lower panel
of Figure \ref{fig:Bandpass}.

For completeness, we note that towards \hii\ regions, free-free
emission may represent a non-trivial fraction of the millimeter flux
density in addition to the thermal dust continuum.  This effect is of
course best assessed on a case-by-case basis by comparing the Bolocam
maps with radio surveys, e.g., the VGPS \citep{stil06}.

\subsection{Comparison to Other Surveys}
\label{sec:FluxComparison}

Comparison of the Bolocam flux densities is easiest for an instrument
with the same or similar bandpass, since the additional complication
of the source spectral index is avoided.  For this reason, we choose
to make an initial comparison of the accuracy of our calibration with
the MAMBO and SIMBA instruments, since their passbands at 1.2 mm are
quite close to that of Bolocam.  MAMBO operates on the IRAM 30-meter
diameter telescope on Pico Veleta in Spain, and SIMBA operated on the
15 m SEST antenna prior to that facility's closure.  An important
future cross-check will be comparison of the BGPS flux density
calibration with SCUBA 850 \mum\ or ATLASGAL 870 \mum\ results.

The study of the Cygnus-X region by \citet{motte07} (hereafter M07)
presents one of the largest fields in the Galactic plane that has been
mapped with MAMBO which overlaps with the Bolocam survey fields.  M07
used both the 37 element MAMBO and 117 channel MAMBO-2
instruments. The FWHM beam size for the observations was 11\arcsec.
The MAMBO beam model resulting from the chopping-scanning observing
strategy is described in the appendix of \citet{motte01}.  A detailed
description of the MAMBO data reduction is given in
\citet{kauffmann08}.  M07 used Uranus and Mars as their fundamental
flux calibration, with an estimated uncertainty in the flux of 20\%.
The effective passband of the images presented by M07 is about 90 GHz
centered at 260 GHz and thus includes the bright J = 2--1 CO
transition (see Figure \ref{fig:Bandpass}).  

A survey of 2 square degrees towards $\ell = 44\arcdeg$ was conducted
by \citet{matthews09} (hereafter M09).  These images have 22\arcsec
resolution, and unlike the M07 survey, which used chopping and
scanning simultaneously, the SIMBA data were obtained in fast scan
mode only.

To compare the BGPS images to the M07 and M09 images, we first
obtained the data in FITS image format.  The M07 data is publicly
available\footnote{{\tt
ftp://cdsarc.u-strasbg.fr/pub/cats/J/A+A/476/1243/fits/}}.  The M09
data was kindly provided by H. Kirk (private communication).  All data
were re-gridded to a common projection and coordinate system grid with
4\arcsec\ pixels using the program Montage\footnote{This research made
use of Montage, funded by the National Aeronautics and Space
Administration's Earth Science Technology Office, Computation
Technologies Project, under Cooperative Agreement Number NCC5-626
between NASA and the California Institute of Technology. Montage is
maintained by the NASA/IPAC Infrared Science Archive.}.  We then
converted to \mjysr\ in a given pixel based on the beam and pixel
sizes.  Finally, the Bolocam map was convolved with either an
11\arcsec\ (M07) or 22\arcsec (M09) FWHM Gaussian and the M07 and M09
maps are convolved with a 33\arcsec\ FWHM Gaussian so that the
effective resolution of the two pairs (BGPS-M07, BGPS-M09) is the
same.  The convolution preserves the total flux in the map, but
changes the aperture in which that flux appears.  This results in maps
which, were the calibration and observed features in each map
identical, can be compared on a pixel-by-pixel basis.

The plots of pixel values versus each other are shown in Figure
\ref{fig:FluxComparison}.  The BGPS values are indeed correlated with
both M07 and M09 over a range of about two orders of magnitude, but
with a large amount of scatter.  We attribute the scatter to both
noise but primarily to the difference in the handling of extended
structure between the different surveys.  However, the average value
of the correlation coefficient deviates significantly from 1 in
comparing to both M07 and M09, with a multiplicative factor of 1.21 to
1.51 required to make the BGPS fluxes match the other surveys (or 0.66
to 0.83 to make the other surveys agree with the BGPS).  This factor
also depends on the range over which the fit is done, as given in
Table \ref{tab:FluxComparison}.  Including all pixels down to
3~\mjysr\ produces systematically larger required correction factor.
Using the average value (to account for variation of the factor with
brightness) yields a factor of 1.35.  For a dust spectral index of
$\sim3.5$, the 1.2 mm measurements should have produced a flux that is
about 12\% lower than that measured by BGPS (see Table
\ref{tab:ColorCorrections}).  Assuming this is a typical spectral
index for these clumps, {\bf we therefore note that one must multiply
by a factor of 1.35 $\times$ 1.12 = 1.5 (with a range of plus or minus
0.15) the Version 1.0 images and catalog released through IPAC to
obtain good agreement with these surveys.}

The source of this discrepancy remains unclear.  It is unlikely to be
due to line contamination in the MAMBO / SIMBA bands.  Line
contamination could increase the continuum flux by at most a few
percent for MAMBO (see Equation \ref{eq:LineContamination}), while
being negligible for Bolocam.  For most sources, it is also not
obvious that the discrepancy is due to filtering of extended
structure.  The filtering does not reach the requisite 50\% needed to
explain the difference until the source FWHM reaches 230\arcsec\
(Figure \ref{fig:PCA_Filter}).  However, the simulations were run on
Gaussian sources, and it is still not certain whether the filter
function differs for other source shapes.  The effective filter
functions of the SIMBA and MAMBO surveys are also not well understood.

Work is ongoing to thoroughly investigate all potential sources of
this discrepancy with further more realistic simulations, a new
treatment of atmosphere removal and application of distortion
corrections.  In addition, the processing for bright \flux\
calibrators in the current pipeline is different from the processing
of the science fields, and attempts are being made to process the
different observation types with an identical pipeline.  This analysis
will be discussed in a future work.

\section{Final Maps and Data Release}
\label{sec:FinalMaps}

The final maps are produced by co-adding all observations of a given
region.  The maps were processed in 29 separate pieces.  The maps are
made in Galactic coordinates using a plate carr\'{e} (FITS header CAR)
projection.  (This is the same projection used by the \spitzer-GLIMPSE
and BU-FCRAO GRS surveys.)  By placing them all on the same equatorial
grid, mosaicking the maps together is straightforward.  As these maps
are near the coordinate system equator, the difference between truly
equiareal pixels and the pixels used is at most 0.4\%, even 5\arcdeg\
out of the plane (as for IC1396).  The pixel size is 7\farcs2, chosen
to be much smaller than the Bolocam beam.  The maps are written to
standard FITS files.  An example of the FITS header for the BGPS
images is given in Appendix \ref{app:FITS_Header}.

All processed maps are available through
IPAC\footnote{\tt http://irsa.ipac.caltech.edu/Missions/bolocam.html}.
IPAC provides a cutout service for the images as well as a searchable
version of the catalog provided in \citet{rosolowsky10}.  The Version
1.0 release includes the following images, each covering the same
regions:

\begin{enumerate}
\item Calibrated maps for all regions observed (MAP).  The maps are
standard astrophysical maps subject to the caveats previously
mentioned, the most important being the spatial filtering function
(Section \ref{sec:Noise}).

\item Maps of integration time per pixel (NHITSMAP).  The nhitsmap
shows how many recorded data points from the timestream have been
assigned to each pixel; it is proportional to the dwell time per pixel
(each `hit' is 0.1s).

\item Maps of the residual time series (NOISEMAP).  The residual map
is the map of the model timestream subtracted from the data
timestream.

\end{enumerate}

Examples of each are given in Figure \ref{fig:SampleImages}.  Future
releases may contain additional data products.

\section{Discussion}
\label{sec:Discussion}

The BGPS has observed a \bgpsarea\ square degree area at an effective
resolution of a Gaussian PSF of \bcamfwhmeff\ FWHM (corresponding to
an area of $2.9\times10^{-8}$ ster).  The positional uncertainty of
the maps is 6\arcsec\ RMS.  The maps are calibrated into \jyb\ using
Mars as a primary flux standard, and accounting for atmosphere opacity
variations in real time.  The BGPS is well suited for studying source
structure smaller than 3 arcminutes in size.  The processing of the
maps removes more than 90\% of flux for features with extents larger
than 5\farcm9, and attenuates the aperture flux of structures
extending to 3\farcm8 by 50\%.  We have compared our flux densities to
those of M07 and M09 and find that we need to multiply our flux
densities up by a factor of 1.5 to match. We have investigated
possible sources for the discrepancy, but do not understand the source
of the difference.  For now, we recommend multiplying the flux
densities in the maps presented here and in the R10 catalog by the
factor $1.5 \pm 0.15$ to obtain consistency with other data sets.  We
are continuing to explore the source of this discrepancy, particularly
in the calibration and filtering stages of the data pipeline.  The
latest data releases and updates on the discrepancy can be found on
the BGPS IPAC website.

Maps from the entire BGPS are presented in Figure
\ref{fig:BGPSMontage}.  Figures \ref{fig:SCUBAPointingComparison} and
\ref{fig:MontD} show the generally excellent morphological agreement
between \bgps\ maps and those at other (sub)millimeter wavelengths.
Filamentary structures with aspect ratios up to the maximum spatial
dynamic range of the survey ($354\arcsec/33\arcsec \approx 10$) are
present throughout the Galactic plane.  The most crowded fields
consist of frothy and clumpy structure.  The majority of sources are
at least moderately resolved.  We suspect that higher resolution
observations will resolve BGPS clumps into clusters of protostellar
objects, as has been seen in Serpens \citep{enoch08,Testi1998}, OMC 1
\citep{beuther04,johnstone99}, and S255N \citep{cyganowski07} where
bolometer observations at low resolution have been complemented by
interferometer observations at high resolution.

Figure \ref{fig:Multiwavelength} shows a comparison of the BGPS with
{\em Spitzer}-GLIMPSE 8 \mum\ \citep{benjamin03} and VGPS 20 cm
continuum images \citep{stil06}.  In Figure \ref{fig:Multiwavelength},
IRDCs appear as the bright white clumps, often filamentary, near the
mid-plane.  BGPS sources are clearly associated with some of the \hii\
regions detected by the VGPS; a zoom in of this region is shown in
Figure \ref{fig:HII}.  The comparison with 8 $\mu$m IRDCs is
particularly striking, with a high degree of correspondence in their
morphologies.  Detection of an IRDC requires that the cold, dense
cloud be on the near-side of a bright mid-IR background, whereas BGPS
sources are subject only to a sensitivity limit.  Comparison of the
GLIMPSE and BGPS data (see Figure \ref{fig:IRDC}) show that some BGPS
sources are clearly associated with IRDCs, a few are associated with
IR-bright features (likely due to the association of free-free
emission in the millimeter and PAH emission at 8 $\mu$m), and some
have no clear IR-counterpart.  The BGPS sources which are neither
bright nor dark at 8 $\mu$m are probable candidates for ``IRDC-like''
clumps (cold, dense, potential proto-clusters) on the far-side or
outer edges of the Galaxy.  Thus, the BGPS could provide candidate
pre-clusters independent of the favorable viewing angle of being on
the near-side of bright mid-IR emission.  An exploration of this
population, as well as a detailed analysis of the physical properties
and star formation tracers of IRDCs corresponding to BGPS sources may
be found in \citet{battersby10}.  It is worth noting that many
features in the BGPS maps have no corresponding counterpart in the
{\em Spitzer} or radio image.

There are clear maxima in the emission filling factor in the Galactic
Center and in the region from $\lon=23$ to $\lon=31$, the line of
sight through the densest part of the molecular ring.  Additional
areas of high source density are seen at $\lon=10$ and $\lon=13$, as
shown in Figure \ref{fig:FillingFactor}.  This result is in contrast
to CO filling factors which are $\sim1$ for $l \lesssim 40$
\citep{dame01,jackson06}.  In parts of the plane where only one CO
emission line is seen along any given line of sight ($\lon\gtrsim40$),
the integrated emission is much lower, but there are still detections
of CO in every pixel of the \citet{dame01} map in the latitude range
$-0.5 < b < 0.5$.  This behavior is consistent with the interpretation
that the BGPS preferentially detects denser material than CO.  The
connection of the millimeter emission with star forming material
awaits the detailed comparison to other star formation tracers.  BGPS
clumps are, however, associated with well known star forming regions:
the brightest clumps visible in our maps are \object{Sgr B2},
\object{G34.3+0.15}, \object{W 51}, \object{W 43}, \object{W 49}, and
\object{M 17}.  These associations suggest that other bright sources
of millimeter emission will prove to be massive star- or
cluster-forming regions.

In some cases, the distance to BGPS clumps can be determined by
matching Galactic Ring Survey $^{13}$CO and 1.1 mm continuum
morphology \citep{IRDCdistance,jackson06}.  However, in other cases
the association between the CO and 1.1 mm data is not clear, either
because of confusion or because the BGPS sources are too compact to
identify in $^{13}$CO morphology.  Heterodyne follow-up observations
using dense gas tracers (NH$_3$, N$_2$H$^+$, HCO$^+$, and CS) are
being conducted to provide radial velocity, density, chemistry, and
temperature measurements \citep{schlingman10,dunham10}.  

We have detected \ncores\ clumps in the surveyed fields
\citep{rosolowsky10}, providing a sample of clumps suitable for
multi-wavelength and high-resolution studies with existing telescopes
and future facilities.  This sample is free of the usual biases
associated with observing only regions with signposts of star
formation, such as \hii\ regions or masers. The survey depth can be
converted into an estimate of limiting mass sensitivity via standard
estimates, i.e., under the assumption of a temperature, opacity, and
distance.  Using the dust opacity from \citet{ossenkopf94}
($\kappa({\rm 1.1~mm}) = 0.0114~{\rm cm^{2}~ g^{-1}}$), a gas-to-dust
mass ratio of 100, the Bolocam bandcenter from Table
\ref{tab:ColorCorrections}, and the beam area as given in Section
\ref{sec:FluxCalibration} the mass sensitivity can be written as
\begin{equation} 
\label{eq:Mass}
M_{gas}\approx
14.3 \left( e^{13.0/T_d}-1 \right) \left({S_\nu\over 1\; {\rm Jy}} \right)
\left(\frac{D}{{\rm 1~kpc}}\right)^{2} \msol 
\end{equation}
 At the coldest temperature expected (10 K), the mass increases by a
factor of 2.9 from the mass assuming 20 K.  While sensitivity limits
will bias the survey against distant, low-mass objects, initial
results suggest that the Bolocam sources lie at a range of distances,
and thus range from cores to clumps \citep{rosolowsky10}.  We have
begun to explore the variation of properties with Galactocentric
radius with detailed analyses of the BGPS Galactic center
\citep{bally10} and anti-center \citep{dunham10} data.

Millimeter-wavelength thermal dust emission reveals the repositories
of the densest molecular gas, ranging in scale from cores to whole
clouds. By pinpointing these regions, the BGPS allows the connection
of this gas to nascent and ongoing star formation to be explored.

\noindent---------------------------------

{\it Facilities:} \facility{CSO (Bolocam)}

\acknowledgments

We would like to acknowledge the staff and day crew of the CSO for their
assistance. The BGPS project is supported by the National Science Foundation
through NSF grant AST-0708403. J.A. was supported by a Jansky Fellowship from
the National Radio Astronomy Observatory (NRAO). The first observing runs for
BGPS were supported by travel funds provided by NRAO. Support for the
development of Bolocam was provided by NSF grants AST-9980846 and AST-0206158.
Team support was provided in part by NSF grant AST-0607793 to the
University of Texas at Austin.

We recognize and acknowledge the cultural role and reverence that the
summit of Mauna Kea has within the Hawaiian community. We are
fortunate to conduct observations from this mountain.

\pagebreak

\begin{appendix}

\section{Calculation of Color Corrections}
\label{app:ColorCorrections}

If an experiment has finite bandwidth ($t(\nu) \ne \delta(\nu -
\nu_c)$), to report a source surface brightness at a single
frequency, one must assume a source spectrum.  The power detected from
that source is assumed to be
\begin{equation}
P_{det} = \eta \; A \Omega \; \int I_0(\nu) t(\nu) d \nu
\end{equation}
(Here $\eta$ and $A \Omega$ are the optical efficiency and throughput
of the instrument, $I_0(\nu)$ is the nominal (assumed) surface
brightness of the source, and $t(\nu)$ is the bandpass transmission
normalized to $1.0$ at its peak.)  Bolocam is single-moded, meaning
that $A \Omega = \lambda^2$.  The effective band center
$\nu_c$ is defined implicitly by the equation
\begin{equation}
\label{eq:BandCenter}
I_0(\nu_c) = \frac{\int I_0(\nu) t(\nu) d \nu}
{\int t(\nu) d \nu} 
\end{equation}
In calculating the color corrections and effective band centers, we
consider two types of source spectra.  The most physically motivated
is a ``greybody'' spectrum parametrized as
\begin{equation}
\label{eq:Greybody}
I_{GB}(\nu) = \epsilon(\nu) B_{\nu}(T)  =
(1-\exp{(-\tau_\nu})) B_{\nu}(T)
\equiv (1-\exp{[-(\nu/\nu_0)^\beta]}) B_{\nu}(T) 
\end{equation}
where $B_{\nu}(T)$ is the Planck function, $\epsilon(\nu)$ is the
frequency-dependent emissivity, and $\nu_0$ is the frequency at which
the optical depth $\tau_\nu$ reaches unity.  For a frequency much
lower than the blackbody peak and in the optically thin limit, this
reduces to a simple power law
\begin{equation}
\label{eq:PowerLaw}
I_\alpha(\nu) = \left(\frac{\nu}{\nu_0}\right)^\beta \frac{2kT\nu^2}{c^2}
\propto \left(\frac{\nu}{\nu_0}\right)^\alpha
\end{equation}

The color correction $K$ is defined by the relative change in the
intensity which would be measured at the same effective band center if
the source had a different spectrum $I_1(\nu)$ than that assumed for
calculating the band center; thus
\begin{eqnarray}
\label{eq:ColorCorrection}
\nonumber I_1(\nu_c) &=& I_0(\nu_c) \frac{I_1(\nu_c)}{\int I_1(\nu) t(\nu) d \nu} \left[\frac{I_0(\nu_c)}{\int I_0(\nu) t(\nu) d \nu} \right]^{-1} \\
&\equiv& K^{-1} I_0(\nu_c)
\end{eqnarray}

We calculate the effective band centers and color corrections for
Bolocam in Table \ref{tab:ColorCorrections}.  The fiducial spectrum
for quoting the bandcenter (chosen because of its closeness to mean
Galactic properties; \eg\ \citet{reach95}) is taken to be that of
Equation \ref{eq:Greybody} with parameter values $T=20$~K,
$\beta=1.8$, and $\nu_0=3000$~GHz (100 \mum).  The bandcenters and
color corrections relative to this fiducial spectrum are also computed
for various values of $\alpha$ in Equation \ref{eq:PowerLaw}.  Over
the frequency range of the Bolocam passband, a power law with
$\alpha=3.5$ is a very good approximation to the assumed greybody.

\section{FITS Header Information}
\label{app:FITS_Header}

Below is a sample FITS header from the version 1.0.2 release of the
BGPS images described in this paper.

\begin{verbatim}

SIMPLE  =                    T / Written by IDL:  Sun Apr 19 17:59:29 2009      
BITPIX  =                  -32 / Bits per pixel                                 
NAXIS   =                    2 / Number of axes                                 
NAXIS1  =                 1519 / Axis length                                    
NAXIS2  =                  651 / Axis length                                    
EXTEND  =                    F / File may contain extensions                    
ORIGIN  = 'NOAO-IRAF FITS Image Kernel July 2003' / FITS file originator        
IRAF-TLM= '13:33:03 (13/04/2009)' / Time of last modification                   
DATE    = '2009-04-13T19:33:03' / Creation UTC (CCCC-MM-DD) date of FITS header 
LONPOLE2=        180.000000000 /lonpole                                         
LATPOLE2=        0.00000000000 /latpole                                         
DECFWHM =              14.4000 /Deconvolution Kernel FWHM                       
MEANDC  =              2.06882 /Mean DC level                                   
STDDC   =             0.499487 /Std. dev. DC level                              
UNITS   = 'Jy/Beam '           /                                                
PPBEAM  =        27.7012530153 /pixels per beam                                 
CALIB_0 =         -3.26472E-15 / 0th coefficient for flux cal (see methods paper
CALIB_1 =             0.398740 / 1st coefficient for flux cal                   
CALIB_2 =              3.32002 / 2nd coefficient for flux cal                   
CTYPE1  = 'GLON-CAR'           / Coordinate Type                                
CTYPE2  = 'GLAT-CAR'           / Coordinate Type                                
EQUINOX =              2000.00 / Equinox of Ref. Coord.                         
CD1_1   =    -0.00199999986216 / Degrees / Pixel                                
CD2_2   =     0.00199999986216 / Degrees / Pixel                                
CRPIX1  =        766.378744566 / Reference Pixel in X                           
CRPIX2  =        345.530969032 / Reference Pixel in Y                           
CRVAL1  =        6.00002238819 / Galactic longitude of reference pixel          
CRVAL2  =   -7.44007626964E-05 / Galactic latitude of reference pixel           
PV2_1   =        0.00000000000 /Projection parameter 1                          
PROTITLE= 'Bolocam Galactic Plane Survey' /                                     
CONTACT = 'John Bally'         /john.bally@colorado.edu                         
MAPTYPE = 'map     '           /                                                
BUNIT   = 'Jy/Beam '           /Units in map                                    
BMAJ    =           0.00988889 /                                                
BMIN    =           0.00988889 /                                                
BPA     =                    0 /                                                
BGPSITER=                   50 /Iteration number                                
BGPSNPCA=                   13 /number of PCA components subtracted             
BGPSVERS= '1.0     '           /BGPS Processing Version Number                  
WAVELENG=              1.12000 /mm (avoids CO 2-1)                              
COMMENT FITS (Flexible Image Transport System) format is defined in 'Astronomy  
COMMENT and Astrophysics', volume 376, page 359; bibcode 2001A&A...376..359H    
COMMENT Made by the Bolocam Galactic Plane Survey (BGPS) pipeline               
COMMENT described in Aguirre et al 2009 (not yet published)                     
COMMENT BGPS data was taken at the Caltech Submillimeter Observatory            
COMMENT Pixel coverage is in the nhitsmap file (each hit represents .1s dwell ti
COMMENT Pixel weighting is in the weightmap file                                
COMMENT Flag counts are in the flagmap file                                     
COMMENT Deconvolved model is in the model file                                  
HISTORY PUTAST: Apr  7 09:22:26 2009 World Coordinate System parameters written 
HISTORY Dates and observation numbers included:                                 
HISTORY 060616_ob3                                                              
HISTORY 060621_o20                                                              
HISTORY 060621_o23                                                              
HISTORY 060622_o15                                                              
HISTORY 070701_o27                                                              
HISTORY 070702_o14                                                              
HISTORY 070704_o16                                                              
HISTORY 070715_ob5                                                              
HISTORY 070706_o22                                                              
HISTORY 070707_o31                                                              
HISTORY 070708_o15                                                              
HISTORY 070708_o16                                                              
HISTORY 070709_o15                                                              
HISTORY 070709_o16                                                              
WCSDIM  =                    2                                                  
CDELT1  =    -0.00199999986216                                                  
CDELT2  =     0.00199999986216                                                  
LTV1    =                 -64.                                                  
LTV2    =                 -20.                                                  
LTM1_1  =                   1.                                                  
LTM2_2  =                   1.                                                  
WAT0_001= 'system=image'                                                        
WAT1_001= 'wtype=car axtype=glon'                                               
WAT2_001= 'wtype=car axtype=glat'                                               
                                                                                
                                                                                
                                                                                
                                                                                
                                                                                
                                                                                
                                                                                
                                                                                
END                                                          

\end{verbatim}

\end{appendix}

\pagebreak


\begin{thebibliography}{61}
\expandafter\ifx\csname natexlab\endcsname\relax\def\natexlab#1{#1}\fi

\bibitem[{{Bally} {et~al.}(2010){Bally}, {Aguirre}, {Battersby}, {Bradley},
  {Cyganowski}, {Dowell}, {Drosback}, {Dunham}, {Evans}, {Ginsburg}, {Glenn},
  {Harvey}, {Mills}, {Merello}, {Rosolowsky}, {Schlingman}, {Shirley},
  {Stringfellow}, {Walawender}, \& {Williams}}]{bally10}
{Bally}, J. {et~al.} 2010, \apj, 721, 137

\bibitem[{{Battersby} {et~al.}(2010){Battersby}, {Bally}, {Jackson},
  {Ginsburg}, {Shirley}, {Schlingman}, \& {Glenn}}]{battersby10}
{Battersby}, C., {Bally}, J., {Jackson}, J.~M., {Ginsburg}, A., {Shirley},
  Y.~L., {Schlingman}, W., \& {Glenn}, J. 2010, \apj, 721, 222

\bibitem[{{Benjamin} {et~al.}(2003){Benjamin}, {Churchwell}, {Babler}, {Bania},
  {Clemens}, {Cohen}, {Dickey}, {Indebetouw}, {Jackson}, {Kobulnicky},
  {Lazarian}, {Marston}, {Mathis}, {Meade}, {Seager}, {Stolovy}, {Watson},
  {Whitney}, {Wolff}, \& {Wolfire}}]{benjamin03}
{Benjamin}, R.~A. {et~al.} 2003, \pasp, 115, 953

\bibitem[{{Beuther} {et~al.}(2004){Beuther}, {Zhang}, {Greenhill}, {Reid},
  {Wilner}, {Keto}, {Marrone}, {Ho}, {Moran}, {Rao}, {Shinnaga}, \&
  {Liu}}]{beuther04}
{Beuther}, H. {et~al.} 2004, \apjl, 616, L31

\bibitem[{{Buckle} {et~al.}(2009){Buckle}, {Hills}, {Smith}, {Dent}, {Bell},
  {Curtis}, {Dace}, {Gibson}, {Graves}, {Leech}, {Richer}, {Williamson},
  {Withington}, {Yassin}, {Bennett}, {Hastings}, {Laidlaw}, {Lightfoot},
  {Burgess}, {Dewdney}, {Hovey}, {Willis}, {Redman}, {Wooff}, {Berry},
  {Cavanagh}, {Davis}, {Dempsey}, {Friberg}, {Jenness}, {Kackley}, {Rees},
  {Tilanus}, {Walther}, {Zwart}, {Klapwijk}, {Kroug}, \& {Zijlstra}}]{buckle09}
{Buckle}, J.~V. {et~al.} 2009, \mnras, 399, 1026

\bibitem[{{Carey} {et~al.}(1998){Carey}, {Clark}, {Egan}, {Price}, {Shipman},
  \& {Kuchar}}]{carey98}
{Carey}, S.~J., {Clark}, F.~O., {Egan}, M.~P., {Price}, S.~D., {Shipman},
  R.~F., \& {Kuchar}, T.~A. 1998, \apj, 508, 721

\bibitem[{{Carey} {et~al.}(2009){Carey}, {Noriega-Crespo}, {Mizuno}, {Shenoy},
  {Paladini}, {Kraemer}, {Price}, {Flagey}, {Ryan}, {Ingalls}, {Kuchar},
  {Pinheiro Gon{\c c}alves}, {Indebetouw}, {Billot}, {Marleau}, {Padgett},
  {Rebull}, {Bressert}, {Ali}, {Molinari}, {Martin}, {Berriman}, {Boulanger},
  {Latter}, {Miville-Deschenes}, {Shipman}, \& {Testi}}]{carey09}
{Carey}, S.~J. {et~al.} 2009, \pasp, 121, 76

\bibitem[{{Chapin} {et~al.}(2008){Chapin}, {Ade}, {Bock}, {Brunt}, {Devlin},
  {Dicker}, {Griffin}, {Gundersen}, {Halpern}, {Hargrave}, {Hughes}, {Klein},
  {Marsden}, {Martin}, {Mauskopf}, {Netterfield}, {Olmi}, {Pascale},
  {Patanchon}, {Rex}, {Scott}, {Semisch}, {Truch}, {Tucker}, {Tucker}, {Viero},
  \& {Wiebe}}]{chapin08}
{Chapin}, E.~L. {et~al.} 2008, \apj, 681, 428

\bibitem[{{Cotton} {et~al.}(2009){Cotton}, {Mason}, {Dicker}, {Korngut},
  {Devlin}, {Aquirre}, {Benford}, {Moseley}, {Staguhn}, {Irwin}, \&
  {Ade}}]{cotton09}
{Cotton}, W.~D. {et~al.} 2009, \apj, 701, 1872

\bibitem[{{Cyganowski} {et~al.}(2007){Cyganowski}, {Brogan}, \&
  {Hunter}}]{cyganowski07}
{Cyganowski}, C.~J., {Brogan}, C.~L., \& {Hunter}, T.~R. 2007, \aj, 134, 346

\bibitem[{{Dame} {et~al.}(2001){Dame}, {Hartmann}, \& {Thaddeus}}]{dame01}
{Dame}, T.~M., {Hartmann}, D., \& {Thaddeus}, P. 2001, \apj, 547, 792

\bibitem[{{Di Francesco}(2008)}]{difrancesco08b}
{Di Francesco}, J. 2008, in Bulletin of the American Astronomical Society,
  Vol.~40, Bulletin of the American Astronomical Society, 271--+

\bibitem[{{Di Francesco} {et~al.}(2008){Di Francesco}, {Johnstone}, {Kirk},
  {MacKenzie}, \& {Ledwosinska}}]{difrancesco08}
{Di Francesco}, J., {Johnstone}, D., {Kirk}, H., {MacKenzie}, T., \&
  {Ledwosinska}, E. 2008, \apjs, 175, 277

\bibitem[{{Dunham} {et~al.}(2010){Dunham}, {Rosolowsky}, {Evans}, {Cyganowski},
  {Aguirre}, {Bally}, {Battersby}, {Bradley}, {Dowell}, {Drosback}, {Ginsburg},
  {Glenn}, {Harvey}, {Merello}, {Schlingman}, {Shirley}, {Stringfellow},
  {Walawender}, \& {Williams}}]{dunham10}
{Dunham}, M.~K. {et~al.} 2010, \apj, 717, 1157

\bibitem[{{Egan} {et~al.}(1998){Egan}, {Shipman}, {Price}, {Carey}, {Clark}, \&
  {Cohen}}]{egan98}
{Egan}, M.~P., {Shipman}, R.~F., {Price}, S.~D., {Carey}, S.~J., {Clark},
  F.~O., \& {Cohen}, M. 1998, \apjl, 494, L199+

\bibitem[{{Enoch} {et~al.}(2008){Enoch}, {Evans}, {Sargent}, {Glenn},
  {Rosolowsky}, \& {Myers}}]{enoch08}
{Enoch}, M.~L., {Evans}, II, N.~J., {Sargent}, A.~I., {Glenn}, J.,
  {Rosolowsky}, E., \& {Myers}, P. 2008, \apj, 684, 1240

\bibitem[{{Enoch} {et~al.}(2007){Enoch}, {Glenn}, {Evans}, {Sargent}, {Young},
  \& {Huard}}]{enoch07}
{Enoch}, M.~L., {Glenn}, J., {Evans}, II, N.~J., {Sargent}, A.~I., {Young},
  K.~E., \& {Huard}, T.~L. 2007, \apj, 666, 982

\bibitem[{{Enoch} {et~al.}(2006){Enoch}, {Young}, {Glenn}, {Evans}, {Golwala},
  {Sargent}, {Harvey}, {Aguirre}, {Goldin}, {Haig}, {Huard}, {Lange},
  {Laurent}, {Maloney}, {Mauskopf}, {Rossinot}, \& {Sayers}}]{enoch06}
{Enoch}, M.~L. {et~al.} 2006, \apj, 638, 293

\bibitem[{{Evans} {et~al.}(2009){Evans}, {Dunham}, {J{\o}rgensen}, {Enoch},
  {Mer{\'{\i}}n}, {van Dishoeck}, {Alcal{\'a}}, {Myers}, {Stapelfeldt},
  {Huard}, {Allen}, {Harvey}, {van Kempen}, {Blake}, {Koerner}, {Mundy},
  {Padgett}, \& {Sargent}}]{evans09}
{Evans}, N.~J. {et~al.} 2009, \apjs, 181, 321

\bibitem[{{Glenn} {et~al.}(2003){Glenn}, {Ade}, {Amarie}, {Bock}, {Edgington},
  {Goldin}, {Golwala}, {Haig}, {Lange}, {Laurent}, {Mauskopf}, {Yun}, \&
  {Nguyen}}]{glenn03}
{Glenn}, J. {et~al.} 2003, in \procspie, Vol. 4855, Millimeter and
  Submillimeter Detectors for Astronomy, 30

\bibitem[{{Griffin} \& {Orton}(1993)}]{griffin93}
{Griffin}, M.~J. \& {Orton}, G.~S. 1993, Icarus, 105, 537

\bibitem[{{Holland} {et~al.}(1999){Holland}, {Robson}, {Gear}, {Cunningham},
  {Lightfoot}, {Jenness}, {Ivison}, {Stevens}, {Ade}, {Griffin}, {Duncan},
  {Murphy}, \& {Naylor}}]{holland99}
{Holland}, W.~S. {et~al.} 1999, \mnras, 303, 659

\bibitem[{{Hollis} {et~al.}(1992){Hollis}, {Dorband}, \&
  {Yusef-Zadeh}}]{hollis92}
{Hollis}, J.~M., {Dorband}, J.~E., \& {Yusef-Zadeh}, F. 1992, \apj, 386, 293

\bibitem[{{Jackson} {et~al.}(2006){Jackson}, {Rathborne}, {Shah}, {Simon},
  {Bania}, {Clemens}, {Chambers}, {Johnson}, {Dormody}, {Lavoie}, \&
  {Heyer}}]{jackson06}
{Jackson}, J.~M. {et~al.} 2006, \apjs, 163, 145

\bibitem[{{Johnstone} \& {Bally}(1999)}]{johnstone99}
{Johnstone}, D. \& {Bally}, J. 1999, \apjl, 510, L49

\bibitem[{{Johnstone} \& {Bally}(2006)}]{johnstone06}
---. 2006, \apj, 653, 383

\bibitem[{{Kauffmann} {et~al.}(2008){Kauffmann}, {Bertoldi}, {Bourke}, {Evans},
  \& {Lee}}]{kauffmann08}
{Kauffmann}, J., {Bertoldi}, F., {Bourke}, T.~L., {Evans}, II, N.~J., \& {Lee},
  C.~W. 2008, \aap, 487, 993

\bibitem[{{Kov{\'a}cs}(2008)}]{kovacs08}
{Kov{\'a}cs}, A. 2008, in Presented at the Society of Photo-Optical
  Instrumentation Engineers (SPIE) Conference, Vol. 7020, Society of
  Photo-Optical Instrumentation Engineers (SPIE) Conference Series

\bibitem[{{Lada} {et~al.}(1991){Lada}, {Bally}, \& {Stark}}]{lada91}
{Lada}, E.~A., {Bally}, J., \& {Stark}, A.~A. 1991, \apj, 368, 432

\bibitem[{{Laurent} {et~al.}(2005){Laurent}, {Aguirre}, {Glenn}, {Ade}, {Bock},
  {Edgington}, {Goldin}, {Golwala}, {Haig}, {Lange}, {Maloney}, {Mauskopf},
  {Nguyen}, {Rossinot}, {Sayers}, \& {Stover}}]{laurent05}
{Laurent}, G.~T. {et~al.} 2005, \apj, 623, 742

\bibitem[{{Matthews} {et~al.}(2009){Matthews}, {Kirk}, {Johnstone},
  {Weferling}, {Cohen}, {Jenness}, {Evans}, {Davis}, {Dent}, {Fuller},
  {Jackson}, {Rathborne}, {Richer}, \& {Simon}}]{matthews09}
{Matthews}, H. {et~al.} 2009, \aj, 138, 1380

\bibitem[{{McKee} \& {Ostriker}(2007)}]{mckee07}
{McKee}, C.~F. \& {Ostriker}, E.~C. 2007, \araa, 45, 565

\bibitem[{{Molinari} {et~al.}(2010){Molinari}, {Swinyard}, {Bally}, {Barlow},
  {Bernard}, {Martin}, {Moore}, {Noriega-Crespo}, {Plume}, {Testi}, {Zavagno},
  {Abergel}, {Ali}, {Andr{\'e}}, {Baluteau}, {Benedettini}, {Bern{\'e}},
  {Billot}, {Blommaert}, {Bontemps}, {Boulanger}, {Brand}, {Brunt}, {Burton},
  {Campeggio}, {Carey}, {Caselli}, {Cesaroni}, {Cernicharo}, {Chakrabarti},
  {Chrysostomou}, {Codella}, {Cohen}, {Compiegne}, {Davis}, {de Bernardis}, {de
  Gasperis}, {Di Francesco}, {di Giorgio}, {Elia}, {Faustini}, {Fischera},
  {Fukui}, {Fuller}, {Ganga}, {Garcia-Lario}, {Giard}, {Giardino}, {Glenn},
  {Goldsmith}, {Griffin}, {Hoare}, {Huang}, {Jiang}, {Joblin}, {Joncas},
  {Juvela}, {Kirk}, {Lagache}, {Li}, {Lim}, {Lord}, {Lucas}, {Maiolo},
  {Marengo}, {Marshall}, {Masi}, {Massi}, {Matsuura}, {Meny}, {Minier},
  {Miville-Desch{\^e}nes}, {Montier}, {Motte}, {M{\"u}ller}, {Natoli}, {Neves},
  {Olmi}, {Paladini}, {Paradis}, {Pestalozzi}, {Pezzuto}, {Piacentini},
  {Pomar{\`e}s}, {Popescu}, {Reach}, {Richer}, {Ristorcelli}, {Roy}, {Royer},
  {Russeil}, {Saraceno}, {Sauvage}, {Schilke}, {Schneider-Bontemps},
  {Schuller}, {Schultz}, {Shepherd}, {Sibthorpe}, {Smith}, {Smith},
  {Spinoglio}, {Stamatellos}, {Strafella}, {Stringfellow}, {Sturm}, {Taylor},
  {Thompson}, {Tuffs}, {Umana}, {Valenziano}, {Vavrek}, {Viti}, {Waelkens},
  {Ward-Thompson}, {White}, {Wyrowski}, {Yorke}, \& {Zhang}}]{molinari10}
{Molinari}, S. {et~al.} 2010, \pasp, 122, 314

\bibitem[{{Motte} \& {Andr{\'e}}(2001)}]{motte01}
{Motte}, F. \& {Andr{\'e}}, P. 2001, \aap, 365, 440

\bibitem[{{Motte} {et~al.}(2007){Motte}, {Bontemps}, {Schilke}, {Schneider},
  {Menten}, \& {Brogui{\`e}re}}]{motte07}
{Motte}, F., {Bontemps}, S., {Schilke}, P., {Schneider}, N., {Menten}, K.~M.,
  \& {Brogui{\`e}re}, D. 2007, \aap, 476, 1243

\bibitem[{{Mueller} {et~al.}(2002){Mueller}, {Shirley}, {Evans}, \&
  {Jacobson}}]{mueller02}
{Mueller}, K.~E., {Shirley}, Y.~L., {Evans}, II, N.~J., \& {Jacobson}, H.~R.
  2002, \apjs, 143, 469

\bibitem[{{Netterfield} {et~al.}(2009){Netterfield}, {Ade}, {Bock}, {Chapin},
  {Devlin}, {Griffin}, {Gundersen}, {Halpern}, {Hargrave}, {Hughes}, {Klein},
  {Marsden}, {Martin}, {Mauskopf}, {Olmi}, {Pascale}, {Patanchon}, {Rex},
  {Roy}, {Scott}, {Semisch}, {Thomas}, {Truch}, {Tucker}, {Tucker}, {Viero}, \&
  {Wiebe}}]{netterfield09}
{Netterfield}, C.~B. {et~al.} 2009, \apj, 707, 1824

\bibitem[{{Nummelin} {et~al.}(1998){Nummelin}, {Bergman}, {Hjalmarson},
  {Friberg}, {Irvine}, {Millar}, {Ohishi}, \& {Saito}}]{nummelin1998}
{Nummelin}, A., {Bergman}, P., {Hjalmarson}, A., {Friberg}, P., {Irvine},
  W.~M., {Millar}, T.~J., {Ohishi}, M., \& {Saito}, S. 1998, \apjs, 117, 427

\bibitem[{{Olmi} {et~al.}(2009){Olmi}, {Ade}, {Angl{\'e}s-Alc{\'a}zar}, {Bock},
  {Chapin}, {De Luca}, {Devlin}, {Dicker}, {Elia}, {Fazio}, {Giannini},
  {Griffin}, {Gundersen}, {Halpern}, {Hargrave}, {Hughes}, {Klein},
  {Lorenzetti}, {Marengo}, {Marsden}, {Martin}, {Massi}, {Mauskopf},
  {Netterfield}, {Patanchon}, {Rex}, {Salama}, {Scott}, {Semisch}, {Smith},
  {Strafella}, {Thomas}, {Truch}, {Tucker}, {Tucker}, {Viero}, \&
  {Wiebe}}]{olmi09}
{Olmi}, L. {et~al.} 2009, \apj, 707, 1836

\bibitem[{{Orton} {et~al.}(1986){Orton}, {Griffin}, {Ade}, {Nolt}, \&
  {Radostitz}}]{orton86}
{Orton}, G.~S., {Griffin}, M.~J., {Ade}, P.~A.~R., {Nolt}, I.~G., \&
  {Radostitz}, J.~V. 1986, Icarus, 67, 289

\bibitem[{{Ossenkopf} \& {Henning}(1994)}]{ossenkopf94}
{Ossenkopf}, V. \& {Henning}, T. 1994, \aap, 291, 943

\bibitem[{{Peretto} \& {Fuller}(2009)}]{peretto09}
{Peretto}, N. \& {Fuller}, G.~A. 2009, \aap, 505, 405

\bibitem[{{Pohl} {et~al.}(2008){Pohl}, {Englmaier}, \&
  {Bissantz}}]{RotationCurve}
{Pohl}, M., {Englmaier}, P., \& {Bissantz}, N. 2008, \apj, 677, 283

\bibitem[{{Rathborne} {et~al.}(2006){Rathborne}, {Jackson}, \&
  {Simon}}]{rathborne06}
{Rathborne}, J.~M., {Jackson}, J.~M., \& {Simon}, R. 2006, \apj, 641, 389

\bibitem[{{Rathborne} {et~al.}(2008){Rathborne}, {Jackson}, {Zhang}, \&
  {Simon}}]{rathborne08}
{Rathborne}, J.~M., {Jackson}, J.~M., {Zhang}, Q., \& {Simon}, R. 2008, \apj,
  689, 1141

\bibitem[{{Reach} {et~al.}(1995){Reach}, {Dwek}, {Fixsen}, {Hewagama},
  {Mather}, {Shafer}, {Banday}, {Bennett}, {Cheng}, {Eplee}, {Leisawitz},
  {Lubin}, {Read}, {Rosen}, {Shuman}, {Smoot}, {Sodroski}, \&
  {Wright}}]{reach95}
{Reach}, W.~T. {et~al.} 1995, \apj, 451, 188

\bibitem[{{Reid} {et~al.}(2009){Reid}, {Menten}, {Zheng}, {Brunthaler},
  {Moscadelli}, {Xu}, {Zhang}, {Sato}, {Honma}, {Hirota}, {Hachisuka}, {Choi},
  {Moellenbrock}, \& {Bartkiewicz}}]{reid09}
{Reid}, M.~J. {et~al.} 2009, \apj, 700, 137

\bibitem[{{Rosolowsky} {et~al.}(2010){Rosolowsky}, {Dunham}, {Ginsburg},
  {Bradley}, {Aguirre}, {Bally}, {Battersby}, {Cyganowski}, {Dowell},
  {Drosback}, {Evans}, {Glenn}, {Harvey}, {Stringfellow}, {Walawender}, \&
  {Williams}}]{rosolowsky10}
{Rosolowsky}, E. {et~al.} 2010, \apjs, 188, 123

\bibitem[{{Roy} {et~al.}(2010){Roy}, {Ade}, {Bock}, {Chapin}, {Devlin},
  {Dicker}, {Griffin}, {Gundersen}, {Halpern}, {Hargrave}, {Hughes}, {Klein},
  {Marsden}, {Martin}, {Mauskopf}, {Miville-Desch{\^e}nes}, {Netterfield},
  {Olmi}, {Patanchon}, {Rex}, {Scott}, {Semisch}, {Truch}, {Tucker}, {Tucker},
  {Viero}, \& {Wiebe}}]{roy10}
{Roy}, A. {et~al.} 2010, \apj, 708, 1611

\bibitem[{{Sayers} {et~al.}(2010){Sayers}, {Golwala}, {Ade}, {Aguirre}, {Bock},
  {Edgington}, {Glenn}, {Goldin}, {Haig}, {Lange}, {Laurent}, {Mauskopf},
  {Nguyen}, {Rossinot}, \& {Schlaerth}}]{sayers10}
{Sayers}, J. {et~al.} 2010, \apj, 708, 1674

\bibitem[{{Sayers} {et~al.}(2009){Sayers}, {Golwala}, {Rossinot}, {Ade},
  {Aguirre}, {Bock}, {Edgington}, {Glenn}, {Goldin}, {Haig}, {Lange},
  {Laurent}, {Mauskopf}, \& {Nguyen}}]{sayers09}
---. 2009, \apj, 690, 1597

\bibitem[{{Schlingman} {et~al.}(2010)}]{schlingman10}
{Schlingman}, W. {et~al.} 2010, \apj, in prep

\bibitem[{{Schuller} {et~al.}(2009){Schuller}, {Menten}, {Contreras},
  {Wyrowski}, {Schilke}, {Bronfman}, {Henning}, {Walmsley}, {Beuther},
  {Bontemps}, {Cesaroni}, {Deharveng}, {Garay}, {Herpin}, {Lefloch}, {Linz},
  {Mardones}, {Minier}, {Molinari}, {Motte}, {Nyman}, {Reveret}, {Risacher},
  {Russeil}, {Schneider}, {Testi}, {Troost}, {Vasyunina}, {Wienen}, {Zavagno},
  {Kovacs}, {Kreysa}, {Siringo}, \& {Wei{\ss}}}]{schuller09}
{Schuller}, F. {et~al.} 2009, \aap, 504, 415

\bibitem[{{Simon} {et~al.}(2006{\natexlab{a}}){Simon}, {Jackson}, {Rathborne},
  \& {Chambers}}]{simon06a}
{Simon}, R., {Jackson}, J.~M., {Rathborne}, J.~M., \& {Chambers}, E.~T.
  2006{\natexlab{a}}, \apj, 639, 227

\bibitem[{{Simon} {et~al.}(2006{\natexlab{b}}){Simon}, {Rathborne}, {Shah},
  {Jackson}, \& {Chambers}}]{IRDCdistance}
{Simon}, R., {Rathborne}, J.~M., {Shah}, R.~Y., {Jackson}, J.~M., \&
  {Chambers}, E.~T. 2006{\natexlab{b}}, \apj, 653, 1325

\bibitem[{{Stark} \& {Lee}(2006)}]{stark06}
{Stark}, A.~A. \& {Lee}, Y. 2006, \apjl, 641, L113

\bibitem[{{Stil} {et~al.}(2006){Stil}, {Taylor}, {Dickey}, {Kavars}, {Martin},
  {Rothwell}, {Boothroyd}, {Lockman}, \& {McClure-Griffiths}}]{stil06}
{Stil}, J.~M. {et~al.} 2006, astro-ph/0605422

\bibitem[{{Testi} \& {Sargent}(1998)}]{Testi1998}
{Testi}, L. \& {Sargent}, A.~I. 1998, \apjl, 508, L91

\bibitem[{{van Dishoeck} \& {Blake}(1998)}]{vandishoeck98}
{van Dishoeck}, E.~F. \& {Blake}, G.~A. 1998, \araa, 36, 317

\bibitem[{{Williams} {et~al.}(2000){Williams}, {Blitz}, \&
  {McKee}}]{williams00}
{Williams}, J.~P., {Blitz}, L., \& {McKee}, C.~F. 2000, Protostars and Planets
  IV, 97

\bibitem[{{Young} {et~al.}(2006){Young}, {Enoch}, {Evans}, {Glenn}, {Sargent},
  {Huard}, {Aguirre}, {Golwala}, {Haig}, {Harvey}, {Laurent}, {Mauskopf}, \&
  {Sayers}}]{young06}
{Young}, K.~E. {et~al.} 2006, \apj, 644, 326

\end{thebibliography}

\pagebreak

\Figure{f1}
{The coverage of the BGPS, showing the continuous coverage in the
first quadrant, and the regions targeted in the outer Galaxy.  The
background greyscale is IRAS 100 \mum.}{fig:Coverage}{1.0}

\Figure{f2}
{Top: the position and ellipticity of the detectors in the Bolocam
focal plane as mapped to the sky.  The size of the pixel corresponds
to its FWHM; it is clear that the beams do not completely overlap on
the sky.  The relative response to source is shown by the greyscale,
with the darkest being the largest response.  The path of the center
of each pixel as the array is scanned from left to right is shown for
a particular orientation of the array relative to the scan direction.
In this case, the sampling of the sky orthogonal to the scan direction
is poor.  Bottom: the same as the top, but with the array rotated to
its optimal angle relative to the scan direction.  It is clear that
the sampling orthogonal to the scan direction is now much more
uniform.}{fig:Array}{0.7}

\Figure{f3}{The distribution of the pointing calibration sources
  across the local sky in Hawaii during July 2007 (Epoch V) when the
  master pointing model was constructed.  Note that the entire local
  sky is sampled, and the residual pointing RMS is valid for any point
  on the sky.}{fig:PointingCalibrators}{1.5}

\begin{figure}
\begin{center}
  \subfigure[The pointing model for Epoch V, from which the master
  reference images were derived for subsequent alignment.  The left
  column is the pointing model correction and the right the residuals
  (both in arcseconds).  The red line indicates the fitted model. 510
  pointing sources were included, and the final RMS was 5.77\arcsec\ in
  altitude and 3.21\arcsec\ in azimuth, or a total RMS offset of
  6.6\arcsec.  No systematic offset with azimuth was found.]
  {\label{fig:PointingModel-a}\includegraphics[scale=0.6]{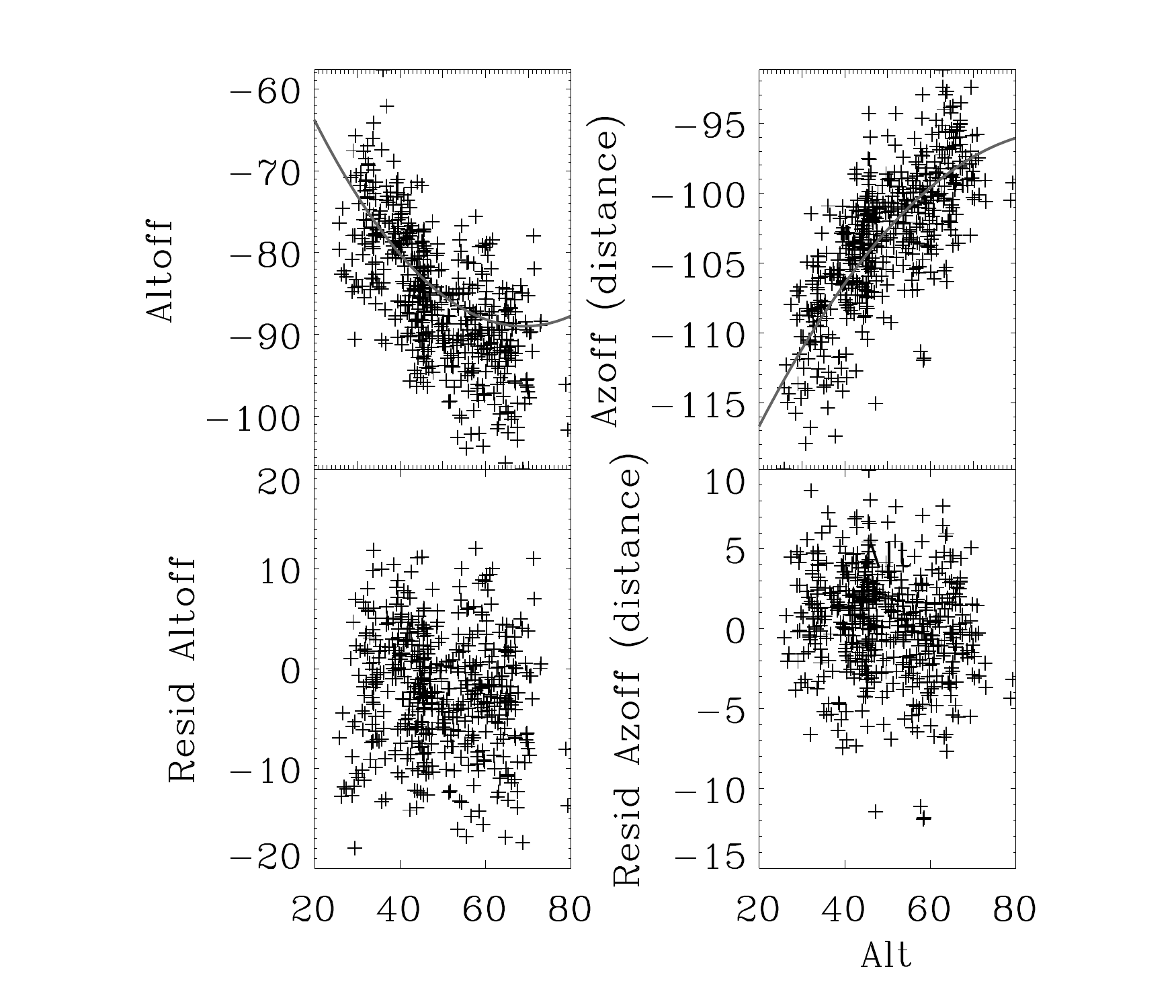}}

  \subfigure[The residuals of the pointing model.  The Gaussian fits have
  $\sigma = 5.94,2.92$ in altitude offset and azimuth offset respectively.]
{\label{fig:PointingModel-b}\includegraphics[scale=0.6]{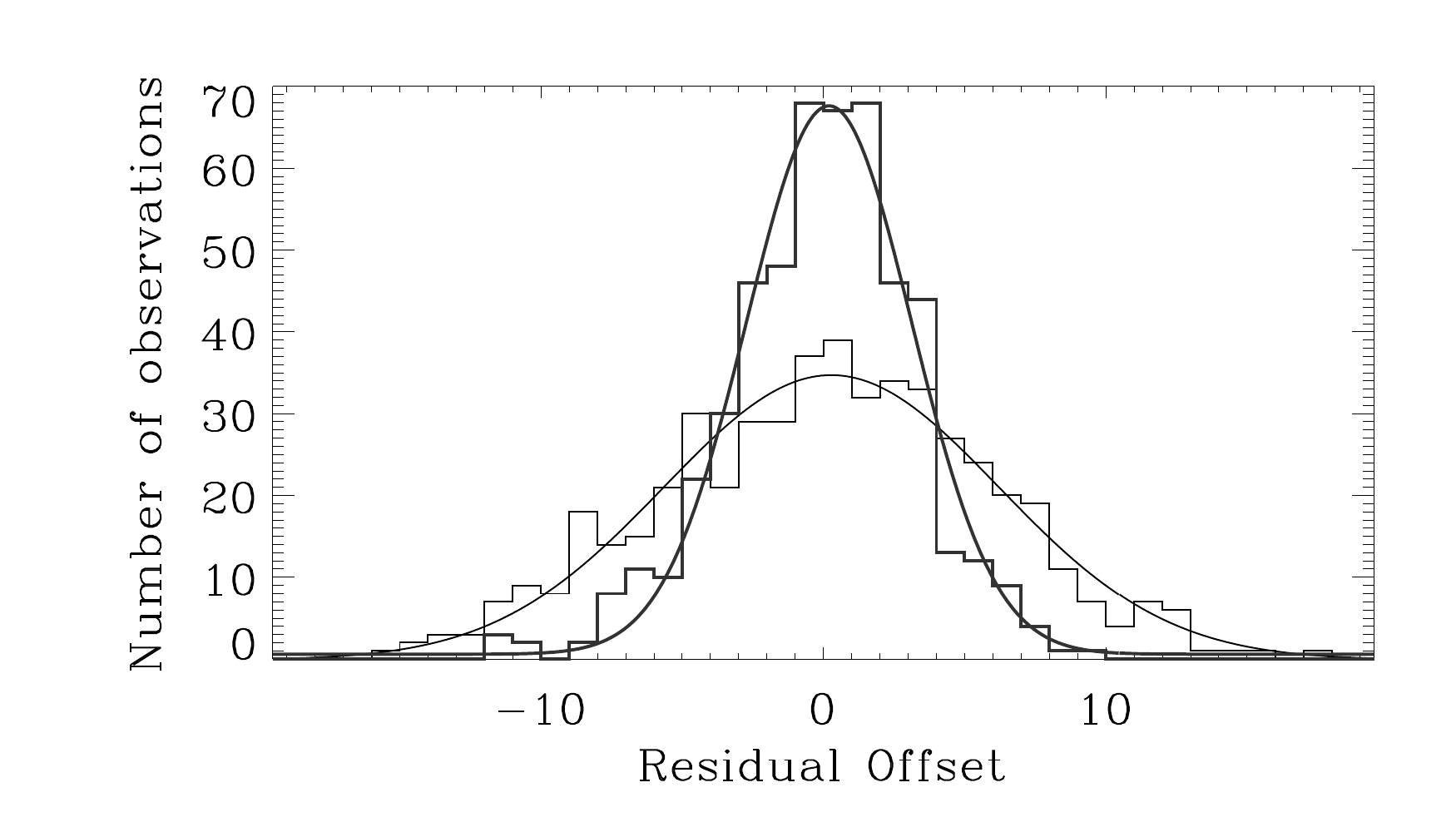}}

\caption{The Bolocam pointing model for Epoch V.}
\label{fig:PointingModel}
\end{center}
\end{figure}

\begin{figure}

  \subfigure[Left: the Bolocam map in a portion of the $\lon=30$
  field.  Right: the SLC maps in the same region.  The green contours
  on the Bolocam map indicate the positions of the SLC sources.  Note
  that the SLC maps only cover a small portion the entire region
  shown, corresponding to the isolated sources indicated by the
  contours.  Negative bowls in both images are forced to zero before
  cross-correlating.]
  {\label{fig:SCUBAPointingComparison-a}\includegraphics[scale=0.8]
  {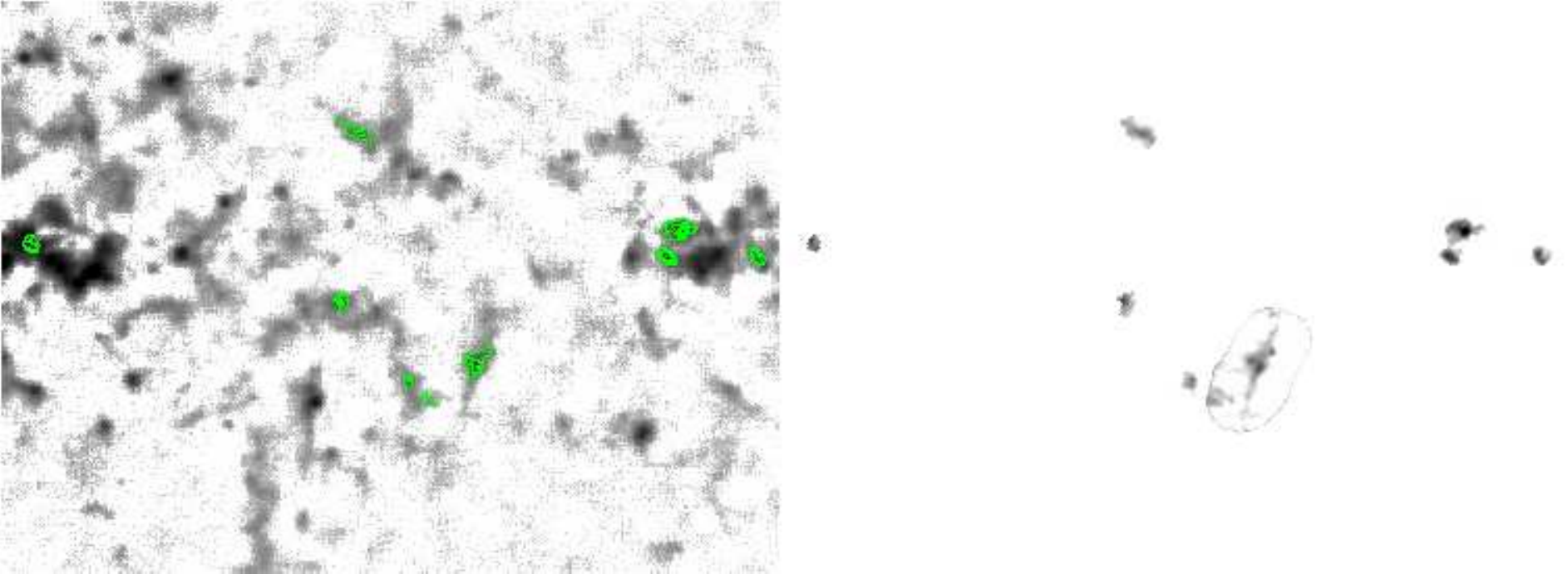}}

  \begin{tabular}{cc}

  \subfigure[The cross-correlation map of the BGPS and SLC maps in
  Figure \ref{fig:SCUBAPointingComparison-a}.] 
  {\label{fig:SCUBAPointingComparison-b}\includegraphics[scale=0.45]
  {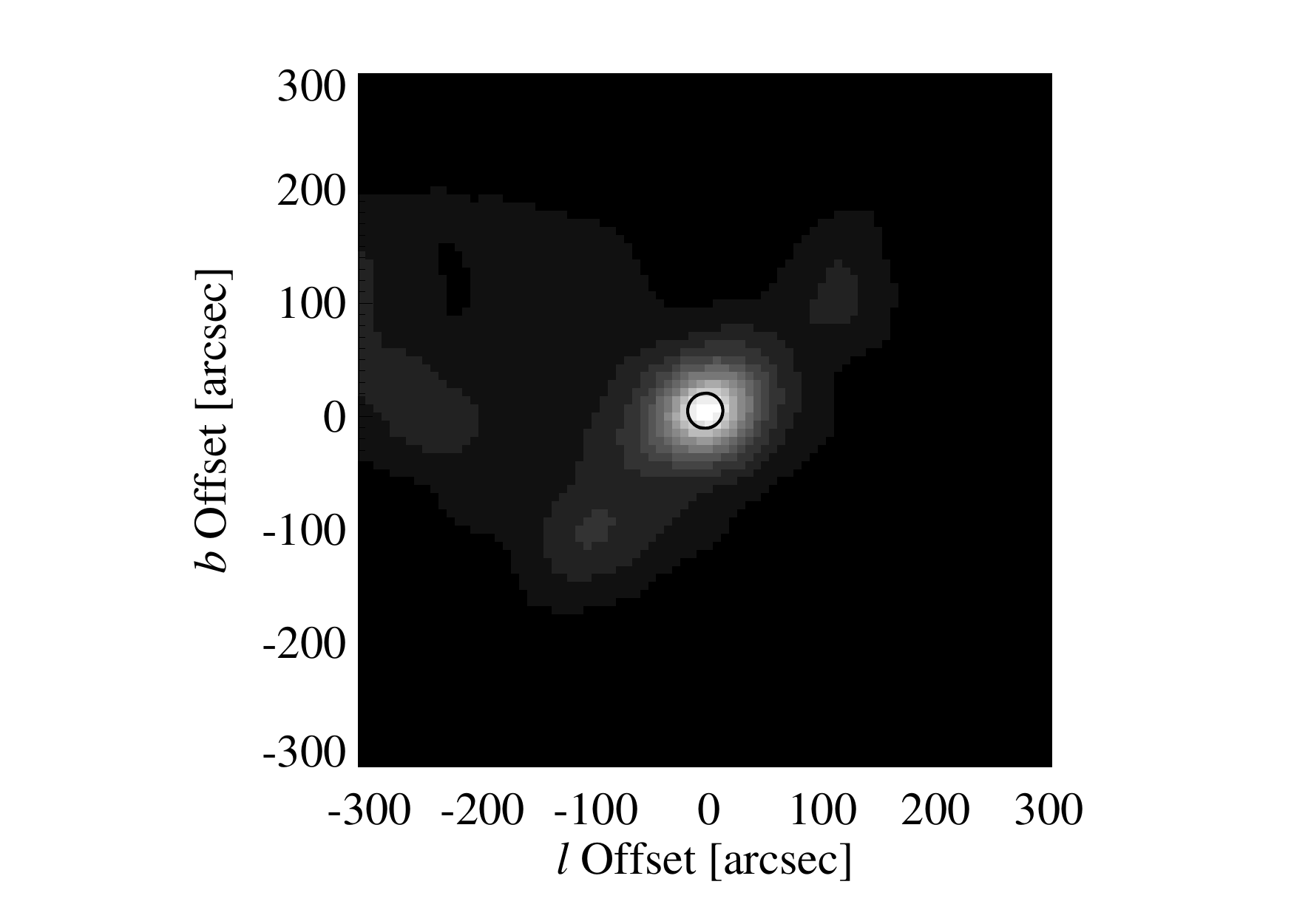}}

  \subfigure[The measured offset between Bolocam and SCUBA Legacy
sources \citep{difrancesco08}, using the method described in Setion
\ref{sec:SCUBAPointingComparison}.  The circle indicates the $1\sigma$
region, which is also consistent with the error derived from the
internal Bolocam pointing model.]
{\label{fig:SCUBAPointingComparison-c}\includegraphics[scale=0.45]
{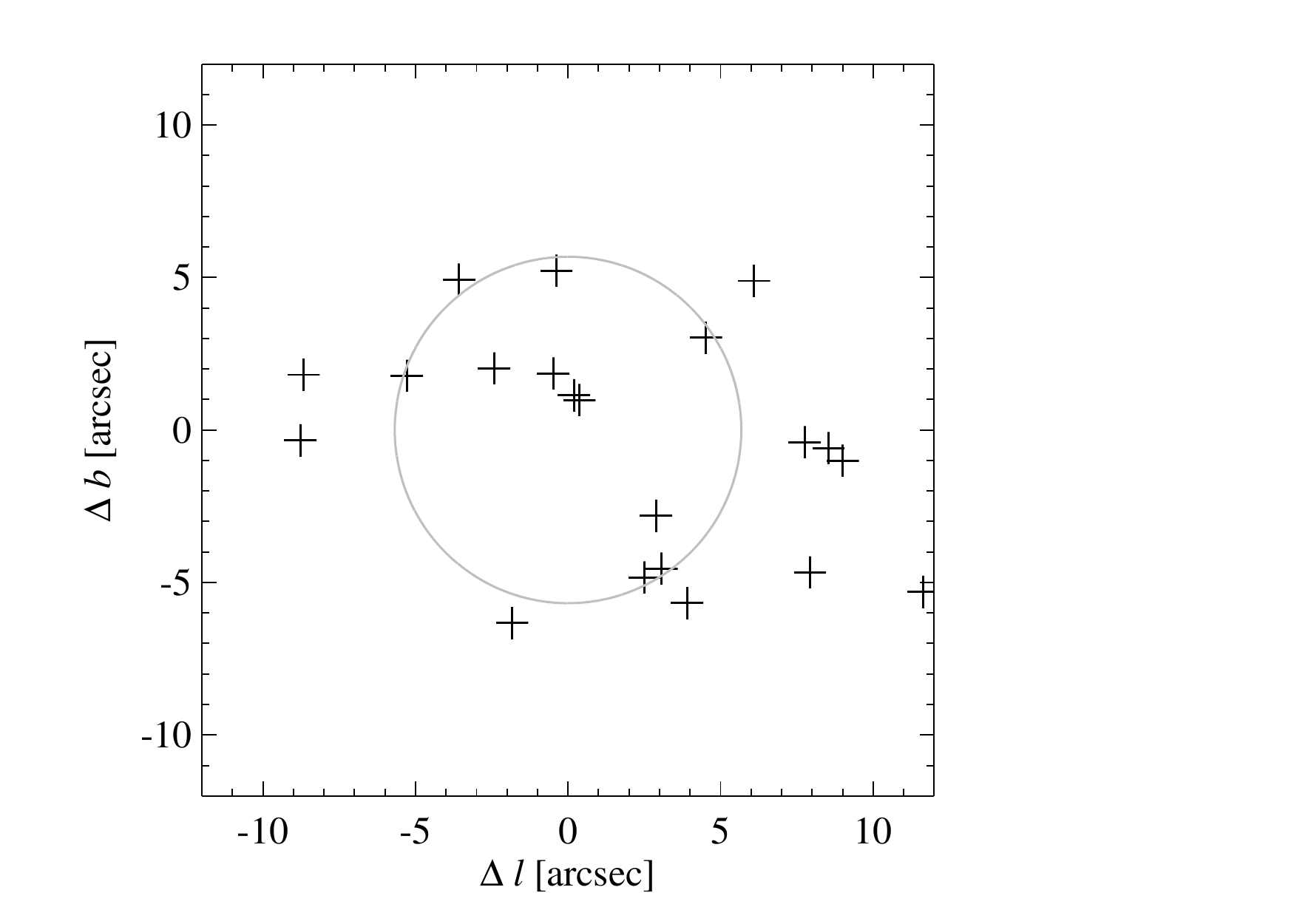}}

\end{tabular}

\caption{Comparing the Bolocam positions to those of the SCUBA Legacy Catalog.}
\label{fig:SCUBAPointingComparison}
\end{figure}

\setcounter{subfig}{1}

\begin{figure}
\caption{The behavior of the time series as iterative mapping proceeds.}
\label{fig:IterativeMapping}

\renewcommand{\thefigure}{\arabic{figure}\alph{subfig}}

\addtocounter{figure}{-1}
  
  \begin{minipage}{6.5in}
    \begin{center}
      \includegraphics[scale=0.9]{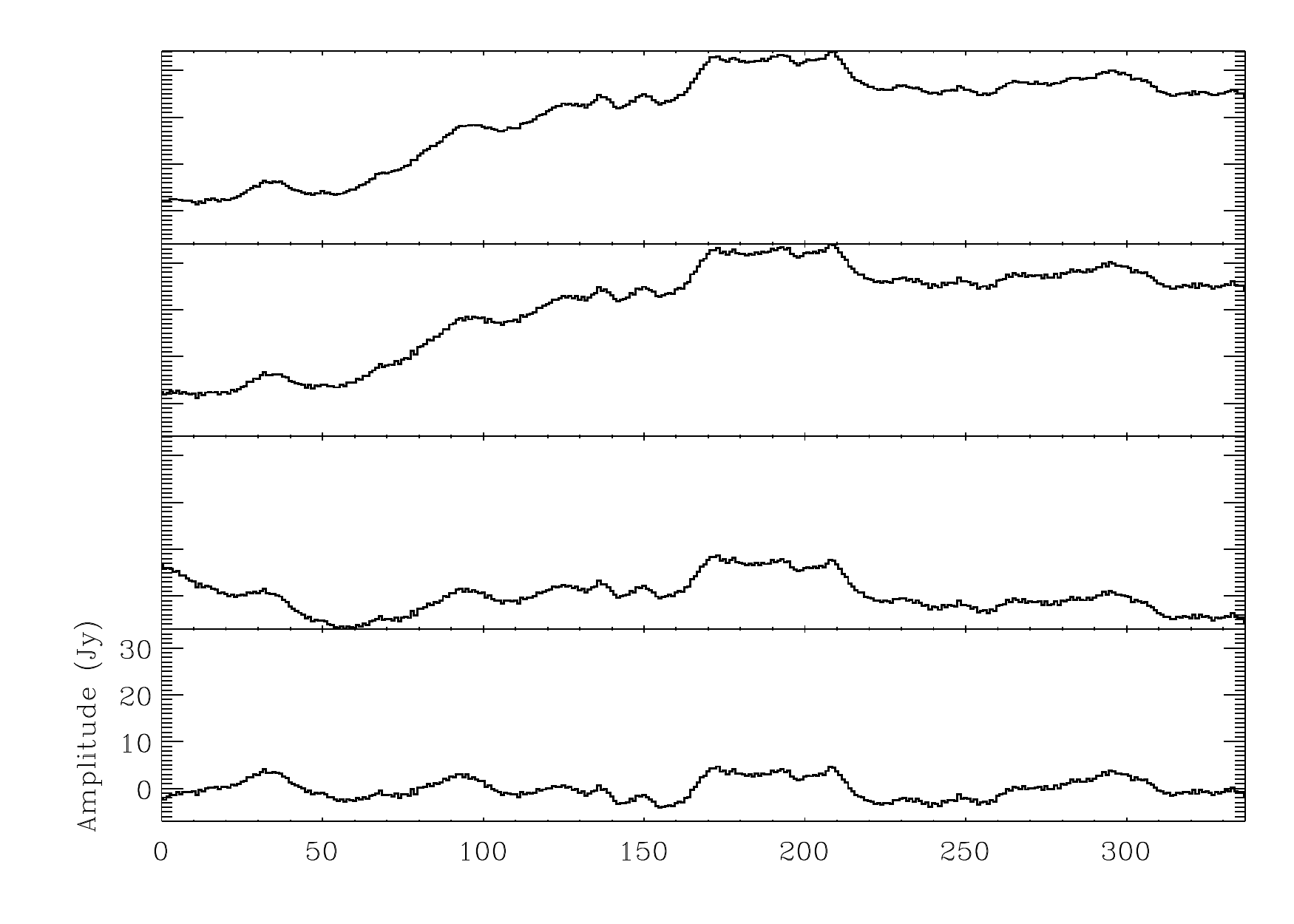}
      \caption{Removal of the $e$ and $p$ models.  From top top
      bottom: 1. Raw time series 2. After removal of residual 60Hz
      signal 3. Exponential decay function at scan turnarounds
      subtracted 4. A polynomial (with astrophysical source rejection)
      is fit to remove large-scale variations across the scan. }
    \label{fig:IterativeMapping-a}
    \end{center}
  \end{minipage}
\end{figure}

\renewcommand{\thefigure}{\arabic{figure}\alph{subfig}}

\addtocounter{figure}{-1}
\addtocounter{subfig}{1}
  
\begin{figure}
  \begin{minipage}{6.5in} \begin{center}
    \includegraphics[scale=0.9]{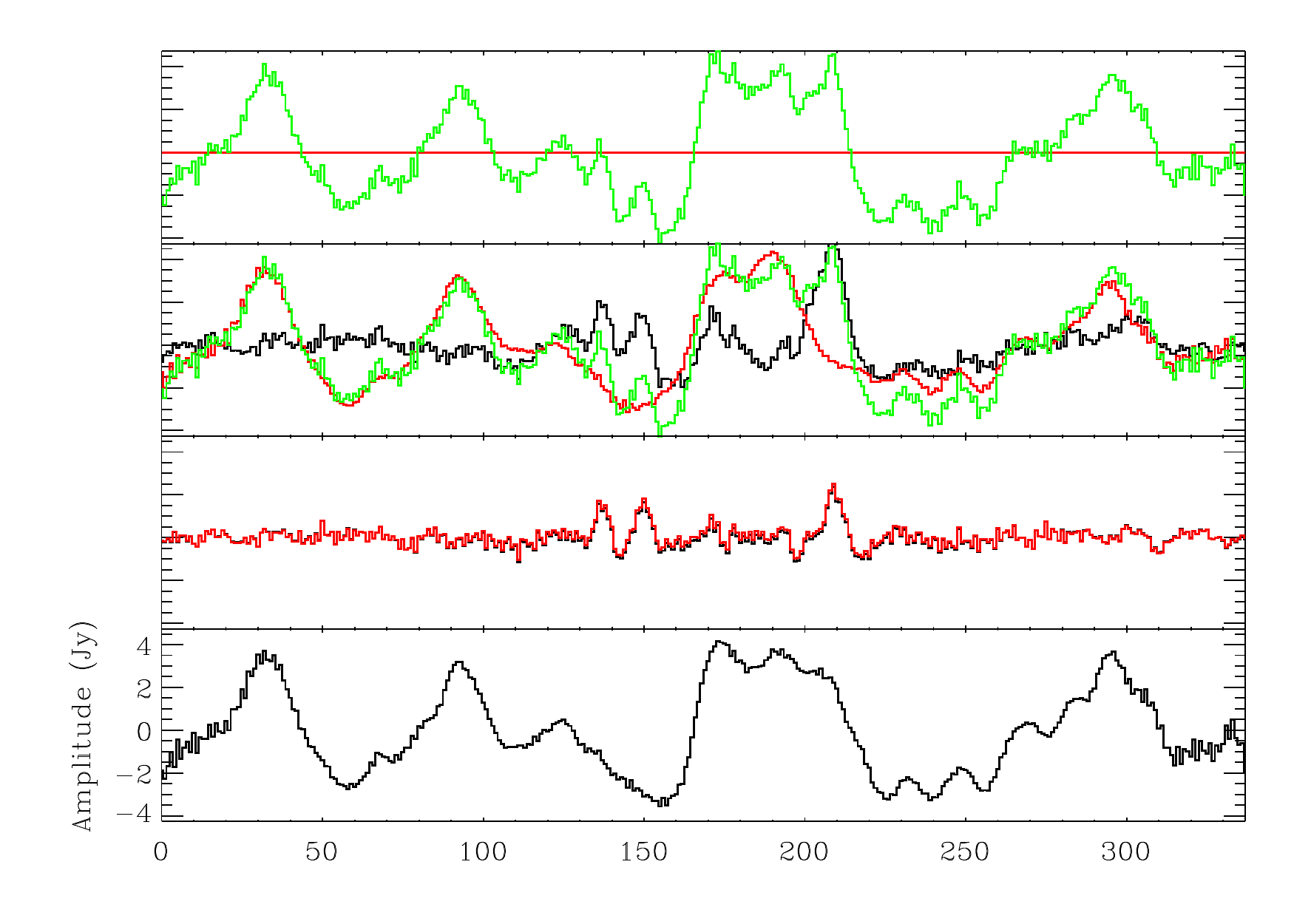}

    \caption{Panel 1 is the same as panel 4 in the previous figure.
    Before the first cleaning, the raw data (black) and remainder data
    (green) are equal because there is no model.  Panel 2 shows the
    median-atmosphere subtraction, which is the first- order
    correction.  Panel 3 shows the astrophysical signal left over from
    the PCA selection.  Panel 4 shows the total atmosphere model from
    PCA and median estimation}

    \end{center}
    
    \label{fig:IterativeMapping-b} \end{minipage}
\end{figure}

\addtocounter{figure}{-1}
\addtocounter{subfig}{1}

\begin{figure}
  \begin{minipage}{6.5in} \begin{center}
    \includegraphics[scale=0.9]{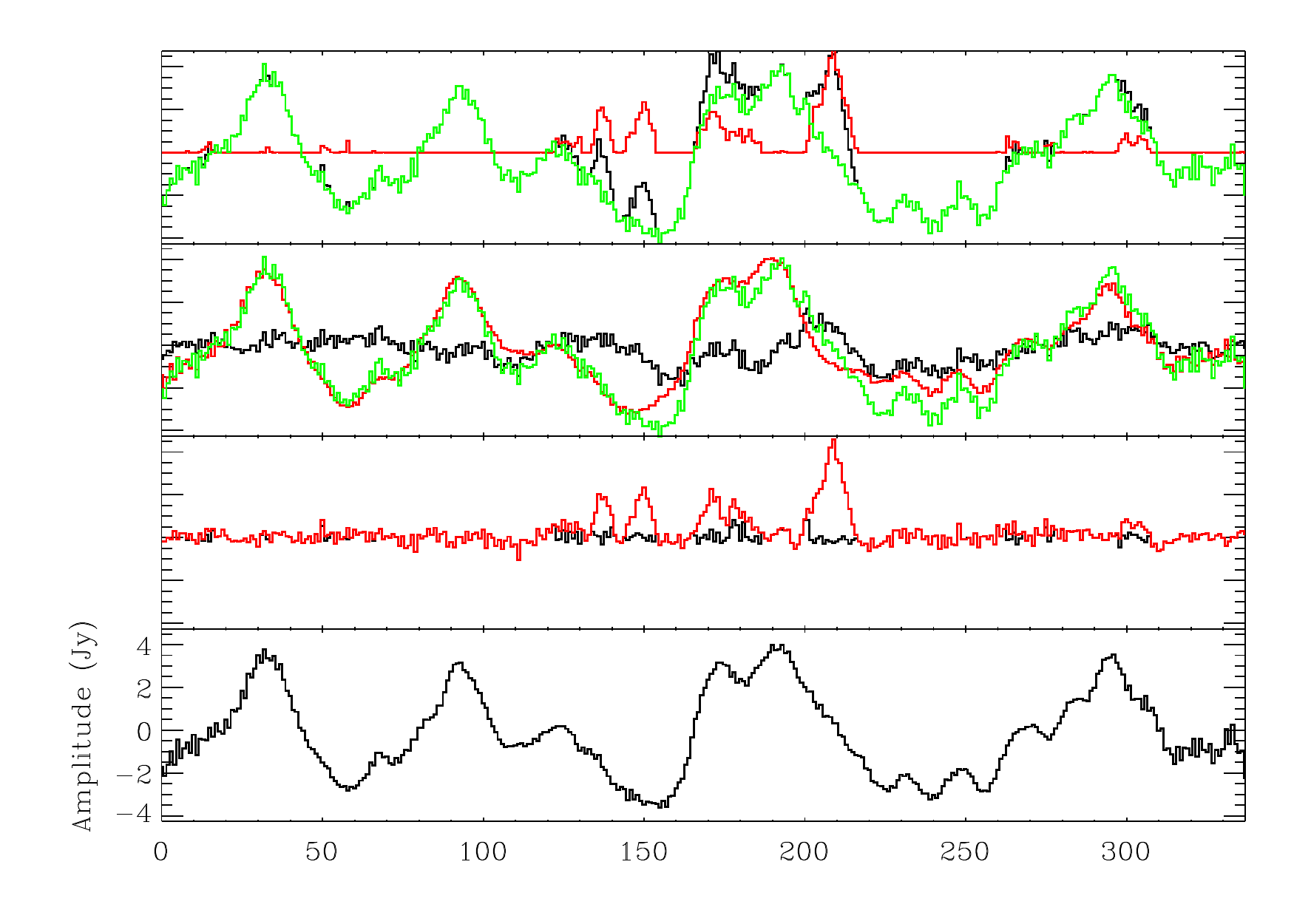}

    \caption{Same as previous figure, except after 20 iterations.
    Panel 1 - The deconvolved astrophysical map is returned to a
    timestream (red) and subtracted from the 'raw' data (black).
    Panel 2 - The median of the remainder (green) is subtracted as the
    first atmosphere estimation Panel 3 - The cumulative astrophysical
    model (red) and the additional astrophysical signal from iteration
    20 (black) Panel 4 - The atmosphere signal after 20 iterations}

    \end{center}
    
    \label{fig:IterativeMapping-c} \end{minipage}
\end{figure}

\renewcommand{\thefigure}{\arabic{figure}}

\begin{figure}
  \begin{minipage}{6.5in}
    \begin{center}
      \includegraphics[angle=270,scale=0.6]{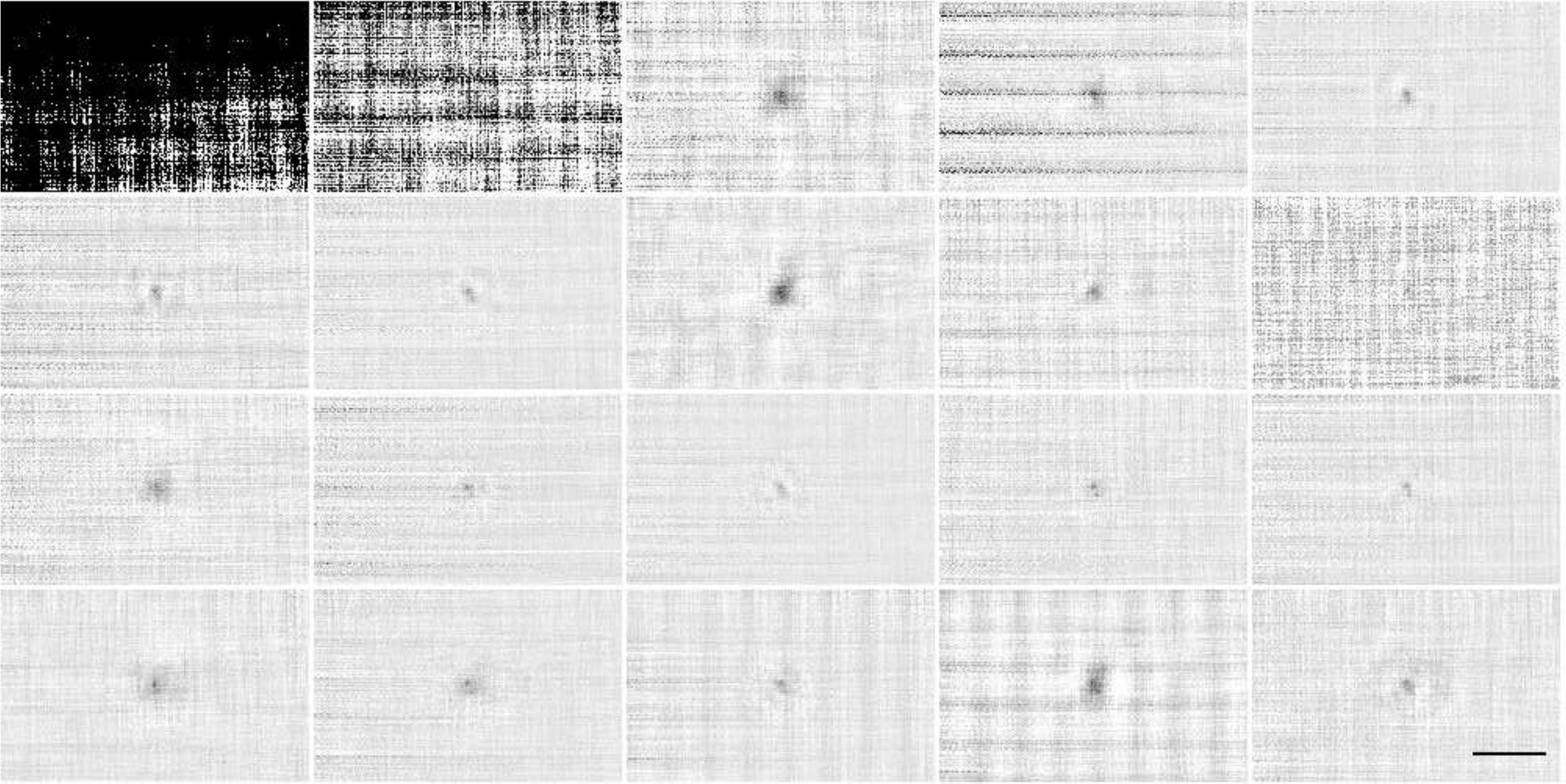}
    \end{center}
  \end{minipage}
\vspace{0.25in}
  \begin{minipage}{6.5in}
    \begin{center}
      \includegraphics[angle=270,scale=0.6]{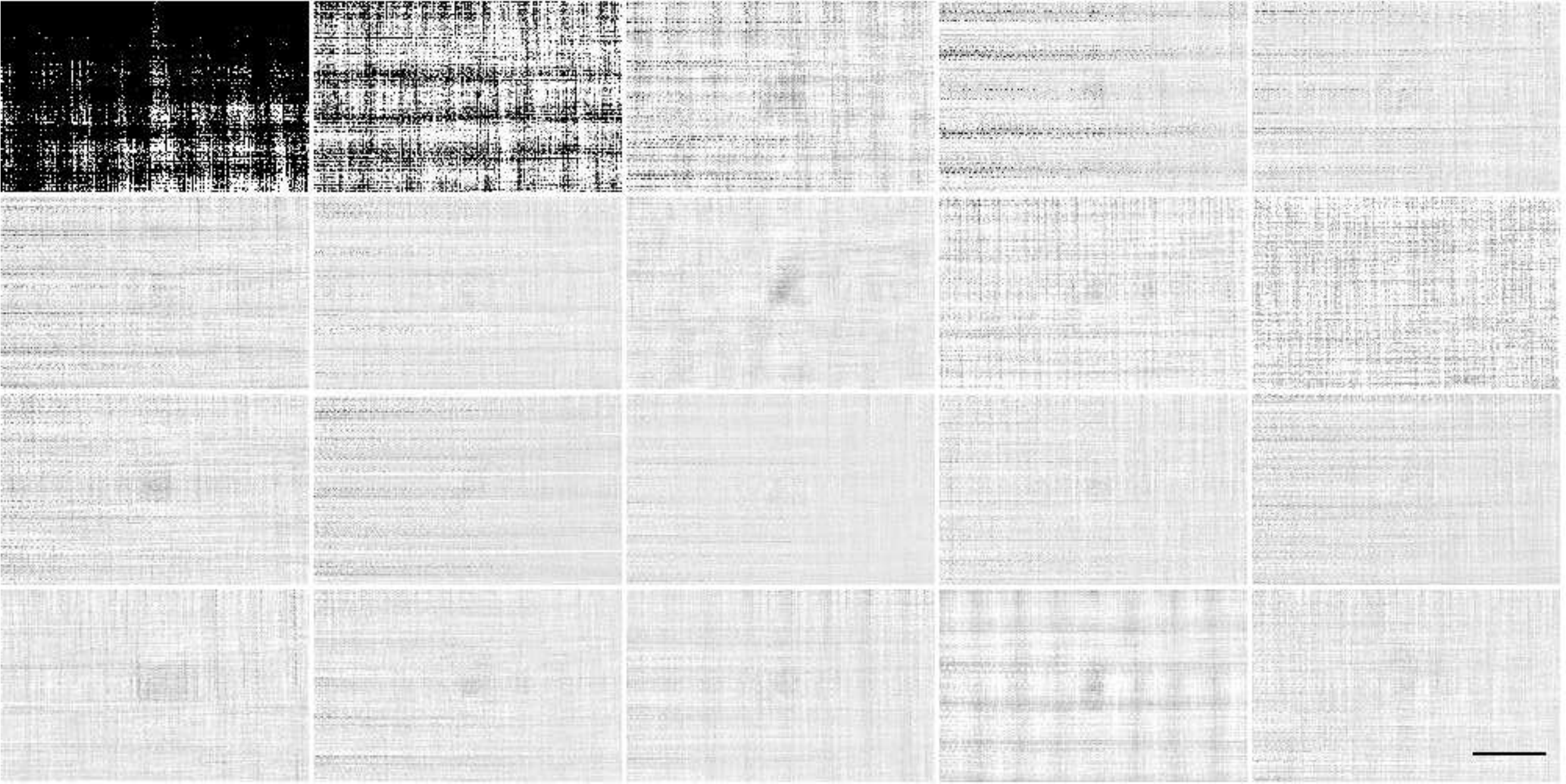}
    \end{center}
  \end{minipage}
  \caption{Each PCA component in time can be turned into a map.  Above
are selected images (in inverted greyscale) to illustrate the PCA
components and the effect of iteration on the process of creating
these components.  The displays are in Galactic coordinates with a
color range from -.1 to 1.0 \jyb.  The bar in the bottom right
indicates a 10\arcmin\ length.  {\it Top 4 rows}: a grid of maps of
each of the first 20 PCA components displayed at the same scale.  It
is clear that NCG 7538, the object imaged, has emission in each
component.  There are varying levels of noise in each component, with
the first and second being the most obvious atmospheric components.
Most of the other noise components are probably detector noise
correlated among a subset of the bolometers.  The streaks are
residuals of the scan-turnaround noise that was not removed by the
exponential model fit.  {\it Bottom 4 rows}: The same figure, but
after 20 iterations.  The figure is the breakdown of the atmosphere
remainder (i.e.  the raw data minus the astrophysical model) into
eigenfunctions.  Very little astrophysical flux remains at any level
of correlation, though there is some at large (few arcminute) scales.}
\label{fig:PCA_Graphical}

\end{figure}

\FigureTwo{f8a}{f8b}
{Left: An illustration of a ``glitch'' in a single bolometer
timestream (red), likely due to a cosmic ray strike.  Note the acausal
ringing due to the application of the downsampling filter.  The time
series for a physically adjacent bolometer is shown in blue, and a
bolometer which does not pass over the source in green.  The PCA
atmosphere estimation has already been subtracted from the
data. Right: The distribution of glitch amplitudes flagged and removed
from the data in the $\lon=111$ field with their corresponding rate of
occurrence.  Positive-going glitches are the solid histogram and
negative-going the dashed.  The negative-going glitches are likely the
ringing from positive-going glitches.  Both distributions are
well-described by a power law above $\sim 3$~\jyb; below this
detection threshold, glitches may contribute to the overall
noise.}{fig:Glitches}{0.8}

\setcounter{subfig}{1}
\begin{figure}
  \begin{minipage}{3.25in}
    \begin{center}
      \includegraphics[scale=0.25]{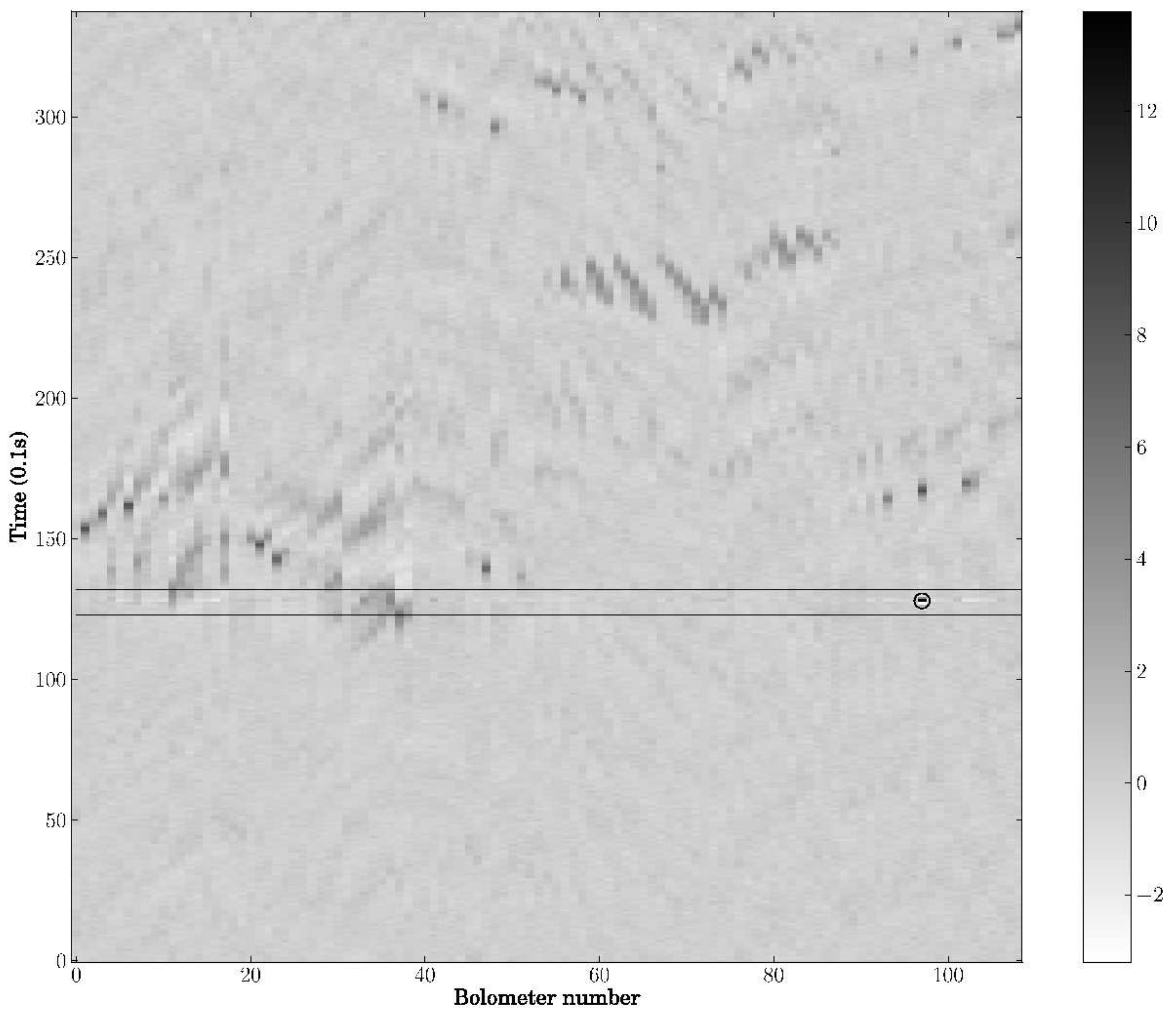}
    \end{center}
  \end{minipage}
  \begin{minipage}{3.25in}
    \begin{center}
      \includegraphics[scale=0.25]{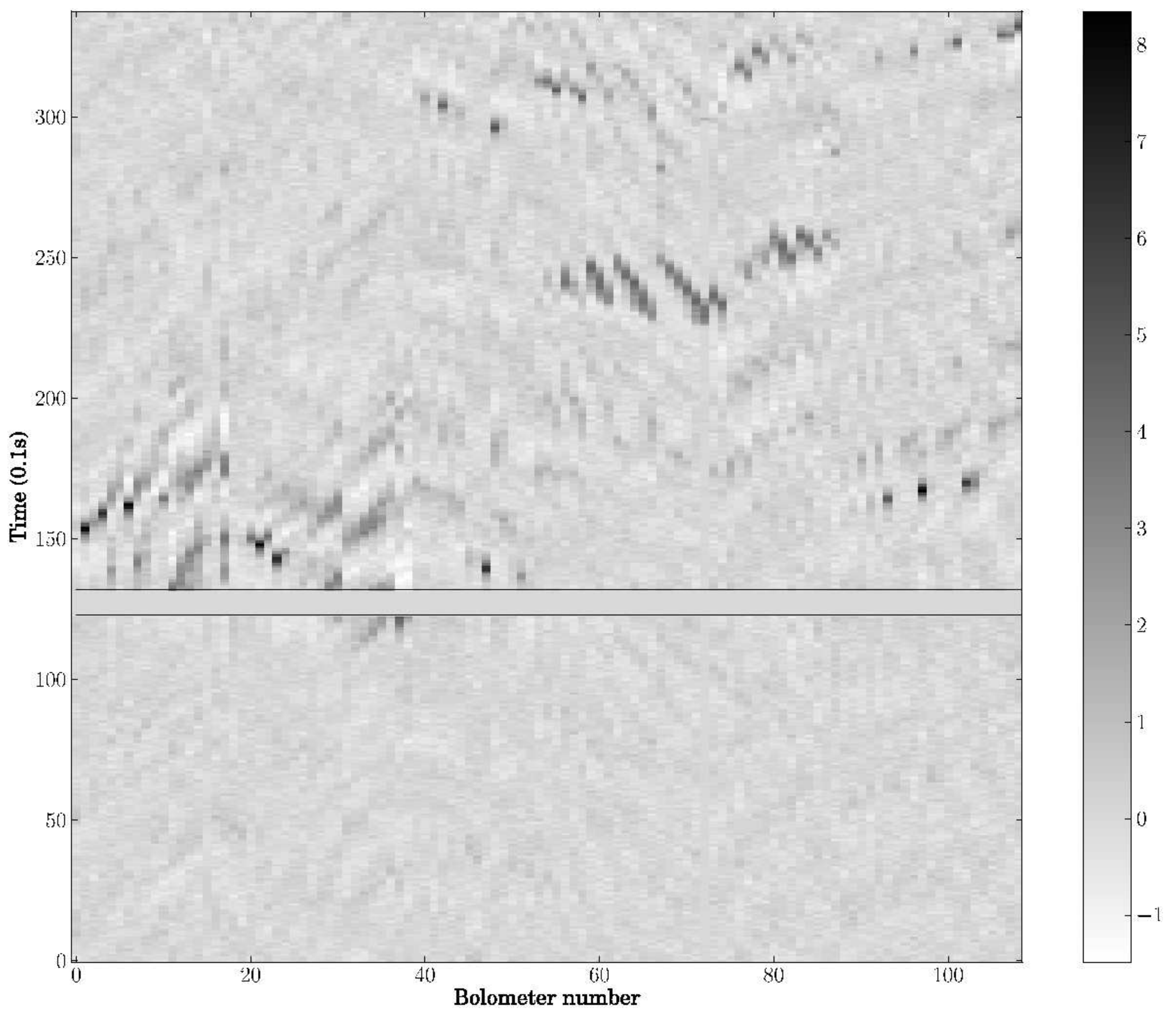}
    \end{center}
  \end{minipage}

  \caption{An illustration of the flagging process using the waterfall
  plot.  Left: A glitch is circled in the waterfall plot (bolometer
  98, near sample 125).  It is most prominent as a single hot pixel.
  Because of the PCA subtraction, however, the effect of the glitch
  also propagates over the other bolometers, appearing as a horizontal
  stripe; to emphasize this, two horizontal dark lines are placed
  above and below the stripe.  Right: The same data displayed after
  flagging out all data affected.  Note that the other features in the
  plot correspond to passages of the bolometers over real emission,
  including NGC 7538, which has a peak of $\sim 7$~\jyb.}

  \label{fig:Flagger}

\end{figure}

\Figure{f10}
{The effect of deconvolution on the iterative process can be seen in
its effect on the residuals in these images of Sgr B2, all displayed
at the same scale. From left to right: map, residual, model. Top to
bottom kernel size: 14\farcs4, 21\farcs6, 31\farcs2, 7\farcs2. The
final version of the pipeline uses 14\farcs4, which has the result of
leaving no flux in the residual at the location of Sgr B2, and does
not ``dig a hole'' in the residual map, as the 7\farcs2\ kernel does.
The 7\farcs2\ kernel also lacks the noise-rejection features of the
larger kernels, as can be seen in the bottom right
panel.}{fig:Deconvolution}{1.0}

\Figure{f11}{The variation of RMS noise as a function of Galatic
longitude for the fields in the inner Galaxy.  The noise reaches a
minimum at $l=31$ at 11 m\jyb, and a maximum at $l=57$ at 57 m\jyb.
The shaded region shows the range within which 68\% of the RMS
measurements in a field lie.}{fig:NoiseVsLongitude}{1.0}

\FigureTwo{f12a}{f12b} {Left: Average calibration curve [V/Jy] versus
the mean detector voltage, a proxy for atmospheric loading.  Asterisks
are observations of Mars and crosses of Uranus.  The black line is a
2nd-order fit with 0,0 forced (no response if no measurable potential
difference) .  Right: Scaling of the relative response to the
atmosphere for a single detector compared to the array median.
Outliers are from noisy scans that are strongly downweighted.  No
systematic trend in the relative calibration is apparent over the
range of atmosphere loading observed; the median value is indicated by
the dashed grey line.}{fig:CalibrationCurves}{1.0}

\Figure{f13}{An illustration of the relative sensitivity calibration
using the atmosphere as a calibrator.  Black is the median over all
bolometers (the 1st-order atmosphere model), red and green are
individual bolometers before (top) and after (bottom) applying the
relative calibration.  Note the improved agreement.}
{fig:Flatfield}{1.0}

\FigureTwo{f14a}{f14b}{Left: The radial profile (solid) of the Bolocam
PSF derived from Neptune and Uranus observations, including all
bolometers.  The main lobe is well approximated by a Gaussian (dotted)
and the sidelobes are due to the Airy pattern from truncation of the
beam at the cold Lyot stop (dashed).  Error bars indicate the
azimuthal standard deviation.  Right: Enclosed flux in an aperture of
a given radius based on the measured PSF.  This curve is the basis of
the point source aperture corrections (Table
\ref{tab:ApertureCorrections}).  Vertical dashed lines indicated 20,
30, and 40 arcsecond radii.}{fig:PSF}{1.0}

\Figure{f15} {The fractional flux recovered as a function of
source size for well-separated Gaussian sources with FWHM as
indicated.  The thin lines show (from top to bottom) the effect of
increasing the number of PCA components subtracted (3, 7, 10, 13, 16,
21, 26, and 31).  The thick line shows the curve for the 13 PCA
component cleaning used in the released data; the flux recovery drops
to 50\% at 3\farcm8\ for this case.  The 3 and 7 PCA cases show bumps
because atmospheric noise is still present at large scales.  The
dashed line shows the approximation of a a linear high pass filter
with ``brick wall'' cutoff below 1/10 inverse arcminutes.}
{fig:PCA_Filter}{1.0}

\clearpage

\Figure{f16}
{Top: the Bolocam 1.1 mm bandpass (thick line).  Also shown is the
atmospheric transmission at Mauna Kea for 1 mm of precipitable water
vapor (thin line).  The grey curve is the MAMBO-2 bandpass (taken from
{\tt
http://www.mpifr-bonn.mpg.de/div/bolometer/mambo\_parameters\_020430.html}).
Note that the Bolocam bandpass is well away from the strong water
absorption features at 183 and 325 GHz.  Bottom: The effect of line
contamination is illustrated with the integrated line intensities from
the \citet{nummelin1998} survey of Sgr B2, who found that 22\% of the
flux density in one pointing was due to line emission.  Note that the
Bolocam passband rejects $>90\%$ of the $^{12}$CO flux.  However, as
shown from the \citet{nummelin1998} survey, SO$_2$ and CH$_3$OH lines
can be strong contributors to line flux in the passband due to their
broad width. Other lines lying in the Bolocam passband include
CS($5\to4$) and ($6\to5$) (245 and 293 GHz), HCN($3\to2$) (265 GHz)
and HCO$^+(3\to2)$ (267 GHz).} {fig:Bandpass}{1.0}

\FigureTwo{f17a}{f17b}
{Left: A plot of the values of the pixels in the M07 and BGPS maps
after both have been convolved to the same resolution on the same
grid.  The solid line is the expected 1:1 correspondence; the other
lines show the slopes obtained with various lower bounds as given in
Table \ref{tab:FluxComparison}.  There is clearly a great of scatter
about the mean.  Right: The same analysis, but for the M09 SIMBA
data.}{fig:FluxComparison}{1.0}

\Figure{f18} {Examples of the three image types in the IPAC V1.0 data
release.  Images are zoomed in to show detail; axes are offsets in
arcminutes from the center position.  Top left: MAP.  Middle: NOISEMAP
Bottom: NHITSMAP.  Note the slight ``ghosts'' in the residual
(NOISEMAP) at the positions of bright sources.  Note also the
``basket-weave'' pattern in the coverage (NHITSMAP) due to gaps in the
detector array combined with the scan
strategy.}{fig:SampleImages}{1.0}

\begin{figure}
  \caption{Images from the BGPS.}
  \label{fig:BGPSMontage}
  \addtocounter{figure}{-1}
  \setcounter{subfig}{1}
  \renewcommand{\thefigure}{\arabic{figure}\alph{subfig}}  
  \begin{minipage}{6.5in}
    \begin{center}
      \includegraphics[scale=0.8]{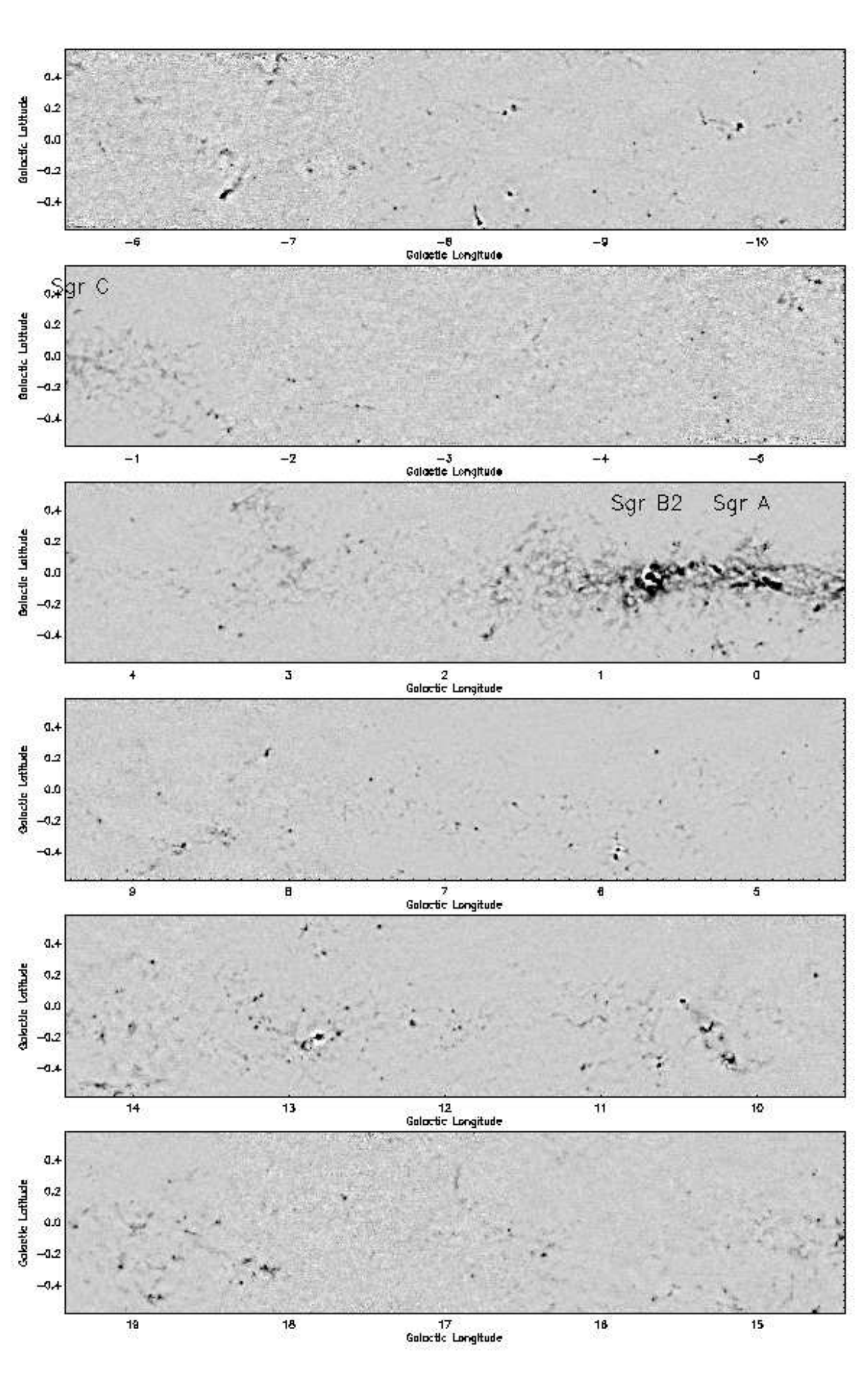}

      \caption{$\lon=-10.5$ to $\lon=19.5$ In this figure and the
      following, the brightest sources, e.g. Sgr B2, Sgr A, and
      sources near $\lon=10$ and $\lon=13$, appear to be saturated,
      but this is only a display artifact.  The astrophysical sources
      are always much fainter than the atmosphere (which is within the
      dynamic range of the detectors) and therefore do not saturate.
      The noise is more pronounced from $\lon=-7$ to $\lon=-2$ because
      this region was observed less.}

    \end{center}
  \end{minipage}
\end{figure}

\renewcommand{\thefigure}{\arabic{figure}\alph{subfig}}
\addtocounter{figure}{-1}
\addtocounter{subfig}{1}

\begin{figure}
  \begin{minipage}{6.5in} 
    \begin{center}
      \includegraphics[scale=0.8]{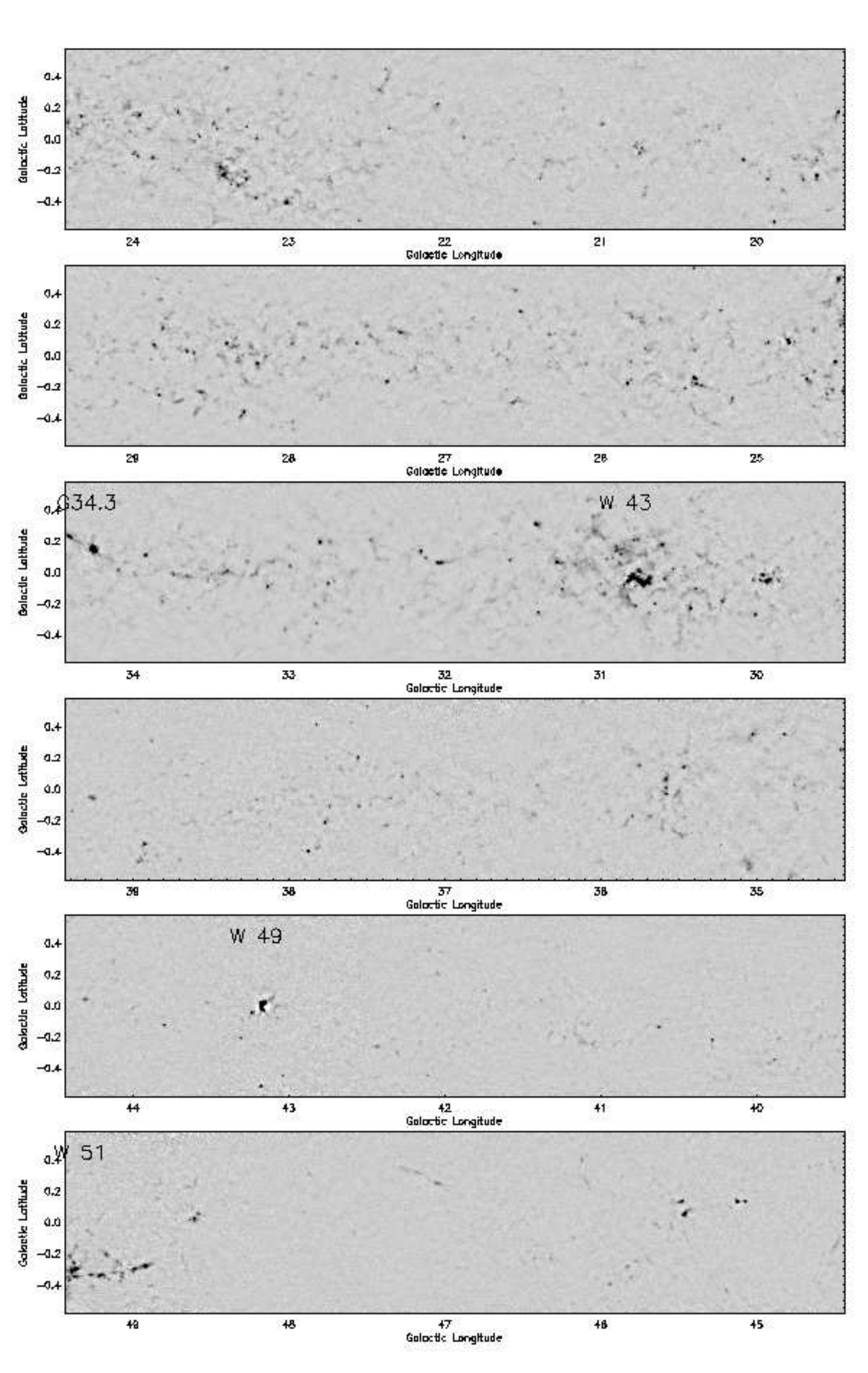} 
      \caption{$\lon=19.5$ to $\lon=49.5$.  G34.3+0.15, W 51, W 43, W
	49, and M 17 appear to be saturated, but this is only a
	display artifact.  The $20 < \ell < 40$ region through the 4-8
	kpc molecular ring and approximately the termination of the
	galactic bar is particularly rich in clumps.}
    \end{center}
  \end{minipage}
\end{figure}

\addtocounter{figure}{-1}
\addtocounter{subfig}{1}

\begin{figure}
  \begin{minipage}{6.5in} \begin{center}
    \includegraphics[scale=0.8]{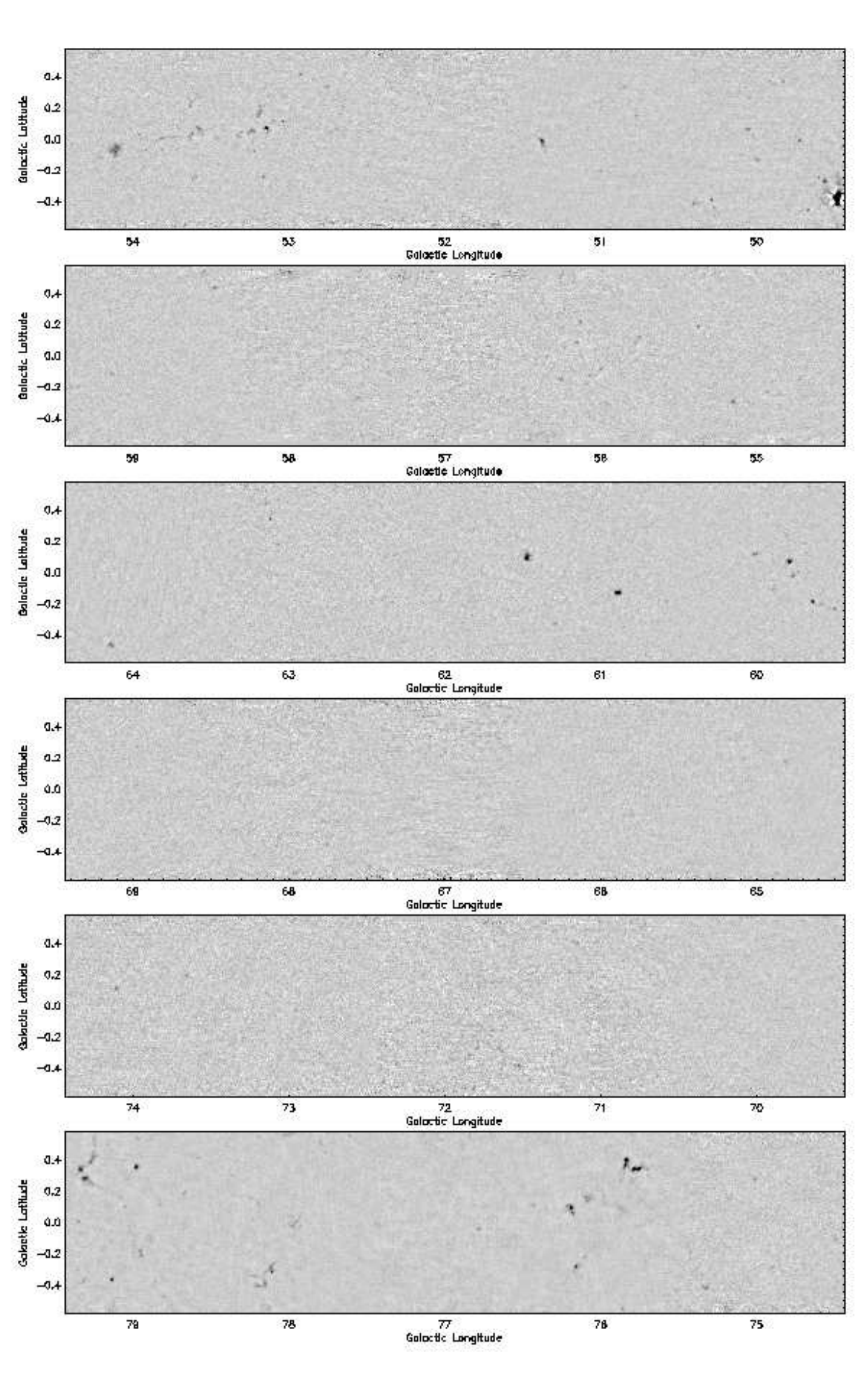}
    \caption{$\lon=49.5$ to $\lon=74.5$.  In comparison to the inner
    galaxy, the $65 < \ell < 75$ has a very sparse population of faint
    clumps} \end{center} \end{minipage}
\end{figure}

\addtocounter{figure}{-1}
\addtocounter{subfig}{1}

\begin{figure}
  \includegraphics[scale=0.7]{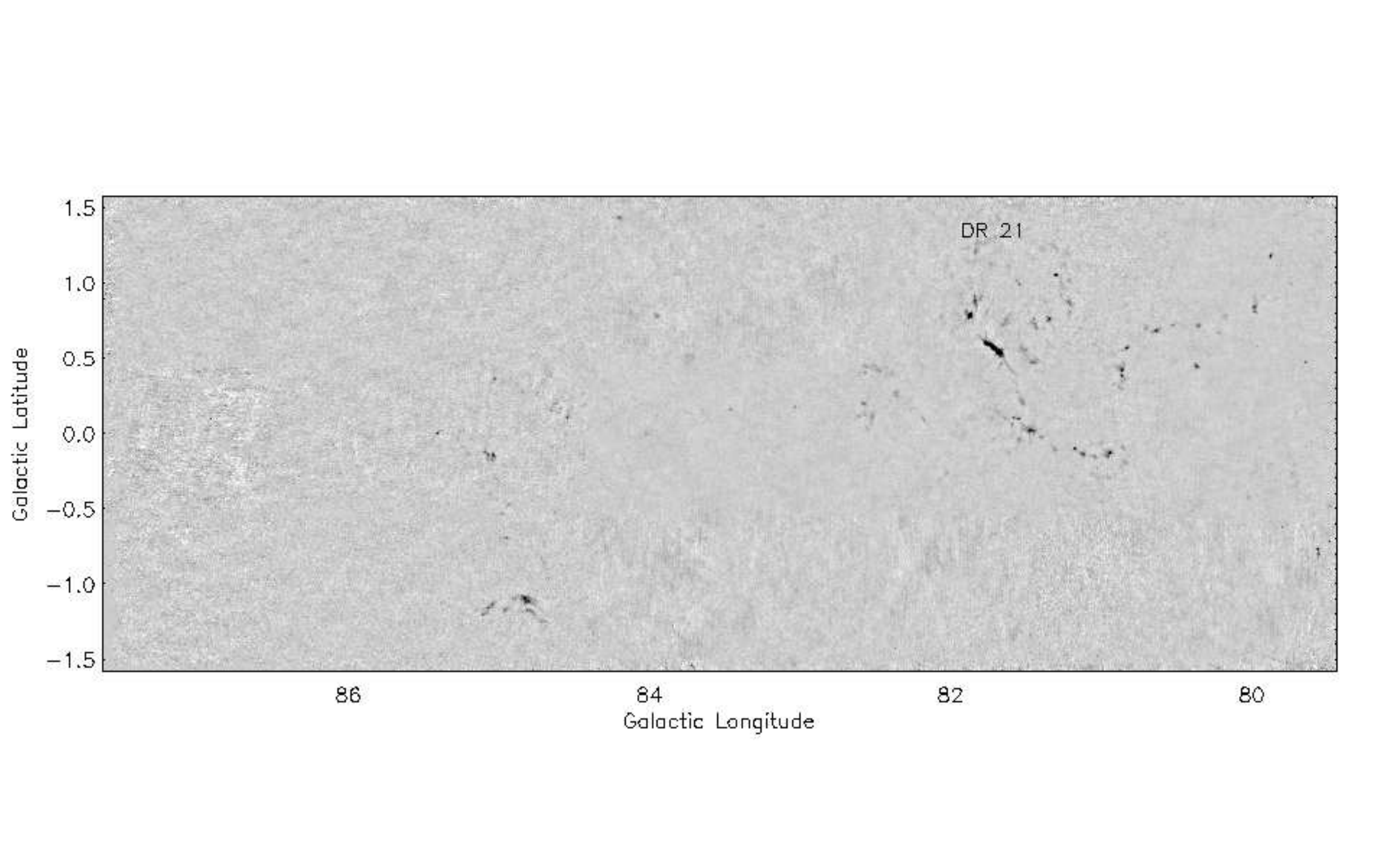} 
  \begin{center}
  \includegraphics[scale=0.8]{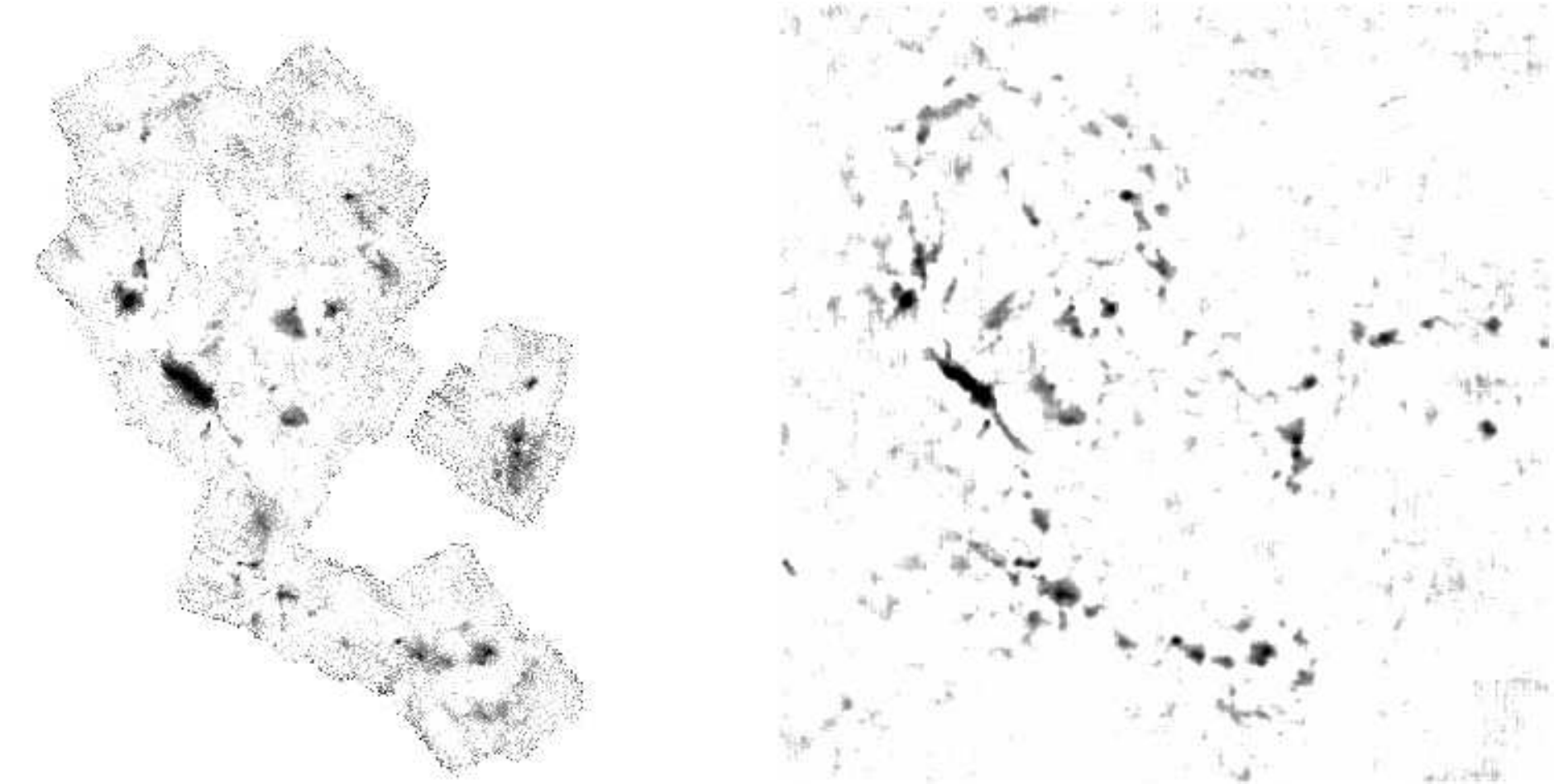}
  \end{center}
  \caption{Top: The Cygnus Arm. Note that coverage in $b$ is extended
  to $\pm 1.5\deg$. Bottom: A zoom-in of the region near DR21, showing
  at left the data from the M07 IRAM study of this region, and at
  right the BGPS map, in which the filamentary nature of the emission
  is more apparent.  Flux densities from the BGPS and M07 are compared
  in Section \ref{sec:FluxComparison}.}

  \label{fig:MontD}

\end{figure}

\addtocounter{figure}{-1}
\addtocounter{subfig}{1}

\begin{figure}
\begin{center}
  \includegraphics[scale=1.0]{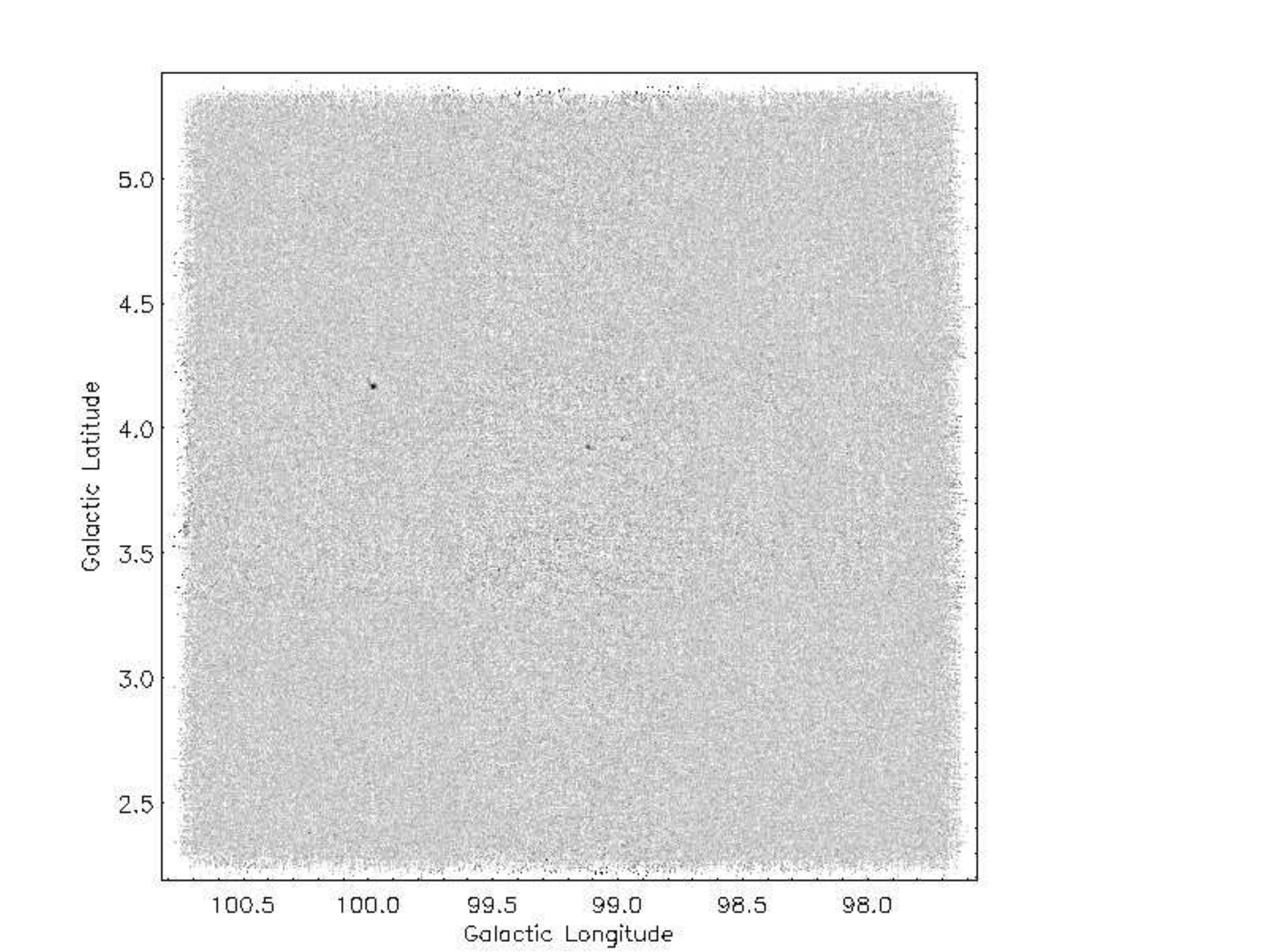} \caption{The IC1396 region.  In
  spite of copious CO emission here, there are only two faint sources
  detected in 9 square degrees.}
\end{center}
\end{figure}

\addtocounter{figure}{-1}
\addtocounter{subfig}{1}

\begin{figure}
\begin{center}
  \includegraphics[scale=1.0]{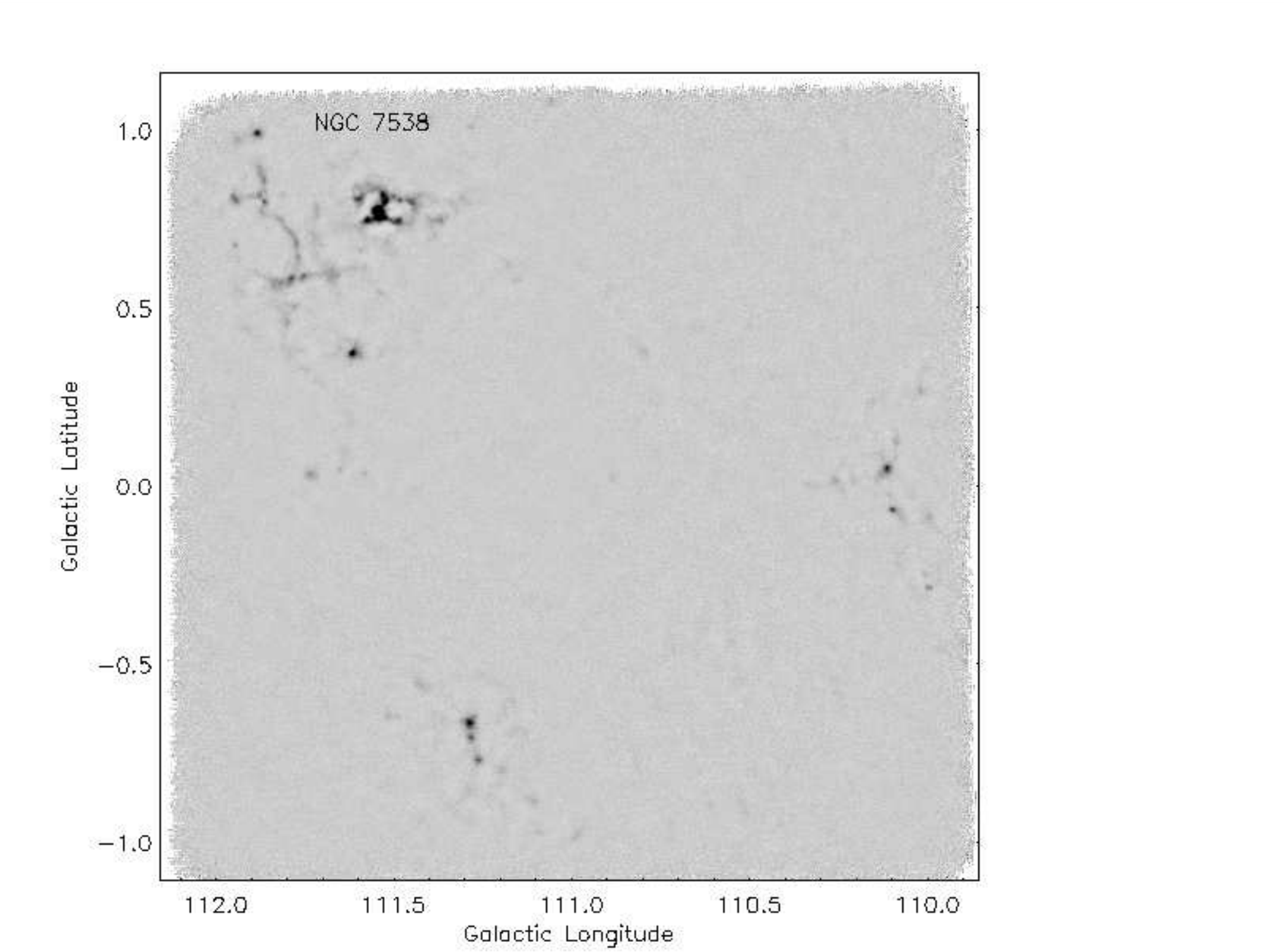}
  \caption{Cloud complexes centered at $\lon=111$ in the Perseus Arm. 
  The NGC 7538 complex is in the upper-left.}
\end{center}
\end{figure}

\addtocounter{figure}{-1}
\addtocounter{subfig}{1}

\begin{figure}
\begin{center}
  \includegraphics[scale=1.0]{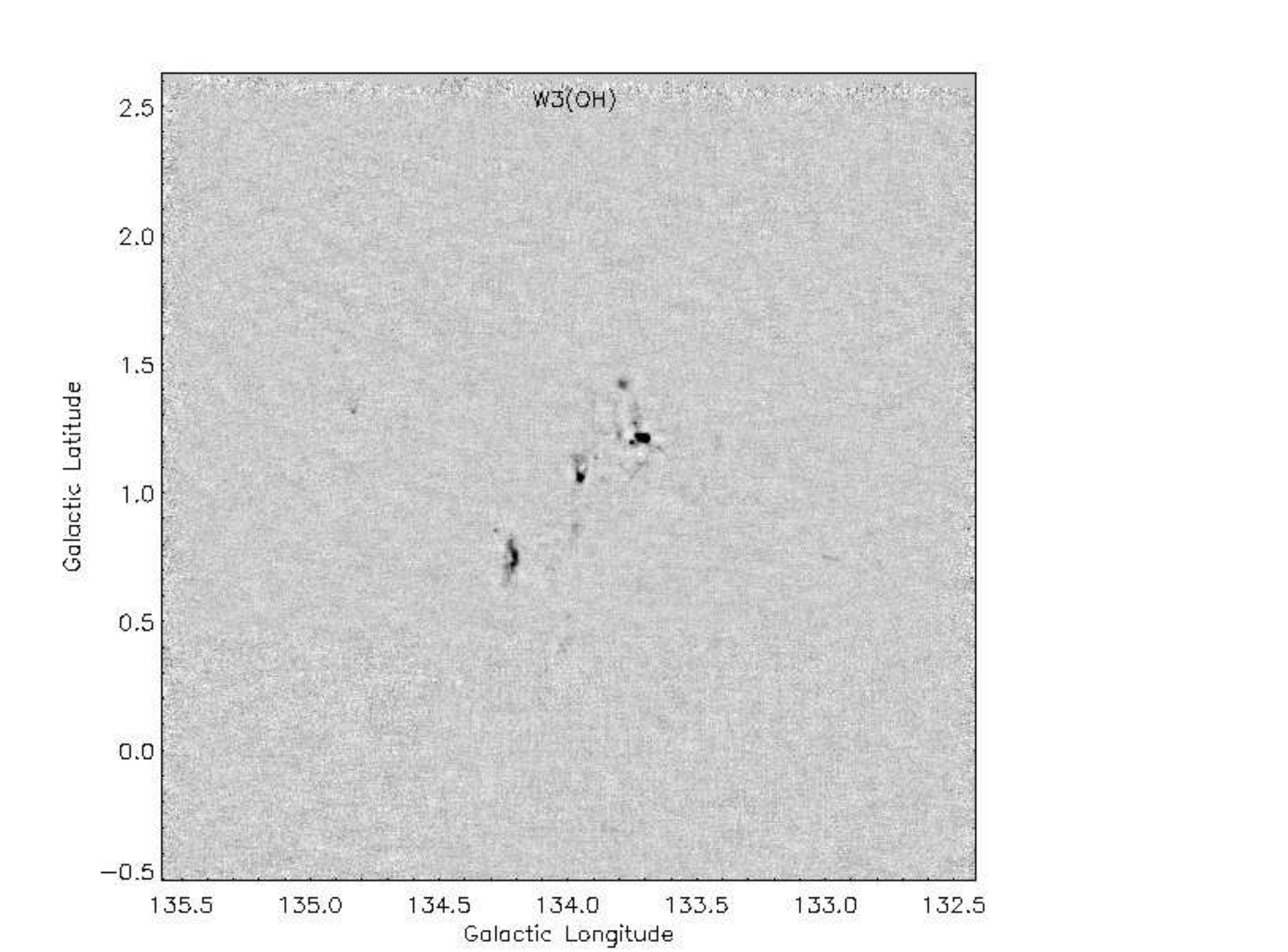}
  \caption{W3.  The W3(OH)/W3 Main complex is the bright source
  on the right side of the image.}
\end{center}
\end{figure}

\addtocounter{figure}{-1}
\addtocounter{subfig}{1}

\begin{figure}
\begin{center}
  \includegraphics[scale=1.0]{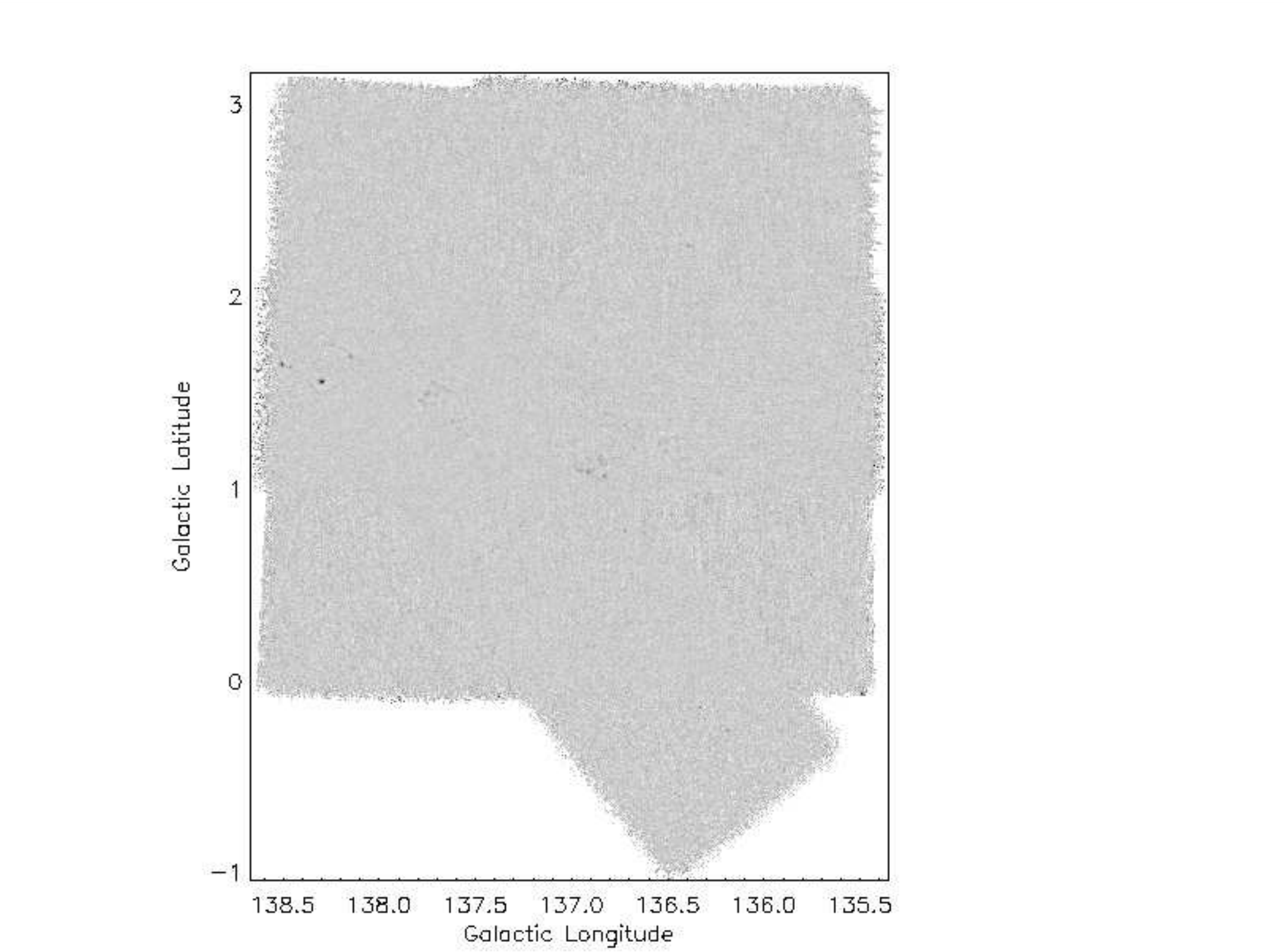} \caption{W4/5.  W5 has one scan
  performed in RA/Dec instead of Galactic coordinates and so has
  non-uniform noise properties in each square degree}
\end{center}
\end{figure}

\addtocounter{figure}{-1}
\addtocounter{subfig}{1}

\begin{figure}
\begin{center}
  \includegraphics[scale=0.7]{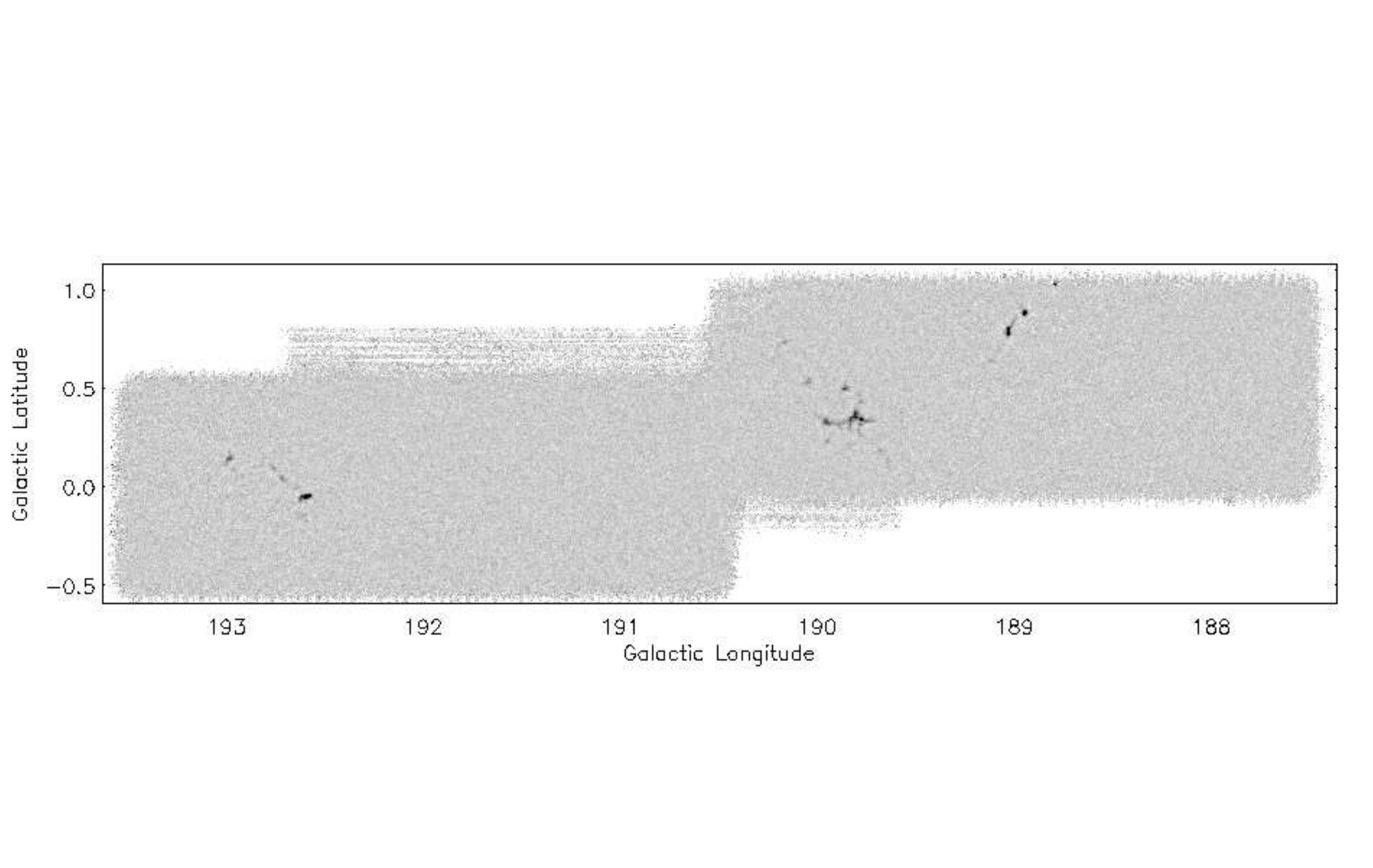} \caption{Gem OB1.  This region
  has been thoroughly surveyed in \ammonia, and is discussed in
  \citet{dunham10}.}
\end{center}
\end{figure}

\renewcommand{\thefigure}{\arabic{figure}}

\clearpage

\begin{figure}
  \begin{center}
  \includegraphics[scale=0.8]{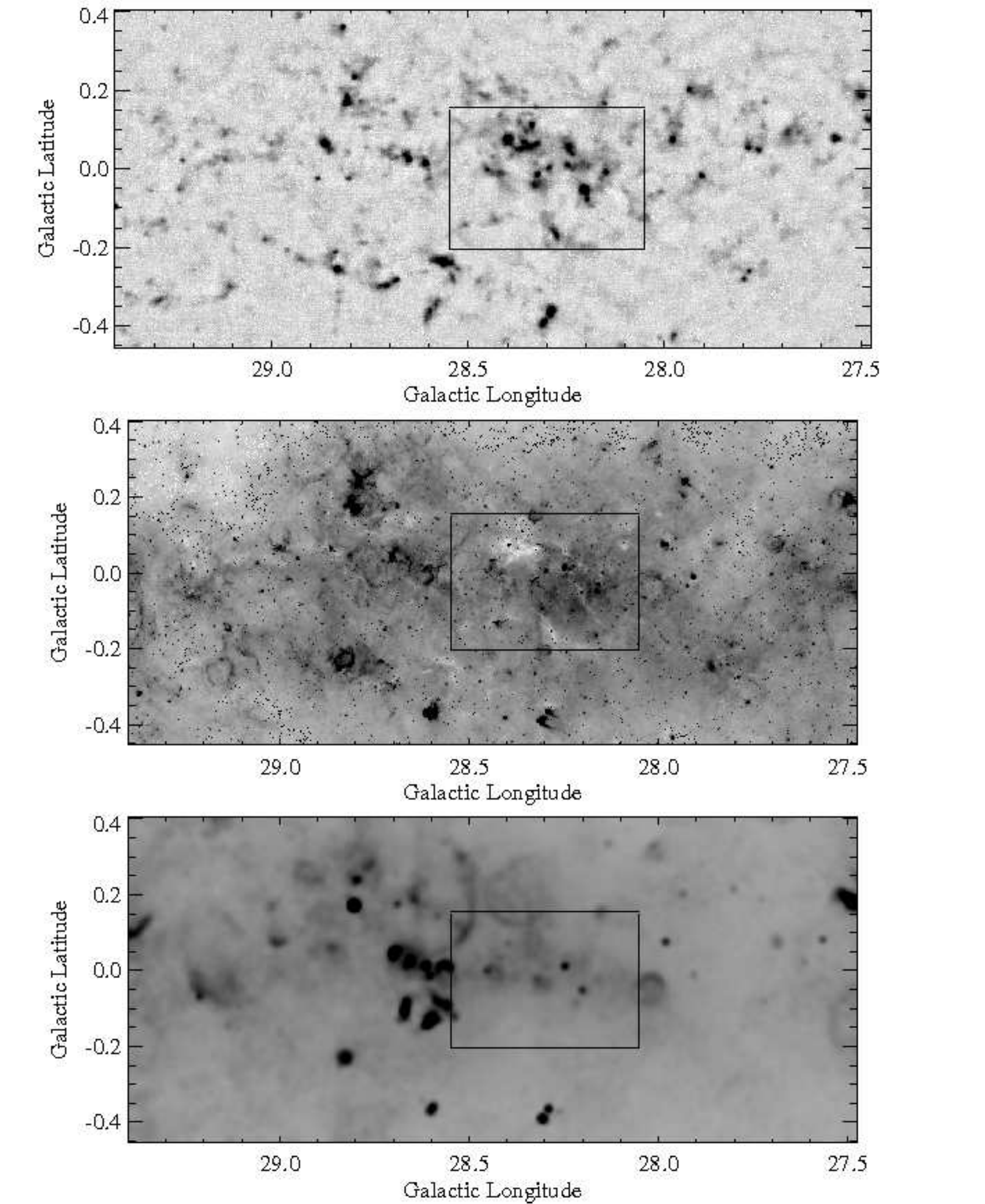}
  \end{center}

   \caption{A view of two square degrees of the Galactic Plane
    centered at $\lon=28.45$, $b=0$.  Top: BGPS, Middle: {\em
    Spitzer}-GLIMPSE 8 \mum\ (1\farcs9), Bottom: VGPS 20 cm continuum
    (40\arcsec).  IRDCs appear in the {\em Spitzer} negative image as
    wispy white regions against the grey emission background.  While
    some of the Bolocam sources are apparent as IRDCs or strong
    infrared or radio sources, many lack any apparent association with
    features at these wavelengths.  The box indicates the zoom-in
    region in Figures \ref{fig:HII} and \ref{fig:IRDC}.}

  \label{fig:Multiwavelength}

\end{figure}

\Figure{f21} {A zoom-in of the region shown in Figure
\ref{fig:Multiwavelength}, showing the VGPS image in inverted
greyscle, with contours of BGPS emission in black.  Selected sources
from the \Ros\ catalog are marked with X's and their catalog numbers.
Note that while \hii\ regions appear at or near the peaks of
millimeter emission for clumps 3913, 3922, 3935, 3955, and 3971,
others have no associated emission (3897, 3921, 3923).}{fig:HII}{1.0}

\Figure{f22} {A zoom-in of the region shown in Figure
\ref{fig:Multiwavelength}, showing the {\em Spitzer}-GLIMPSE image in
inverted greyscle, with contours of BGPS emission in black.  IRDCs
appear as white in this image.  Selected sources from the \Ros\
catalog are marked with X's and their catalog numbers.  Note the high
degree of correspondence between the IRDC morphology and the
millimeter emission for BGPS sources 3923 and 3955.}{fig:IRDC}{1.0}

\Figure{f23} {The filling factor of millimeter continuum emission from
the BGPS compared to CO($1\to0$) emission from \citet{jackson06}.  The
axis on the left side gives the filling fraction above 3$\sigma$ of
the local RMS from the BGPS (solid line).  The right axis gives the CO
filling fraction above 50 (dotted) and 100 (dashed) K km
s$^{-1}$.}{fig:FillingFactor}{1.0}

\Table
{lll}
{Observing Epochs for the BGPS}
{Number & Begin (UT) & End (UT)}
{tab:Observing}
{
I   & 2005 Jul 03 & 2005 Jul 09 \\ 
II  & 2005 Sep 05 & 2005 Sep 12 \\ 
III & 2006 Jun 02 & 2006 Jun 30 \\ 
IV  & 2006 Sep 03 & 2006 Sep 19 \\ 
V   & 2007 Jul 01 & 2007 Jul 25 \\ 
VI  & 2007 Sep 04 & 2007 Sep 09    
}

\Table
{ll}
{Point Source Aperture Corrections}
{\Ros Aperture & Recommended Correction}
{tab:ApertureCorrections}
{
40\arcsec & 1.46 \\
80\arcsec & 1.04 \\
120\arcsec & 1.00 \\
object integrated & 1.00 
}

\Table
{llll}
{Flux comparison with M07 and M09}
{Comparison & Factor & Factor & Factor \\
Survey & $>3~\mjysr$ & $>10~\mjysr$ & $>20~\mjysr$}
{tab:FluxComparison}
{
M07 & 1.32 & 1.25 & 1.21 \\
M09 & 1.51 & 1.44 & 1.38 
}

\TableNote
{lllll} {Comparison of band centers, color corrections, and flux ratios
for Bolocam and MAMBO.}
{         & Bolocam       & $K$                          & MAMBO         & $S_{\rm BOLOCAM}/$ \\
 $\alpha$\tablenotemark{a}  & $\nu_c$ [GHz] & Eq. \ref{eq:ColorCorrection} & $\nu_c$ [GHz] & $S_{\rm MAMBO}$\tablenotemark{b}}
{tab:ColorCorrections}
{---\tablenotemark{c} & 271.1 & 1.000 & 262.0 & 1.124 \\
\tableline
\rule{-3pt}{8pt}
1.0 & 267.8 & 0.988 & 254.8 & 1.051 \\
1.5 & 268.6 & 0.986 & 256.4 & 1.072 \\
2.0 & 269.4 & 0.987 & 258.0 & 1.090 \\
2.5 & 270.0 & 0.990 & 259.5 & 1.104 \\
3.0 & 270.7 & 0.996 & 261.1 & 1.115 \\
3.5 & 271.4 & 1.004 & 262.5 & 1.124 \\
4.0 & 272.2 & 1.015 & 264.0 & 1.130 \\
4.5 & 272.9 & 1.030 & 265.4 & 1.134 \\
5.0 & 273.8 & 1.049 & 266.8 & 1.137
}
{
\tablenotetext{a}{The spectral index as defined by Equation \ref{eq:PowerLaw}.}
\tablenotetext{b}{The expected ratio of flux densities, assuming
identical apertures, the passbands of Figure \ref{fig:Bandpass}, and
the SED indicated.}
\tablenotetext{c}{The fiducial SED is taken to be that of Equation
\ref{eq:Greybody} with $T=20$~K, $\beta=1.8$, and unity optical depth
at 100 \mum.} 
}
\clearpage

\end{document}